\documentclass[tightenlines,twocolumn,superscriptaddress,preprintnumbers,amssymb,nofootinbib]{revtex4-1}
\usepackage{amsmath, amsthm, amssymb, mathrsfs, bm, mathtools} 
\usepackage{graphics}
\usepackage{epsfig}
\usepackage{graphicx}
\usepackage{wrapfig}
\usepackage{color}
\usepackage{tikz}
\usepackage[hidelinks,colorlinks=true,linkcolor=blue,citecolor=blue,colorlinks = true,urlcolor  = blue,]{hyperref}

\RequirePackage{subfigure}
\usetikzlibrary{arrows,shapes}
\usetikzlibrary{trees}
\usetikzlibrary{matrix,arrows} 			
\usetikzlibrary{positioning}				
\usetikzlibrary{calc,through}				
\usepackage{pgffor}							

\usetikzlibrary{decorations.pathmorphing}	
\usetikzlibrary{decorations.markings}
\tikzset{
    sigmaCT/.style={draw=black, postaction={decorate},
        decoration={markings,mark=at position .99 with {\arrow[draw=black]{>}},mark=at 		 position .99 with {\arrow[
draw=black]{<}}}},
    pionCT/.style={dashed,draw=black, postaction={decorate},
        decoration={markings,mark=at position .99 with {\arrow[draw=black]{>}},mark=at position .99 with {\arrow[draw=black]{<}}}},    
    fermionCT/.style={draw=black, postaction={decorate},
        decoration={markings,mark=at position .5 with {\arrow[draw=black]{>}},mark=at position .99 with {\arrow[draw=black]{>}},mark=at position .99 with {\arrow[draw=black]{<}}}},    
    fermion/.style={draw=black, postaction={decorate},
        decoration={markings,mark=at position .55 with {\arrow[draw=black]{>}}}},
    fermionbar/.style={draw=black, postaction={decorate},
        decoration={markings,mark=at position .55 with {\arrow[draw=black]{<}}}},
    pion/.style={dashed,draw=black, postaction={decorate}},
    sigma/.style={draw=black, postaction={decorate}}
}

\newcommand{\beq}{\begin{equation}}
\newcommand{\eeq}{\end{equation}}
\newcommand{\bqa}{\begin{eqnarray}}
\newcommand{\eqa}{\end{eqnarray}}

\newcommand{\os}{\text{\tiny OS}}
\newcommand{\ms}{\overline{\text{\tiny MS}}}
\newcommand{\stb}{\overline{\text{\tiny SB}}}
\newcommand{\pqmvt}{\text{\tiny PQMVT}}

\newcommand{\cep}{\text{\tiny CEP}}

\newcommand{\Tr}{{\mathrm{Tr}}}

\begin{document}

\title{Thermodynamics and phase diagrams of the Polyakov quark-meson model with on-shell versus curvature mass parameter fixing}
\author{Suraj Kumar Rai}
\email{surajrai050@gmail.com}
\affiliation{Department of Physics, University of Allahabad, Prayagraj, India-211002}
\affiliation{Department of Physics, Acharya Narendra Deo Kisan P.G. College, Babhnan Gonda, India-271313}
\author{Vivek Kumar Tiwari}
\email{vivekkrt@gmail.com}
\affiliation{Department of Physics, University of Allahabad, Prayagraj, India-211002}
\date{\today}

\begin{abstract}
    The Quantum Chromodynamics (QCD) phase structure has been studied using the Polyakov-loop augmented quark-meson model (PQM) in the extended mean field approximation (e-MFA) where the quark one-loop vacuum term is included.~When the divergent vacuum term is regularized in the minimal subtraction scheme and the curvature meson masses are used to fix the parameters,~the Polyakov quark-meson model with the vacuum term (PQMVT) becomes inconsistent as the curvature masses are determined by calculating the self energies at zero momentum.~The above inconsistency is remedied by the on-shell parameter fixing when the pion decay constant and the pole masses of the mesons are put into the relation of the couplings and running mass parameter by using the on-shell and the minimal subtraction renormalization scheme.~Combining the modified chiral effective potential of the on-shell renormalized quark-meson model (RQM) with the Polyakov-loop potential that mimics the physics of the confinement-deconfinement transition,~we get the renormalized Polyakov quark-meson (RPQM) model.~The phase diagrams and the thermodynamics details for the PQM, PQMVT and RPQM model, have been computed and compared for different forms of the Polyakov-loop potentials with and without the quark back-reaction.~The results have also been compared with the available lattice QCD data.~The so called quarkyonic phase region in the phase diagram, where the chiral symmetry is restored but the quarks and anti-quarks are still confined,~gets reduced by the quark back-reaction in the unquenched Polyakov-loop potential.~It altogether disappears for the chemical potential dependent parameter $T_{0} \equiv T_{0} (\mu)$ in the Log or the PolyLog-glue form of the Polyakov-loop potential in the RPQM model.  
\end{abstract}
\maketitle
\section{Introduction}
The strong interaction theory, quantum chromodynamics (QCD) indicates that under the extreme conditions of high temperatures and/or densities, there should be a phase transition from the normal hadronic matter to a collective form of matter known as the Quark Gluon Plasma (QGP) \cite{Cabibbo75,SveLer,Mull,Ortms,Riske}.~The general properties of such a hot and dense matter are summarized in the QCD phase diagram \cite{Cabibbo75} which can be probed by the ultrarelativistic heavy ion collision experiments like the RHIC (BNL), LHC (CERN) and the upcoming CBM experiments at the FAIR facility (GSI-Darmstadt).~The study of the QCD thermodynamics and its phase structure is a very active area of current research as several issues are not yet settled. One gets important information and insights regarding the QCD phase transition from the first-principle lattice QCD simulations \cite{AliKhan:2001ek,Digal:01,Karsch:02,Fodor:03,Allton:05,Karsch:05,Aoki:06,Cheng:06,Cheng:08,JLange} but these calculations get seriously compromised as the QCD action becomes complex due to the fermion sign problem \cite{Karsch:02} when the baryon density/chemical potential is nonzero. Hence one turns to the effective models \cite{Alf,Fukhat} for the study of the QCD thermodynamics and its phase diagram. 

The QCD Lagrangian has the global $ SU_{\tiny{L+R}}(2) \times SU_{L-R}(2) $ symmetry for the two flavor of massless quarks.~In the low energy hadronic vacuum of the QCD, the chiral (axial $A=L-R$) symmetry is spontaneously broken and one gets the chiral condensate as order parameter and the three massless pions as Goldstone modes.~The small explicit  
breaking of the chiral symmetry due to the light quark masses, gives a little mass to the physical pions which are the lightest among hadrons.~The $ SU_{L}(2) \times SU_{R}(2) $ linear sigma model \cite{Roder,fuku11,grahl} is a good framework to study the chiral symmetry breaking and restoring phase transition.~Coupling the iso-singlet $\sigma$, iso-triplet $\vec{a_{0}}$ scalar mesons together with the iso-singlet $\eta$, iso-triplet $\vec{\pi}$ pseudo-scalar mesons of the sigma model with the two flavor of quarks, one gets the QCD-like framework of the quark-meson (QM) model to study the QCD thermodynamics and its phase diagram in great detail.

The QCD phase structure/phase diagram, has already been studied in the chiral models \cite{Rischke:00,jakobi,Herpay:05,Herpay:06,Herpay:07,Kovacs:2006ym,kahara,Bowman:2008kc,Fejos,Jakovac:2010uy,koch,marko}, two and three flavor QM model \cite{scav, mocsy,bj,Schaefer:2006ds,Schaefer:09}. But the QM model in the standard mean field approximation (s-MFA) gives inconsistent result as the chiral phase transition at zero baryon densities becomes first-order in the chiral limit  which is at odds with the general theoretical considerations  \cite{rob,hjss}. This inconsistency is remedied by the proper treatment of the Dirac sea \cite{vac} after including the  quark one-loop vacuum fluctuation.~In the modified framework of the quark-meson model with the vacuum term (QMVT), several QCD phase structure studies      \cite{lars,guptiw,schafwag12,chatmoh1,TranAnd,vkkr12,chatmoh2,vkkt13,Herbst,Weyrich,kovacs,zacchi1,zacchi2,Rai}, regularized the divergent one-loop vacuum term in the minimal substraction scheme and after identifing the pion decay constant with the vacuum expectation value of the sigma mean field, fixed the model parameters using the curvature masses of the mesons. The above parameter fixing procedure turns out to be inconsistent as one notes that the effective potential is the generator of the n-point functions of the theory at vanishing external momenta and the curvature mass is defined by the evaluation of the self-energy at zero momentum \cite{laine,Adhiand1,BubaCar,Naylor,fix1}.


The radiative corrections to the physical quantities change their tree level relations to the parameters of the Lagrangian. Hence the use of tree level values of the parameters for the calculation of effective potential becomes inconsistent. One has to account for the renormalization scale $\Lambda$ dependence of the running parameters in the (modified) minimal subtraction $\overline{\text{MS}}$ scheme while the on-shell parameters have their tree-level values. Following the correct renormalization procedure one needs to calculate the counterterms both in the $\overline{\text{MS}}$ scheme and in the on-shell scheme and then connect the renormalized parameters of the two schemes. The effective potential is then calculated using the $\overline{\text{MS}}$ procedure where the relations between the running parameters and the on-shell parameters (physical quantities) are used as input \cite{Adhiand1}.~Using the abovementioned renormalization prescription Adhikari and collaborators \cite{Adhiand1,Adhiand2,Adhiand3,asmuAnd} correctly accounted for the effect of Dirac sea in the QM model whose $O(4)$ sigma model has the  $\sigma$ and  $\vec{\pi}$ as meson degrees of freedom.~In a very recent work \cite{RaiTiw}, we have applied the exact renormalization method for the on-shell parameter fixing to that version of the QM model in which the two flavor of quarks are coupled to the  
eight mesons (iso-singlet and iso-triplet combination of the $ \sigma$ and $ \vec{a_{0}} $  as scalars and the  $\eta$ and $ \vec{\pi} $ as pseudo-scalars ) of the $ SU_{L}(2) \times SU_{R}(2) $ sigma model and termed this setting as the renormalized quark-meson (RQM) model. 

The physics of quark confinement in the hadrons at low temperatures and densities is implemented by the introduction of the Polyakov-loop where the QCD confinement is mimicked in a statistical sense by coupling the chiral models to a  constant background $SU(N_{c})$ gauge field $A_{\mu}^{a}$ \cite{SveLer,Polyakov:78plb,
benji,BankUka,Pisarski:00prd,fuku,Vkt:06}. In such studies, the free energy density from the gluons is added to the QM model using the phenomenological Polyakov-loop potential \cite{ratti,Roesnr,fuku2}, and it becomes the PQM model \cite{SchaPQM2F,SchaPQM3F,Mao,TiPQM3F}.~In the present work, the modified chiral parts of the effective potentials for both the on-shell renormalized quark-meson model (RQM) and the curvature mass parameterized QMVT model, have been augmented with the physics of confinement/deconfinement transition by including the Polyakov-loop potential and the respective settings have been termed as the renormalized Polyakov quark-meson model (RPQM) and the PQMVT model. In this study, we have considered the important improvement of the Polyakov-loop potential from a pure gauge potential to a unquenched glue potential in which backreaction effects of the quarks are included \cite{Haas,Redlo,TkHerbst,BielichP}.~It is worthwhile to explore the consequences of coupling the improved chiral effective potential with the unquenched Polyakov-loop potential because it leads to the linkage of the chiral and deconfinement phase transitions also at small temperatures and large chemical potentials.~The abovementioned attribute  is also seen in the functional renormalization (FRG) improvement of the PQM model when Yang-Mills Polyakov-loop potential is used \cite{Herbst,THerbst2}.~We have computed the relative shift of the critical end point (CEP) and made the qualitative and quantitative  comparisons of the phase diagrams and thermodynamics in the RPQM and PQMVT models when different forms of the Polyakov-loop potentials are considered with and without quark back-reaction.

The paper is arranged as follows.~Section~\ref{sec:II} presents a brief formulation of the $ SU_{L}(2) \times SU_{R}(2) $ PQM model.~Section~\ref{sec:IIA} 
describes the different forms of the Polyakov-loop potentials while Section~\ref{sec:IIB} presents the thermodynamic grand potential in the PQM model.~Section~\ref{sec:III} gives a brief account of the Polyakov quark-meson model with the vacuum term (PQMVT).~The renormalized Polyakov quark-meson model (RPQM) effective potential has been presented in Section~\ref{sec:IV}.~Section~\ref{sec:V} gives results and discussion.~Section~\ref{sec:VA} discusses the order parameters and their temperature derivatives,~Section~\ref{sec:VB} describes the sigma mass and the model dependence of the phase diagrams,~Section~\ref{sec:VC} reports the results for the quarkyonic phase in the RPQM model,~Section~\ref{sec:VD} explains the computed results for the thermodynamic observables while the specific heat and the speed of sound are discussed in  Section~\ref{sec:VE}.~Section~\ref{sec:VI} presents the summary and conclusion.~ Appendix~\ref{appenA} presents the parameter fixing for the QMVT model.~The essential steps of the exact on-shell parameter fixing and the calculation of the quark one-loop effective potential in the large $N_{c}$ limit for the  renormalized quark-meson (RQM) model,~have been presented in Appendix~\ref{appenB}.~The integrals are presented in Appendix~\ref{appenC}.    


\section{Model Formulation}
\label{sec:II}
We will be combining the Polyakov-loop potential with the $ SU_{L}(2) \times SU_{R}(2) $ quark-meson model.
In this model a spatially constant temporal gauge field,~two light quarks and  $SU_V(2)\times SU_A(2)$ symmetric meson fields, are coupled together.~The Polyakov-loop field $\Phi$ is defined as the thermal expectation value of color trace of the Wilson loop in temporal direction.

\begin{equation}
\Phi= \frac{1}{N_c} \langle \Tr_c L(\vec{x})\rangle, \qquad \qquad  \bar\Phi =
\frac{1}{N_c} \langle \Tr_c L^{\dagger}(\vec{x}) \rangle\;,
\end{equation}
where $L$ is a matrix in the fundamental representation of the $SU_c(3)$ color gauge group,
\begin{equation}
\label{eq:Ploop}
L(\vec{x})=\mathcal{P}\mathrm{exp}\left[i\int_0^{\beta}d \tau
A_0(\vec{x},\tau)\right].
\end{equation}
Here $\mathcal{P}$ is path ordering,  $A_0$ is the temporal component of vector 
field and $\beta = T^{-1}$ \cite{Polyakov:78plb}. In accordance of
Ref. \cite{ratti,Roesnr},~we have considered a homogeneous Polyakov-loop field $\Phi(\vec{x})=\Phi$=constant and
$\bar\Phi(\vec{x})=\bar\Phi$=constant.
 
The model Lagrangian has quarks, mesons, couplings 
and Polyakov-loop potential ${\cal U} \left( \Phi, \bar\Phi, T \right)$ as :

\begin{equation}
\label{eq:Lag}
{\cal L}_{PQM} = {\cal L}_{QM} - {\cal U} \big( \Phi , \bar\Phi , T \big). 
\end{equation}

The Lagrangian of the model \cite{Roder,fuku11,grahl} is written as
\bqa
\nonumber
{\cal L}_{QM}&=&\bar{\psi}[i\gamma^\mu \partial_\mu-g t_0(\sigma+i\gamma_5 \eta)\\
&&-g \vec t\cdot(\vec a+i\gamma_5 \vec\pi)]\psi+\cal{L(M)},
\label{lag}
\eqa
where $\psi$ is a color $N_c$-plet, a four-component Dirac spinor and a flavor doublet, 
\bqa
\psi&=&
\left(
\begin{array}{c}
u\\
d
\end{array}\right)\;.
\eqa
The Lagrangian for meson fields is \cite{Roder} 
\bqa
\nonumber
\label{lag11}
\cal{L(M)}&=&\text{Tr} (\partial_\mu {\cal{M}}^{\dagger}\partial^\mu {\cal{M}}-m^{2}({\cal{M}}^{\dagger}{\cal{M}}))\\
\nonumber
&&-\lambda_1\left[\text{Tr}({\cal{M}}^{\dagger}{\cal{M}})\right]^2-\lambda_2\text{Tr}({\cal{M}}^{\dagger}{\cal{M}})^2\\
&&+c[\text{det}{\cal{M}}+\text{det}{\cal{M}}^\dagger]+\text{Tr}\left[H({\cal{M}}+{\cal{M}}^\dagger)\right]\;,
\label{lag1}
\eqa
here, the field ${\cal{M}}$ is a complex $2\times2$ matrix
\bqa
\label{mesmat}
{\cal{M}}=t_0(\sigma+i\eta)+\vec t\cdot(\vec a+i\vec\pi)
\eqa
\ \ \ \ \ \ \ \ \ \ \ \ \ \ \ \ \ \  with $t_0=\dfrac{1}{2}\left(
\begin{array}{c c}
1 & 0 \\
0 & 1
\end{array}\right),$ $t_1=\dfrac{1}{2}\left(
\begin{array}{c c}
0 & 1 \\
1 & 0
\end{array}\right),$ \\ 
\ \ \ \ \ \ \ \ \ \  \ \ \ \ \ \ \ \ \ \ $ t_2=\dfrac{1}{2}\left(
\begin{array}{c c}
0 & -i \\
i & 0
\end{array}\right),$ $t_3=\dfrac{1}{2}\left(
\begin{array}{c c}
1 & 0 \\
0 & -1
\end{array}\right).$ 

One can rewrite the Lagrangian (\ref{lag11})  in the form \cite{fuku11}
\bqa
\nonumber
{\cal{L(M)}}&=&\mbox{$1\over2$}(\partial_\mu\sigma\partial_\mu\sigma+\partial_\mu\vec{\pi}\cdot\partial_\mu\vec{\pi}+\partial_\mu\eta\partial_\mu\eta \\
&&+\partial_\mu\vec{a_0}\cdot\partial_\mu\vec{a_0})-U\;,
\eqa
further 
\bqa
\nonumber
U&=&\dfrac{m^2}{2}(\sigma^2+{\vec\pi}^2+\eta^2+{\vec a_0}^2)-\dfrac{c}{2}(\sigma^2-\eta^2+{\vec\pi}^2-{\vec a_0}^2)\\
\nonumber
&&+\dfrac{1}{4}\left(\lambda_1+\dfrac{1}{2}\lambda_2\right)(\sigma^2+{\vec\pi}^2+\eta^2+{\vec a_0}^2)^2\\
\nonumber
&&+\dfrac{\lambda_2}{2}\left((\sigma^2+{\vec\pi}^2)(\eta^2+{\vec a_0}^2)-(\sigma\eta-{\vec\pi}\cdot{\vec a_0})^2\right)\\
&&-h\sigma.
\eqa
The $2\times2$ matrix $H$ explicitly breaks the chiral symmetry and is chosen as 
\bqa
H=t_a h_a\;,
\eqa
where $h_a$ are external fields.
The field $\sigma$ acquires nonzero vacuum expectation value (VEV),~$\overline \sigma$, due to the spontaneous breaking of the chiral symmetry,~while the other scalar and pseudo-scalar fields ($\vec a_0 ,\vec \pi,\eta$) assume zero VEV.~Here the two parameters $h_0$ and $h_3$ may give rise to the explicit breaking of chiral symmetry.~We are neglecting the isospin symmetry breaking,~hence we choose $h_0\neq0$ and $h_3 = 0$.

The field $\sigma$ has to be shifted to $\sigma \xrightarrow \  \overline \sigma + \sigma  $ as it acquires nonzero VEV. At the tree level, the expression of the meson masses are \cite{Roder}
\bqa
\label{ch5_m1}
m^2_\sigma&=&m^2-c+3(\lambda_1+\frac{\lambda_2}{2})\overline  \sigma^2\;,\\
m^2_{a_0}&=&m^2+c+(\lambda_1+\frac{3\lambda_2}{2})\overline \sigma^2\;,\\
m^2_\eta&=&m^2+c+(\lambda_1+\frac{\lambda_2}{2})\overline \sigma^2\;,\\
m^2_\pi&=&m^2-c+(\lambda_1+\frac{\lambda_2}{2})\overline \sigma^2\;,\\
m_q&=& \frac{g \ \overline \sigma}{2}.
\label{ch5_m4}
\eqa
Using (\ref{ch5_m1})--(\ref{ch5_m4}),~the parameters of the Lagrangian (\ref{lag1}) are obtained as  
\bqa
\label{para1}
\lambda_1&=&\dfrac{m^2_\sigma+m^2_\eta-m^2_{a_0}-m^2_\pi}{2\overline \sigma^2}\;,\\
\lambda_2&=&\dfrac{m^2_{a_0}-m^2_{\eta}}{\overline \sigma^2}\;,\\
m^2&=&m^2_\pi+\dfrac{m^2_\eta-m^2_\sigma}{2}\;,\\
c&=&\dfrac{m^2_\eta-m^2_\pi}{2}\;,\\
\frac{g^2}{4}&=&\frac{m_{q}^2}{ \overline \sigma^2}\;.
\label{para4}
\eqa
and the tree level effective potential is written as,
\bqa
\label{effpot}
U(\overline \sigma)=\frac{1}{2}(m^2-c)\overline \sigma^2+\frac{1}{4}\left(\lambda_1+\frac{1}{2}\lambda_2\right)\overline \sigma^4-h\overline \sigma\;, \ \
\eqa
  $\overline \sigma=f_\pi$ gives the minimum of the effective potential at the tree level and
the stationarity condition for the potential (\ref{effpot}), gives 
\bqa
\label{para5}
h&=&f_\pi m^2_\pi.
\eqa

\subsection{Polyakov-loop potentials}
\label{sec:IIA}
There are different possibilities for the functional form of the effective Polyakov-loop potential $\mathcal{U}(\Phi,\bar{\Phi},T)$. Its simplest form is constructed by finding a potential which respects all given symmetries and includes the spontaneously broken $Z(3)$ symmetry for the system in the deconfined phase \cite{SveLer,benji,BankUka}. Thus the minimal content of a Polyakov-loop potential \cite{ratti} is given by the following polynomial form

\bqa
\label{plykov_poly}
\hspace{-0.5 cm}\frac{\mathcal{U_{\rm Poly}}}{T^4}&=&-\frac{b_2(T)}{2}\Phi\bar{\Phi}-\frac{b_3}{6}(\Phi^3+\bar{\Phi}^3)+\frac{b_4}{4}(\Phi\bar{\Phi})^2\;,
\eqa
the coefficients of the Eq.~(\ref{plykov_poly}) are given by
\bqa
b_2(T)=a_0+a_1\left(\frac{T_0}{T}\right)+a_2\left(\frac{T_0}{T}\right)^2+a_3\left(\frac{T_0}{T}\right)^3\;,
\eqa
where $a_0=6.75$, $a_1=-1.95$, $a_2=2.625$, $a_3=-7.44$, $b_3=0.75$ and $b_4=7.5$ .\\

The above Polyakov-loop potential ansatz has been enhanced by adding the contribution that results from the integration of the $SU(3)$ group volume in the generating functional for the Euclidean action. This integration is done by using the Haar measure and takes the form of a Jacobian determinant. Its logarithm is added as an effective potential to the action in the generating functional. The positive coefficient of the logarithm term bounds the potential from below for large $\Phi$ and $\bar{\Phi}$ and the logarithmic form of the Polyakov-loop potential is written as \cite{fuku,Roesnr}:
\bqa
\label{plykov_log}
\nonumber
\hspace{-0.5 cm}\frac{\mathcal{U_{\rm Log}}}{T^4}&=&b(T)\ln[1-6\Phi\bar{\Phi}+4(\Phi^3+\bar{\Phi}^3)-3(\Phi\bar{\Phi})^2]\;\\&&-\frac{1}{2}a(T)\Phi\bar{\Phi}\;.
\eqa
The parameters of the polynomial and log form of the Polyakov-loop potential were determined \cite{ratti,Roesnr} by fitting the lattice data for pressure, entropy density  as well as energy density and the evolution of Polyakov-loop $<\Phi>$ on the lattice in pure gauge theory. The coefficients of the Eq.~(\ref{plykov_log}) are the following \cite{Roesnr},
\bqa
&&a(T)=a_0+a_1\left(\frac{T_0}{T}\right)+a_2\left(\frac{T_0}{T}\right)^2\;,\\
&&b(T)=b_3\left(\frac{T_0}{T}\right)^3\;,
\eqa
where $a_0=3.51$, $a_1=-2.47$, $a_2=15.2$, $b_3=-1.75$. 
Note that the log potential has qualitative consistency with the leading order result of the strong-coupling expansion \cite{JLange}. Also, since the potential diverges for $\Phi$,~$\bar{\Phi} \longrightarrow$  1, the Polyakov-loop always remains smaller than 1 and approaches this value asymptotically as $T \longrightarrow \infty$.

Ref.~\cite{Redlo} took into account the Polyakov-loop fluctuations and constructed the new Polyakov-loop effective potential in which the parameters are so adjusted that apart from the other existing lattice data, the lattice data for the longitudinal as well as the transverse susceptibilities are also reproduced. They enhanced the polynomial form of the Polyakov-loop potential with the addition of the logarithmic term to  arrive at the following expression of the PolyLog Polyakov-loop potential,
\bqa
\label{plykov_polylog}
\nonumber
\hspace{-0.5 cm}\frac{\mathcal{U_{\rm PolyLog}}}{T^4}&=&b(T)\ln[1-6\Phi\bar{\Phi}+4(\Phi^3+\bar{\Phi}^3)-3(\Phi\bar{\Phi})^2]\;\\ \nonumber
&&+a_2(T)\Phi\bar{\Phi}+a_3(T)(\Phi^3+\bar{\Phi}^3)+a_4(T)(\Phi\bar{\Phi})^2.\\
\eqa
The coefficients of the Eq.~(\ref{plykov_polylog}) PolyLog parametrization are defined as
\bqa
&&a_i(T)=\frac{a^{(i)}_0+a^{(i)}_1\left(\frac{T_0}{T}\right)+a^{(i)}_2\left(\frac{T_0}{T}\right)^2}{1+a^{(i)}_3\left(\frac{T_0}{T}\right)+a_4^{(i)}\left(\frac{T_0}{T}\right)^2}\;\\
&&b(T)=b_0\left(\frac{T_0}{T}\right)^{b_1}\left[1-e^{b_2\left(\frac{T_0}{T}\right)^{b_3}}\right].
\eqa
\\
The parameters are summarized in the Table~\ref{tab:plglg}.
\begin{table}[!htbp]
    \caption{Parameters of the PolyLog Polyakov-loop potential have been taken from the Ref.~\cite{Redlo}.}
    \label{tab:plglg}
    \resizebox{0.48\textwidth}{!}{
    \begin{tabular}{p{1.5cm} p{1.5cm} p{1.5cm} p{1.5cm} p{1.5cm} p{1.5cm} p{0.5 cm}}
      \toprule 
      PolyLog& $a^{(2)}_0$ & $a^{(2)}_1$ & $a^{(2)}_2$ & $a^{(2)}_3$ & $a^{(2)}_4$&\\
      & 22.07 & -75.7 & 45.03385 & 2.77173 & 3.56403&\\
      & $a^{(3)}_0$ & $a^{(3)}_1$ & $a^{(3)}_2$ & $a^{(3)}_3$ & $a^{(3)}_4$&\\
      &-25.39805&57.019&-44.7298&3.08718&6.72812\\
      & $a^{(4)}_0$ & $a^{(4)}_1$ & $a^{(4)}_2$ & $a^{(4)}_3$ & $a^{(4)}_4$&\\
      &27.0885&-56.0859&71.2225&2.9715&6.61433&\\
      &$b_0$&$b_1$&$b_2$&$b_3$& &\\
      &-0.32665&5.8559&-82.9823&3.0& &\\
      \hline      
      \hline 
    \end{tabular}}
\end{table}

The deconfinement phase transition is first order for the pure gauge Yang-Mills theory and $T_c^{\rm YM}=T_0=270$ MeV. The first order transition turns to a crossover in the presence of dynamical quarks. The parameter $T_0$ depends on the number of quark flavors and chemical potential in the full dynamical QCD~\cite{SchaPQM2F,Haas,kovacs,BielichP,THerbst2} as it is linked to the mass-scale $\Lambda_{\rm QCD}$ which gets modified by the effect of the fermionic matter fields.  $T_0 \longrightarrow T_0(N_f,\mu)$ is written as,

\bqa
\label{t0_mu}
T_0(N_f,\mu)=\hat{T} \ e^{-1/(\alpha_0 b(N_f,\mu))}\;,
\eqa
with 
\bqa
b(N_f,\mu)=\frac{1}{6\pi}(11N_c-2N_f)-b_\mu\frac{\mu^2}{(\hat{\gamma}\hat{T})^2}.
\eqa
where the parameter $\hat{T}$ is fixed at the scale $\tau$,  $\hat{T}=T_\tau=1.77$ GeV and $\alpha_0=\alpha(\Lambda)$ at a UV scale $\Lambda$. The $T_0(N_f=0)$ = 270 MeV gives  $\alpha_0$ = 0.304 and $b_\mu\simeq\frac{16}{\pi}N_f$. The parameter $\hat{\gamma}$ governs the curvature of $T_0(\mu)$ with the systematic error estimation range $0.7\lesssim\hat{\gamma}\lesssim1$ \cite{SchaPQM2F,THerbst2}. In our calculation, we have taken $\hat{\gamma}=1$. The $N_f$, $\mu$ dependence of the $T_0$ accounts only partially for the unquenching of the pure gauge Polyakov-loop potential to an effective glue potential in QCD \cite{BielichP}.    

For the full QCD with dynamical quarks, the Polyakov-loop potential should be replaced by the QCD glue potential that accounts for the back-reaction of quarks into the Polyakov-loop effective potential. Applying the FRG equations to the QCD, Ref.~\cite{Haas} compared the pure gauge potential $\mathcal{U}_{\rm YM}$ to the ``glue'' potential $\mathcal{U}_{\rm glue}$ where quark polarization was included in the gluon propagator and they found significant differences between the two potentials.~However, it was observed in their study that the two potentials have the same shape and they can be mapped into each other by relating the temperatures of the two systems, $T_{\rm YM}$ and $T_{\rm glue}$.~Denoting the previous equations of the Polyakov-loop potential by $\mathcal{U}_{\rm YM}$, the improved Polyakov-loop potential $\mathcal{U}_{\rm glue}$ can be constructed as \cite{Haas}     

\bqa
\frac{\mathcal{U}_{\rm glue}}{T^4_{\rm glue}}(\Phi, \bar{\Phi}, T_{\rm glue})&=&\frac{\mathcal{U}_{\rm YM}}{T^4_{\rm YM}}(\Phi, \bar{\Phi}, T_{\rm YM})
\eqa
here the temperature $T_{\rm glue}$ is related to $T_{\rm YM}$ as 
\bqa
\frac{T_{\rm YM}-T^{\rm YM}_{\rm c}}{T^{\rm YM}_{\rm c}}=0.57 \
\frac{T_{\rm glue}-T^{\rm glue}_{\rm c}}{T^{\rm glue}_{\rm c}}
\eqa
The $T^{\rm glue}_{\rm c}$ is the transition temperature for the unquenched case. The coefficient 0.57 comes from the comparison of the two effective potentials. $T^{\rm glue}_{\rm c}$ lies within a range $T^{\rm glue}_{\rm c} \in [180,270]$. In practice, we use in the right-hand side of the Polyakov-loop potentials where $T_0$ means $T^{\rm YM}_{\rm c}$, the replacement $T  \longrightarrow T^{\rm YM}_{\rm c}(1+0.57(\frac{T}{T^{\rm glue}_{\rm c}}-1))$ where ($T\sim T_{\rm YM}$) on the left side of the arrow and ($T\sim T_{\rm glue}$) on the right side. In our calculations, we will keep $T^{\rm glue}_{\rm c}$ and $T_0$ both fixed at 208 MeV and also considered the chemical potential dependence of $T_0$ and $T^{\rm glue}_{\rm c}$. The $\mu$ dependence of the $T^{\rm glue}_{\rm c}$ can be found from the Eq.~(\ref{t0_mu}) after making the replacement $T_0 (N_f,\mu)  \longrightarrow {T^{\rm glue}_{\rm c}}(N_f,\mu)$. 

\subsection{Thermodynamic grand potential in PQM model}
\label{sec:IIB}
In the mean-field approximation, the thermodynamic grand
potential for the PQM model is given as~\cite{SchaPQM2F}

\bqa
\nonumber
\Omega_{\text{\tiny MF}}( T,\mu;\overline {\sigma},\Phi,\bar{\Phi})  &=& U(\overline {\sigma} )+
  \Omega_{q\bar{q}} ( T,\mu;\overline {\sigma},\Phi,\bar{\Phi})\;\\
&&+\mathcal{U}( T,\Phi,\bar{\Phi})\;.
\label{Omega_MF}
\eqa
The quark/antiquark contribution is given by
\bqa
\label{vac1}
&&\Omega_{q\bar{q}} (T,\mu;\overline{\sigma},\Phi,\bar{\Phi}) = \Omega_{q\bar{q}}^{vac}+\Omega_{q\bar{q}}^{T,\mu}(\overline{\sigma},\Phi,\bar{\Phi})\;,\\
\label{vac2}
&&\Omega_{q\bar{q}}^{vac} =- 2 N_c\sum_q  \int \frac{d^3 p}{(2\pi)^3} E_q \theta( \Lambda_c^2 - \vec{p}^{2})\;,\\
&&\Omega_{q\bar{q}}^{T,\mu}(\overline{\sigma},\Phi,\bar{\Phi})=- 2 N_c\sum_q \int \frac{d^3 p}{(2\pi)^3} T \left[ \ln g_q^{+}+\ln g_q^{-}\right].\;
\label{vac3}
\eqa
The first term of the Eq.~(\ref{vac1}) denotes the fermion vacuum
contribution, regularized by the ultraviolet cutoff $\Lambda_c$. In  presence of the Polyakov-loop potential, the $g^{+}_q$ and $g^{-}_q$  are specified by the trace in the color space.
\bqa
\hspace{-1 cm}g^{+}_q&=&\left[1+3\Phi e^{-E_{q}^{+}/T}+3\bar{\Phi}e^{-2E_{q}^{+}/T}+e^{-3E_{q}^{+}/T}\right]\;,\qquad\\
g_q^{-}&=&\left[1+3\bar{\Phi} e^{-E_{q}^{-}/T}+3\Phi e^{-2E_{q}^{-}/T}+e^{-3E_{q}^{-}/T}\right]\;.
\eqa
E$_{q}^{\pm} =E_q \mp \mu_{q} $ and $E_q=\sqrt{p^2 + m{_q}{^2}}$ is the
flavor dependent single particle energy of quark/antiquark and $m_q=\dfrac{g \overline{\sigma}}{2}$ is the mass of the given quark flavor.~$\mu_q$ is quark chemical potential.

Neglecting the quark one-loop vacuum term of the Eq.~(\ref{vac1}) in the the standard mean-field approximation (s-MFA),~the PQM model grand potential is written as,

\bqa
\nonumber
 \Omega_{\text{\tiny PQM}}(T,\mu;\overline{\sigma},\Phi,\bar{\Phi})&=&U(\overline {\sigma} )+\Omega_{q\bar{q}}^{T,\mu}(\overline{\sigma},\Phi,\bar{\Phi})+\mathcal{U}( T,\Phi,\bar{\Phi})\;.
 \\
 \eqa
 
 \begin{equation}
     \frac{\partial \Omega_{\text{\tiny PQM}}}{\partial
      \overline{\sigma}} =\frac{\partial \Omega_{\text{\tiny PQM}}}{\partial {\Phi}}=\frac{\partial \Omega_{\text{\tiny PQM}}}{\partial
      \bar{\Phi}} \bigg|_{\overline{\sigma},\Phi,\bar{\Phi}}=0.
\label{PQMsta}
\end {equation}
The global minima of the grand potential in Eq.~(\ref{PQMsta}) gives the $\overline{\sigma}$, $\Phi$ and $\bar{\Phi}$ as a function of temperature and chemical potential. 
\section{PQM model with Vacuum Term}
\label{sec:III}
Here, we give a brief description of the effective potential calculation when the quark one-loop vacuum divergence of the Eq.~(\ref{vac1}) is regularized in the minimal subtraction scheme and the $\sigma$ and $\pi$ meson curvature masses are used for fixing the model parameters. The quark one-loop vacuum contribution of Eq.~(\ref{vac2}) is written as \cite{RaiTiw}  

\bqa
\label{vacdiv}
\Omega^{vac}_{q\bar q}&=&\frac{N_c}{(4 \pi)^2}\sum_q m^4_q\left[\frac{1}{\epsilon}+\frac{3}{2}+\ln\left(\Lambda^2\over m^2_q \right)\right]\;,
\eqa
where $\Lambda$ is renormalization scale.

Adding the counterterm $\delta \mathcal{L}=\frac{N_c}{(4 \pi)^2}\sum_q \frac{m^4_q}{\epsilon}$ to the Lagrangian (\ref{lag}), the thermodynamic potential gets renormalized. After replacing the first term of Eq.~(\ref{vac1}) by the $\Omega^{vac}_{q\bar q}=\frac{N_c}{(4 \pi)^2}\sum_q m^4_q\left[\frac{3}{2}+\ln\left(\Lambda^2\over m^2_q \right)\right]$, one gets the renormalization scale dependent chiral part of the vacuum ($\mu=0$ and $T=0$) effective potential as
\bqa
\label{qmvtomegavac}
\Omega^{\Lambda}(\overline{\sigma})&=&U(\overline{\sigma})+\Omega^{vac}_{q\bar q}\;.
\eqa
The fixing of the model parameters $m^2$, $c$, $\lambda_1$, $\lambda_2$ and $h$ is presented in the  Appendix (\ref{appenA}). When the calculated new parameters are substituted in the Eq.(\ref{qmvtomegavac}) and terms are rearranged, one finds the expression of renormalization scale $\Lambda$ independent vacuum effective potential as: 
\bqa
\nonumber
\Omega(\overline{\sigma})&=&\frac{1}{2}\left(m^2_{s}-\dfrac{N_cg^4f^2_\pi}{2(4\pi)^2}\right)\overline{\sigma}^2-\frac{1}{2}c\overline{\sigma}^2+\frac{1}{4}\left(\lambda_1+\frac{\lambda_{2s}}{2}\right. \\
&&\left.+\dfrac{3N_cg^4}{4(4\pi)^2}\right)\overline{\sigma}^4-h\overline{\sigma}+\frac{N_cg^4\overline{\sigma}^4}{8(4\pi)^2}\ln\left(\dfrac{f^2_\pi}{\overline{\sigma}^2}\right)\;.
\eqa
After quark one-loop vacuum correction, the thermodynamic grand potential for the Polyakov-loop enhanced quark-meson model with vacuum term (PQMVT) can be written as :

\bqa
\label{qmvtomega}
\nonumber
\Omega_{\rm \pqmvt}(T,\mu;\overline{\sigma},\Phi,\bar{\Phi})&=&\Omega(\overline{\sigma})+\Omega^{T,\mu}_{q\bar q}(\overline{\sigma},\Phi,\bar{\Phi})+\mathcal{U}(\Phi,\bar{\Phi})\;,\\
\eqa
\begin{equation}
\frac{\partial \Omega_{\pqmvt}}{\partial
      \overline{\sigma}} =\frac{\partial \Omega_{\pqmvt}}{\partial {\Phi}}=\frac{\partial \Omega_{\pqmvt}}{\partial
      \bar{\Phi}} \bigg|_{\overline{\sigma},\Phi,\bar{\Phi}}=0.
\label{EoMMF2}
\end {equation}
One gets the quark condensate $\overline{\sigma}$  and the Polyakov-loop expectation values $\Phi$, $\bar{\Phi}$ by searching the global minima of the grand potential at a given $T$ and $\mu$. It is pointed out that the pion decay constant $f_\pi $ does not get renormalized because the dressing of the meson propagator is not considered in fixing of the model parameters using  the curvature mass of the pion. The minimum of the effective potential in vacuum ($T=0$, $\mu=0$) remains fixed at $ \overline{\sigma}=f_\pi $.

\section{Renormalized PQM Model}
\label{sec:IV}
The PQMVT/QMVT model investigations \cite{lars,guptiw,schafwag12,chatmoh1,TranAnd,vkkr12,chatmoh2,vkkt13,Herbst,Weyrich,kovacs,zacchi1,zacchi2,Rai} use the curvature (or screening) masses of the mesons to fix the parameters while the pion decay constant $f_{\pi}$ remains unrenormalized.~We know that the poles of the mesons propagators,~give their physical masses and the $f_{\pi}$ gets correction as it is related to the residue of the pion propagator at its pole \cite{BubaCar,Naylor,fix1}.~The definition of the meson curvature masses, involves the evaluation of their self-energies at zero momentum as the effective potential is the generator of the n-point functions of the theory at zero external momenta \cite{laine,Adhiand1,Adhiand2,Adhiand3}.~Furthermore the pole definition is the physical and gauge invariant one \cite{Kobes,Rebhan}.~If the Dirac sea contributions are neglected,~the pole masses and the curvature  masses for mesons become equivalent
but when the quark one-loop vacuum correction is considered,~the curvature masses of the mesons become different from their pole masses \cite{BubaCar,fix1}.~The above inconsistency is removed in the exact on-shell parameter fixing method of the renormalized quark-meson (RQM) model where the physical (pole) masses of the mesons and the pion decay constant,~are put into the relation of the running mass parameter and couplings by using the on-shell and the minimal subtraction renormalization schemes.~The derivation of the quark one-loop effective potential in the large $N_{c}$ limit and 
the mathematical details of the on-shell parameter fixing for the  renormalized quark-meson (RQM) model,~have been presented in the Appendix~\ref{appenB}.~Combining the RQM model chiral 
effective potential in $\overline{\text{MS}}$ scheme  with the Polyakov-loop potential, we get grand thermodynamic potential for the  RPQM model as,
\bqa
\label{omegarqm}
\nonumber
\Omega_{\text{\tiny RPQM}}(T,\mu;\Delta,\Phi,\bar{\Phi})&=&\Omega_{vac}(\Delta)+\Omega^{T,\mu}_{q\bar q}(\Delta,\Phi,\bar{\Phi})+\mathcal{U}(\Phi,\bar{\Phi}),\\
\eqa
using the expression $\Omega_{vac}(\Delta)$ of Eq.~\ref{rqmeff} given in the Appendix~\ref{appenB},~we write the full thermodynamic potential for the RPQM model as,
\begin{widetext}
\bqa
\nonumber
\label{rqmomega}
\Omega_\text{RPQM}(T,\mu;\Delta,\Phi,\bar{\Phi})&=&\frac{(3m^2_\pi-m^2_\sigma)f^2_{\pi}}{4}\left\lbrace 1-\frac{N_cg^2}{(4\pi)^2}\left(\mathcal{C}(m^2_\pi)+m^2_\pi\mathcal{C}^\prime(m^2_\pi)\right)\right\rbrace\frac{\Delta^2}{m^2_q}\; \\ \nonumber
&&+\frac{N_cg^2f^2_\pi}{2(4\pi)^2}\left\lbrace\frac{3m^2_\pi\mathcal{C}(m^2_\pi)-(m^2_\sigma-4m^2_q)\mathcal{C}(m^2_\sigma)}{2}-2m^2_q\right\rbrace\frac{\Delta^2}{m^2_q}\; \\ \nonumber 
&&+\frac{(m^2_\sigma-m^2_\pi)f^2_\pi}{8}\left\lbrace1-\frac{N_cg^2}{(4\pi)^2}\left(\mathcal{C}(m^2_\pi)+m^2_\pi\mathcal{C}^\prime(m^2_\pi)\right)\right\rbrace\frac{\Delta^4}{m^4_q}\; \\ \nonumber
&&+\frac{N_cg^2f^2_\pi}{(4\pi)^2}\left[(m^2_\sigma-4m^2_q)\mathcal{C}(m^2_\sigma)-m^2_\pi\mathcal{C}(m^2_\pi)\over 8\right]\frac{\Delta^4}{m^4_q}+\frac{2N_c\Delta^4}{(4\pi)^2}\left\lbrace\frac{3}{2}-\ln\left(\frac{\Delta^2}{m^2_q}\right)\right\rbrace\; \\ \nonumber
&&-m^2_\pi f^2_\pi\left\lbrace 1-\frac{N_cg^2}{(4\pi)^2}m^2_\pi\mathcal{C}^\prime(m^2_\pi)\right\rbrace\frac{\Delta}{m_q}+\Omega_{q\bar{q}}^{T,\mu}(\Delta,\Phi,\bar{\Phi})+\mathcal{U}(\Phi,\bar{\Phi}); \\ 
\eqa
\end{widetext}
 \begin{equation}
     \frac{\partial \Omega_{\text{\tiny RPQM}}}{\partial
      \Delta} =\frac{\partial \Omega_{\text{\tiny RPQM}}}{\partial {\Phi}}=\frac{\partial \Omega_{\text{\tiny RPQM}}}{\partial
      \bar{\Phi}} \bigg|_{\Delta,\Phi,\bar{\Phi}}=0.
\label{}
\end {equation}
Searching the global minima of the grand potential for a given $\mu$ and $T$,~one gets the quark condensate 
$\overline{\sigma}$ ($\Delta=\frac{g_{\ms} \ \overline{\sigma}_{\ms} }{2}$),~the Polyakov-loop fields $\Phi$ and $\bar{\Phi}$.~It has been explained in the Appendix~\ref{appenB} that the  minimum of the vacuum ($T=0$, $\mu=0$) effective potential remains at $\overline{\sigma}_{\ms}=f_{\pi}$.~In our calculations, we have used the $m_\pi=138.0$ MeV, $m_{a_0}=984.7$ MeV and $m_\eta=547.0$ MeV.~The Yukawa coupling $g=6.5$ and pion decay constant $f_\pi=93.0$ MeV.~The constituent quark mass in the vacuum $m_q=\frac{gf_\pi}{2} = 302.25 $ MeV.



\section{Results and Discussion}

The temperature axis chiral crossover and confinement-deconfinement crossover transition at $\mu=0$,~has been thoroughly investigated and compared with, in different model scenarios of combining, the chiral effective potential computed in different parameter fixing schemes to the different forms of the Polyakov-loop potential with and without quark back-reaction.~The subsection A compares the results for the temperature variations of the chiral condensate,~the Polyakov-loop condensate and their derivatives.~The chiral  and deconfinement crossover transition temperatures for different model scenarios have been computed and presented for comparison in Table~\ref{tab:tablefig1} and Table~\ref{tab:tablefig2}.~In the subsection B,~the PQM, PQMVT and RPQM model phase diagrams have been plotted and compared with each other for different values of $m_\sigma$ and different forms of the Polyakov-loop potentials with and without quark back-reaction.~The appearance and disappearance of the quarkyonic phase in different model scenarios, have been discussed in the subsection C.~In order to understand the $\mu=0$ chiral and deconfinement transition occurring at the temperature axis, we have computed and compared the reduced scale temperature variations of the thermodynamic quantities namely the pressure, entropy density,~energy density and interaction measure in the subsection D while the results for the specific heat, $C^2_s$ and $P/\epsilon$ have been presented in the subsection E.~We have also compared the results obtained for thermodynamic observables with the available lattice QCD data.      

\label{sec:V}
\begin{table}[!htbp]
    \caption{Pseudo-critical temperatures for the $m_\sigma=500 \text{MeV}$ at $\mu=$0.}
    \label{tab:tablefig1}
    \begin{tabular}{p{2cm} p{2.5cm} p{1.5cm} p{1.5cm}}
      \toprule 
      Polyakov-loop & Models & $T^\chi_c(\text{MeV})$ & $T^\Phi_c(\text{MeV})$ \\
      \hline 
      \hline
       & PQM & $172.1$ & $168.9$ \\
      Log & PQMVT & $187.1$ & $168.9$ \\
       & RPQM & $168.6$ & $167.9$ \\ \hline
       & PQM & $165.6$ & $165.4$ \\
      PolyLog-glue & PQMVT & $186.3$ & $156.1$ \\
      & RPQM  & $175.8$ & $154.8$\\
      \hline 
    \end{tabular}
\end{table}

\begin{table}[!htbp]
    \caption{Pseudo-critical temperatures for the $m_\sigma=500 \text{MeV}$ in the RPQM  model at $\mu=$0.}
    \label{tab:tablefig2}
    \begin{tabular}{p{3cm} p{2cm} p{2cm}}
      \toprule 
      Models &$T^\chi_c(\text{MeV})$ & $T^\Phi_c(\text{MeV})$ \\
      \hline 
      \hline
      Log & $168.6$ & $167.9$ \\
      Log-glue & $174.6$ & $145.2$ \\
      Poly & $181.6$ & $176.6$ \\
      Poly-glue & $174.8$ & $159.8$ \\
      PolyLog & $180.1$ & $175.1$ \\
      PolyLog-glue & $175.8$ & $154.8$\\
      \hline 
    \end{tabular}
\end{table}

\subsection{Order parameters and their derivatives}
\label{sec:VA}

\begin{figure*}[htb]
\subfigure[\ Normalized chiral condensate.]{
\label{fig1a} 
\begin{minipage}[b]{0.48\textwidth}
\centering \includegraphics[width=\linewidth]{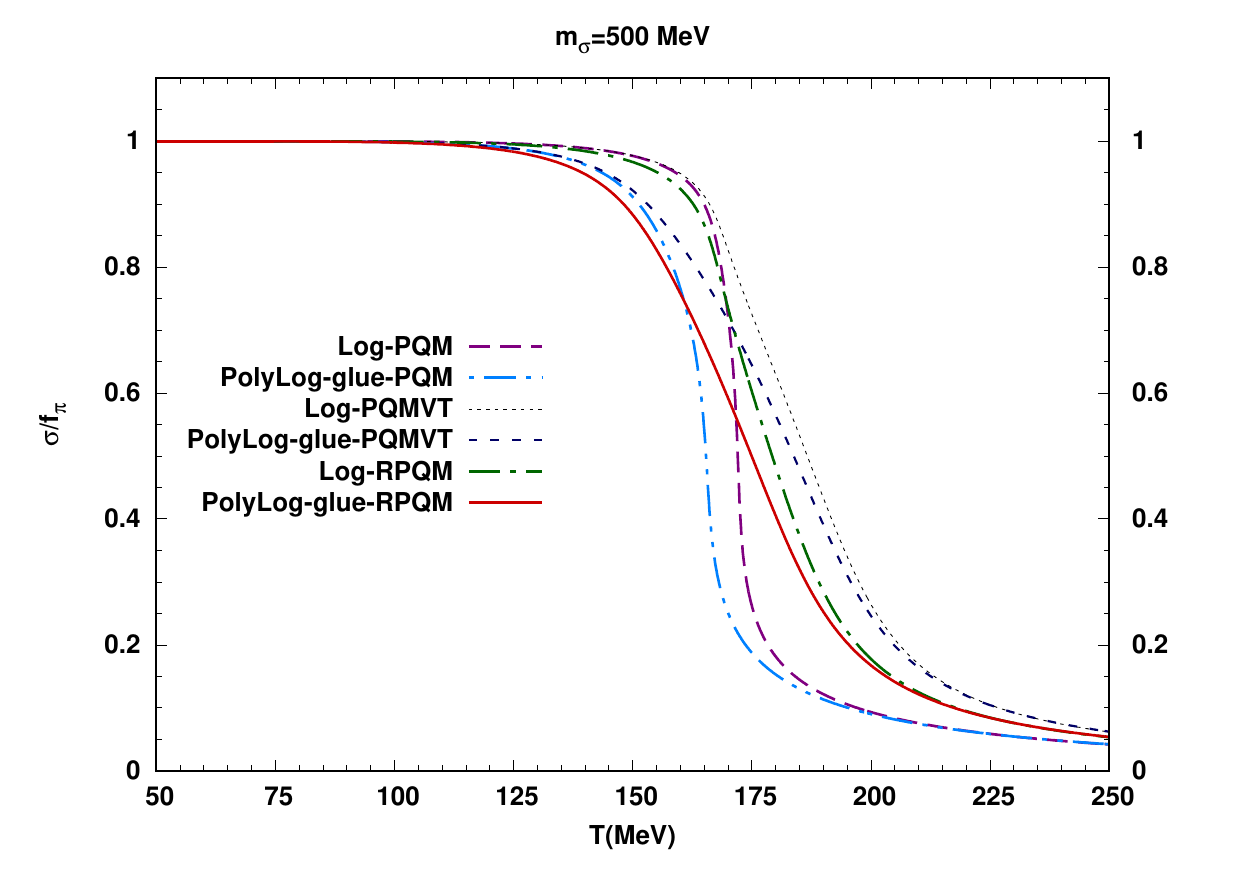}
\end{minipage}}
\hfill
\subfigure[\ Polyakov-loop order parameter.]{
\label{fig1b} 
\begin{minipage}[b]{0.48\textwidth}
\centering \includegraphics[width=\linewidth]{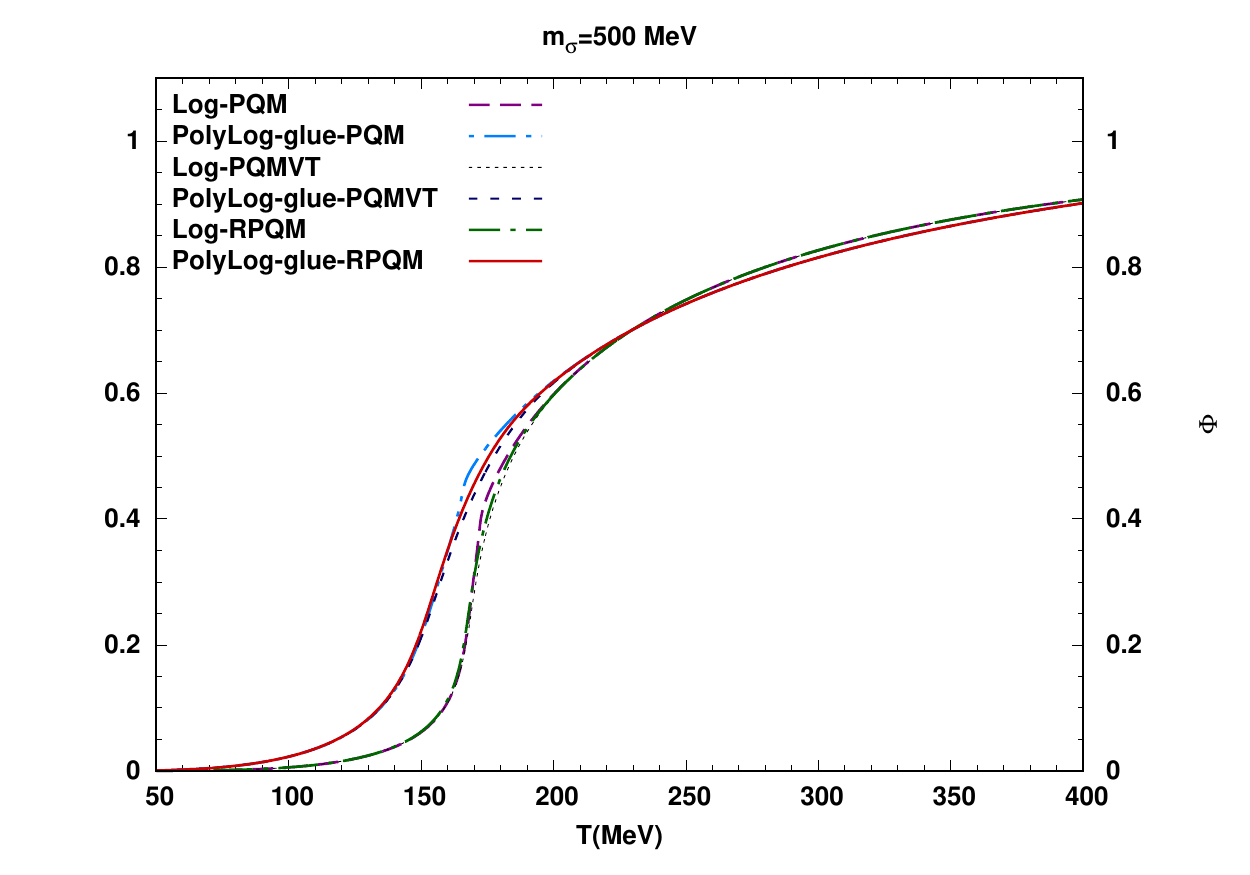}
\end{minipage}}
\caption{The normalized chiral and Polyakov-loop order parameters for the $m_\sigma$ =500 MeV and $\mu=0$ at $T^{\rm glue}_c = T_0=208 \ \text{MeV} \ $.}
\label{fig:mini:fig1} 
\end{figure*}

\begin{figure*}[htb]
\subfigure[\ ]{
\label{fig2a} 
\begin{minipage}[b]{0.32\textwidth}
\centering \includegraphics[width=\linewidth]{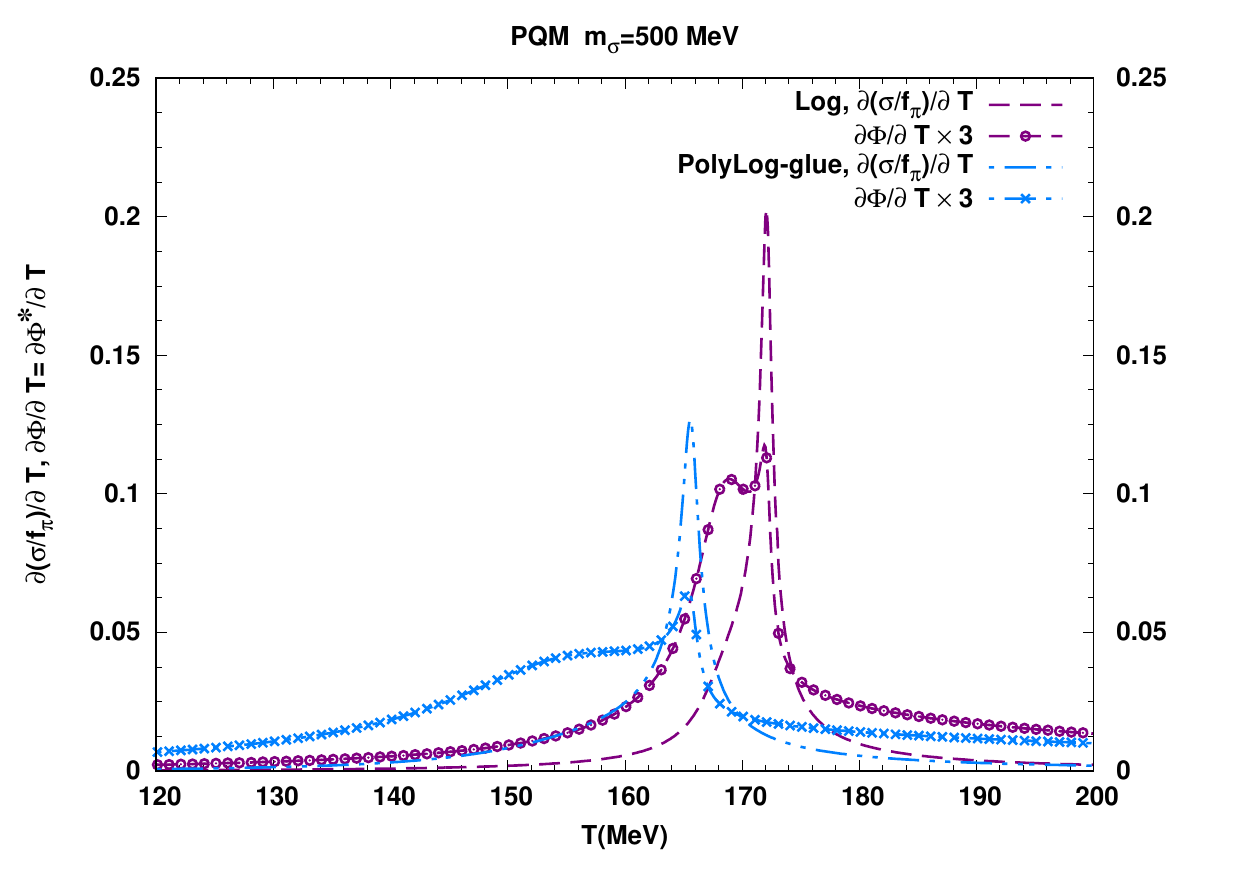}
\end{minipage}}
\hfill
\subfigure[\ ]{
\label{fig2b} 
\begin{minipage}[b]{0.32\textwidth}
\centering \includegraphics[width=\linewidth]{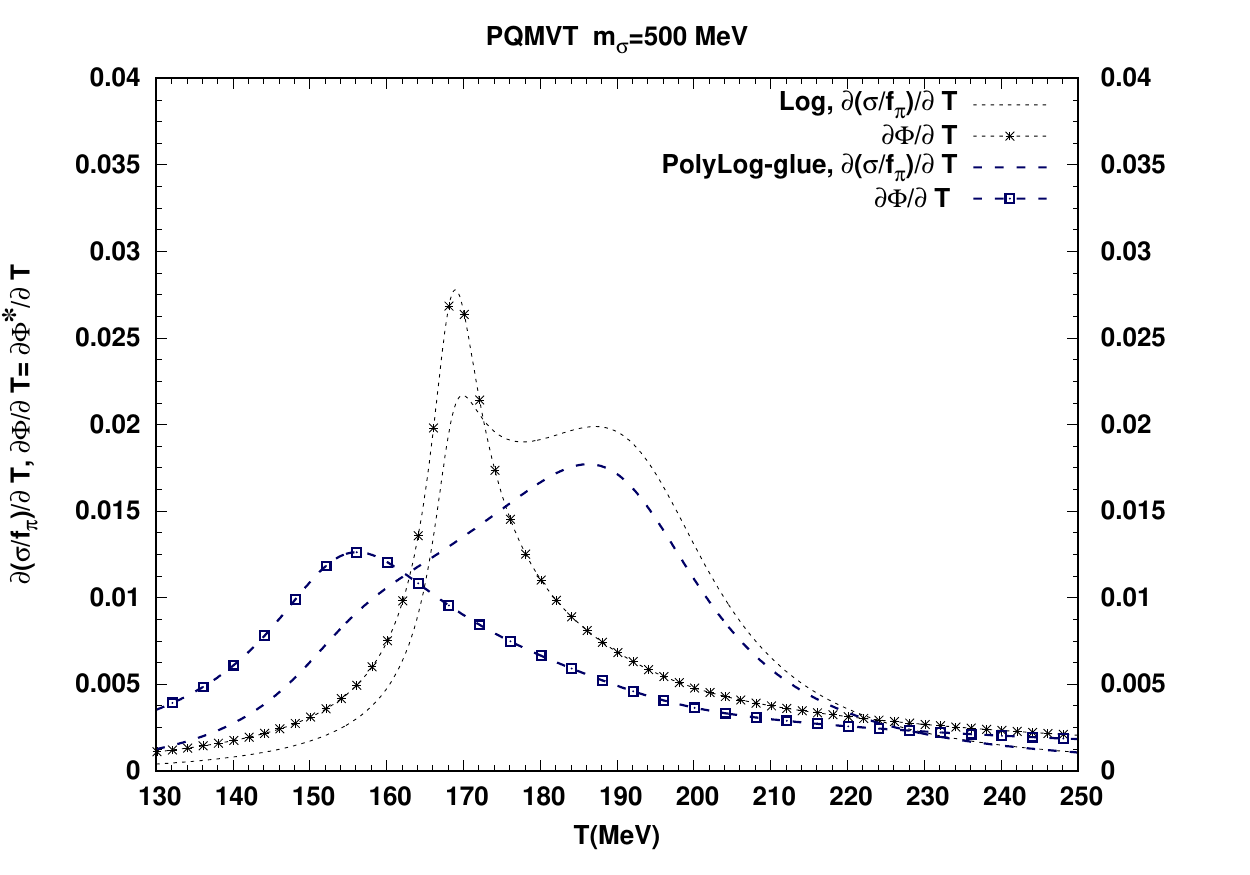}
\end{minipage}}
\hfill
\subfigure[\ ]{
\label{fig2c} 
\begin{minipage}[b]{0.32\textwidth}
\centering \includegraphics[width=\linewidth]{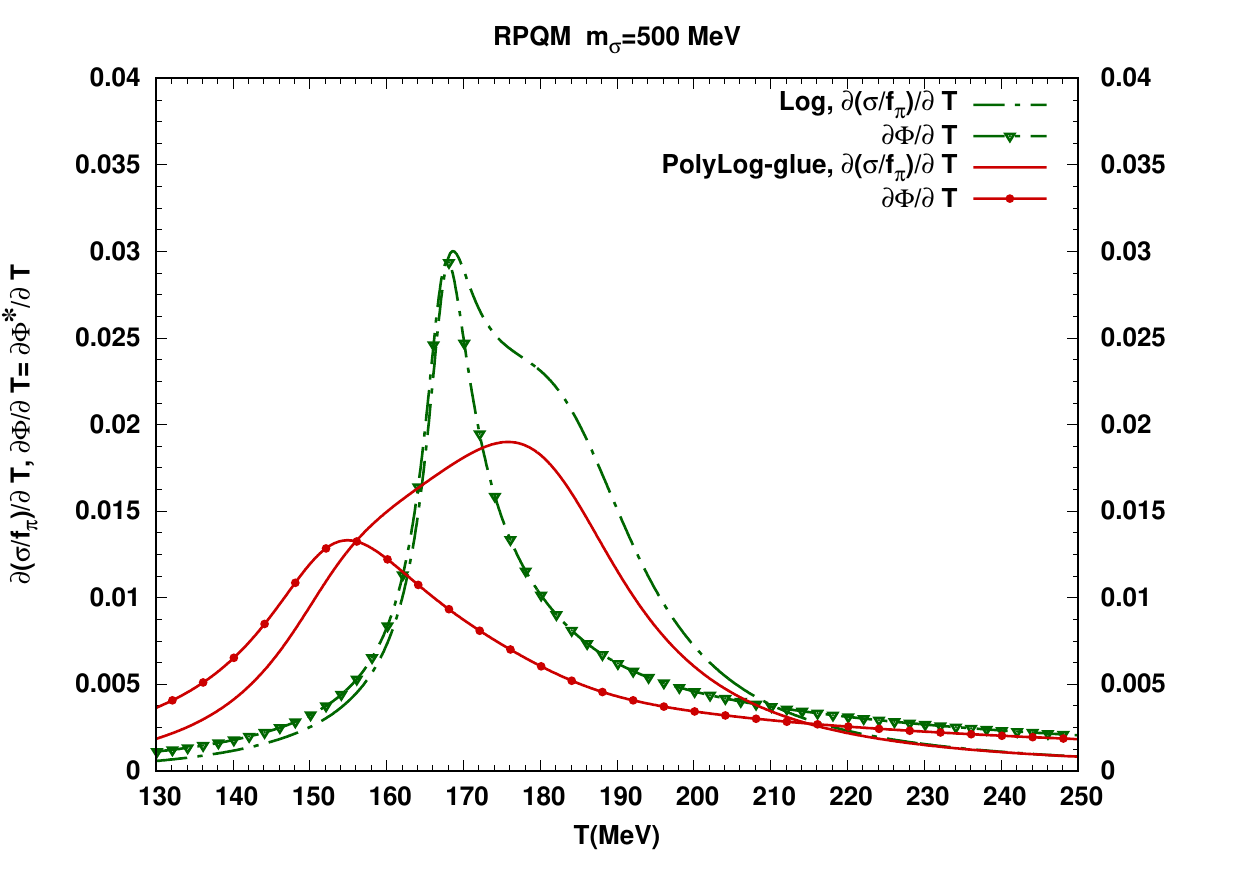}
\end{minipage}}
\caption{The derivatives of the normalized chiral and Polyakov-loop order parameters for the $\mu=$0 and $T^{\rm glue}_c = T_0=208 \ \text{MeV} \ $.}
\label{fig:mini:fig2} 
\end{figure*}
Fig.~\ref{fig1a} and Fig.~\ref{fig1b} present the respective temperature variations of the normalized chiral condensate ($\sigma/f_\pi$) and the Polyakov-loop condensate ($\Phi$) at $\mu=0$ for the sigma mass $m_\sigma$ = 500 MeV in the PQM, PQMVT and RPQM model for the Log and the PolyLog-glue form of the Polyakov-loop potential.~The term glue denotes the unquenching of the Polyakov-loop potential when the quark back-reaction has been taken into account.~Confirming the expected pattern when the models are augmented with the Log form of the Polyakov-loop potential,~the sharper Log-PQM model chiral transition becomes quite smooth and delayed  on account of the quark one-loop vacuum correction in the on-shell renormalized Log-RPQM model while the curvature mass parametrization for the Log-PQMVT model gives rise to an excessively smooth and very delayed variation of the chiral order parameter on the temperature axis in the Fig.~\ref{fig1a}.~When the physics of the confinement-deconfinement transition is coupled with the physics of the chiral transition,~the unquenching of the Polyakov-loop potential in the presence of the quark back-reaction,~leads to the significant smoothing effect on the chiral condensate and shifts its temperature variations early on the temperature scale for the $T<190$ MeV in the PolyLog-glue : PQM, RPQM and PQMVT model.~Furthermore,~the Polyakov-loop condensate temperature variations are lifted up and get shifted early on the temperature scale for the $T<220$ MeV,~due to the effect of the quark back-reaction when compared to the corresponding PQM, RPQM and PQMVT model temperature variations with Log form of the Polyakov-loop potential in the Fig.~\ref{fig1b}.

Presenting the plots of the $\frac{\partial 
(\sigma/f_{\pi})}{\partial {T}}$ and $\frac{\partial 
\Phi}{\partial {T}}$ versus $T$ when the $\mu=0$ and the $m_{\sigma} =500$ MeV in the Fig.~\ref{fig2a}, Fig.~\ref{fig2b} and Fig.~\ref{fig2c} respectively for the PQM, PQMVT and RPQM model with the Log and PolyLog-glue form of the Polyakov-loop potential,~we have compared, how the PQM model results are changed by the effect of the quark one-loop vacuum correction in the curvature mass parameterized PQMVT model versus the on-shell parameterized RPQM model.~We have also compared how the presence of the quark back-reaction changes the results in one particular model.~For the Log-PQM model in the Fig.~\ref{fig2a},~the very sharp variation of the temperature derivative of the $\sigma / f_{\pi}$, drives a double peak structure in the temperature variation of the $\frac{\partial \Phi}{\partial {T}}$ whose first peak at lower temperature gives pseudocritical temperature $T^{\Phi}_c=168.9$ MeV for the confinement-deconfinement transition and its second peak coincides with the very sharp and high peak in the $\frac{\partial 
(\sigma/f_{\pi})}{\partial {T}}$ variation which gets located at the higher pseudocritical temperature for the chiral transition $T^{\chi}_c=172.1$ MeV and the two transitions are separated by the difference $T^{\chi}_c-T^{\Phi}_c=3.2$ MeV as given in the Table \ref{tab:tablefig1}.~One notices that the double peak structure for 
the $\frac{\partial \Phi}{\partial {T}}$ is smoothed out and the sharper chiral transition also becomes quite smooth as the peak heights get significantly reduced due to the effect of the quark back-reaction in the PolyLog-glue PQM model and the two transitions stand very close to each other having a difference of only 0.2 MeV as the $T^{\chi}_c=165.6$ MeV and the  $T^{\Phi}_c=165.4$ MeV.~As the fermionic vacuum correction with the curvature mass parametrization leads to 
excess smoothing of the chiral transition,~the $\frac{\partial(\sigma/f_{\pi})}{\partial {T}}$ variation for 
the Log PQMVT model shows a very smooth double peak structure (similar to Ref.~\cite{guptiw}) in the Fig.~\ref{fig2b}. Here,~in contrast to the Fig.~\ref{fig2a},~the influence of the Log form of the Polyakov-loop potential becomes dominant and the sharper temperature variation of the $\frac{\partial \Phi}{\partial {T}}$ generates a double peak for the $\frac{\partial (\sigma/f_{\pi})}{\partial {T}}$ variation and the separation between the chiral crossover and the confinement-deconfinement transition is equal to the 18.2 MeV as the $T^{\chi}_c=187.1$ MeV and the $T^{\Phi}_c=168.9$ MeV in Table \ref{tab:tablefig1}.~It is to be noted that due to the smoothing influence of the 
quark back-reaction for the unquenched PloyLog-glue PQMVT model,~the double peak of the 
$\frac{\partial (\sigma/f_{\pi})}{\partial {T}}$ variation, gets completely washed out and one gets largest separation of 30.2 MeV between the confinement-deconfinement and the chiral crossover transition temperatures as the $T^{\chi}_c=186.3$ MeV while the $T^{\Phi}_c=156.1$ MeV.~The smoothing influence of the 
quark one-loop vacuum correction,~becomes moderate due to the consistent on-shell parameter fixing for the 
RPQM model and one notices that the temperature variations 
of the $\frac{\partial \Phi}{\partial {T}}$ and the 
$\frac{\partial (\sigma/f_{\pi})}{\partial {T}}$ rise to almost the same height in the Fig.~\ref{fig2c} for the Log form of the Polyakov-loop potential in the RPQM model.~The $\frac{\partial (\sigma/f_{\pi})}{\partial {T}}$ variation falls short of developing a second peak and one finds 
that the chiral crossover transition at the $T^{\chi}_c=168.6$ MeV is very close to the confinement-deconfinement transition occurring at the $T^{\Phi}_c=167.9$ MeV.~Here also in the unquenched PolyLog-glue RPQM model,~the robust effect of the quark back-reaction leads to sufficient smoothing and reduction in the heights of the peaks of the $\frac{\partial \Phi}{\partial {T}}$ and $\frac{\partial (\sigma/f_{\pi})}{\partial {T}}$ temperature variations and a moderate separation of 21.0 MeV is noted between the confinement-deconfinement and the chiral crossover
transition temperatures as the $T^{\chi}_c=175.8$ MeV while the  $T^{\Phi}_c=154.8$ MeV.

\begin{figure*}[htb]
\subfigure[\ Normalized chiral condensate.]{
\label{fig3a} 
\begin{minipage}[b]{0.48\textwidth}
\centering \includegraphics[width=\linewidth]{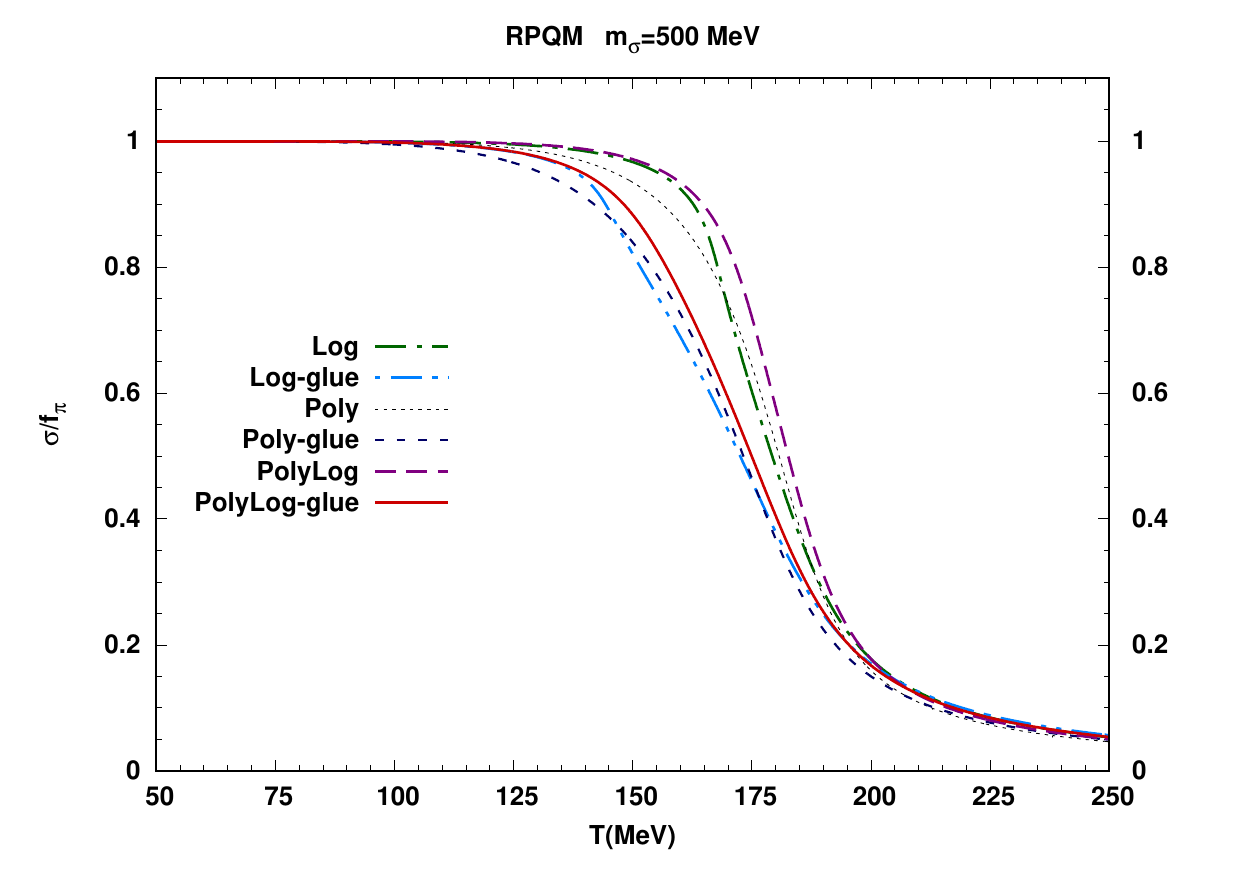}
\end{minipage}}
\hfill
\subfigure[\ Polyakov-loop order parameter.]{
\label{fig3b} 
\begin{minipage}[b]{0.48\textwidth}
\centering \includegraphics[width=\linewidth]{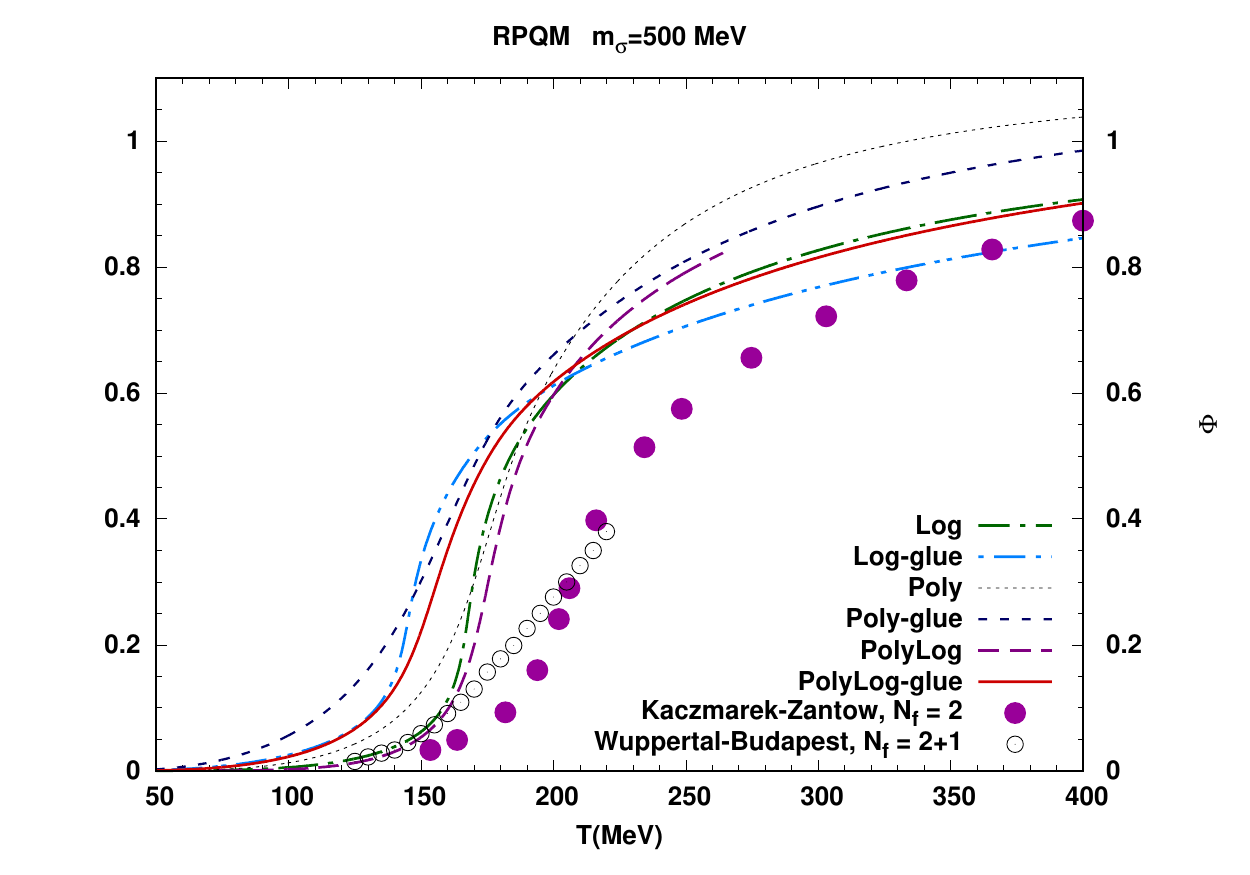}
\end{minipage}}
\caption{The RPQM model normalized chiral and Polyakov-loop order parameters  when $\mu=0$ and $T^{\rm glue}_c = T_0=208 \ \text{MeV} \ $.}
\label{fig:mini:fig3} 
\end{figure*}

\begin{figure*}[htb]
\subfigure[\ ]{
\label{fig4a} 
\begin{minipage}[b]{0.32\textwidth}
\centering \includegraphics[width=\linewidth]{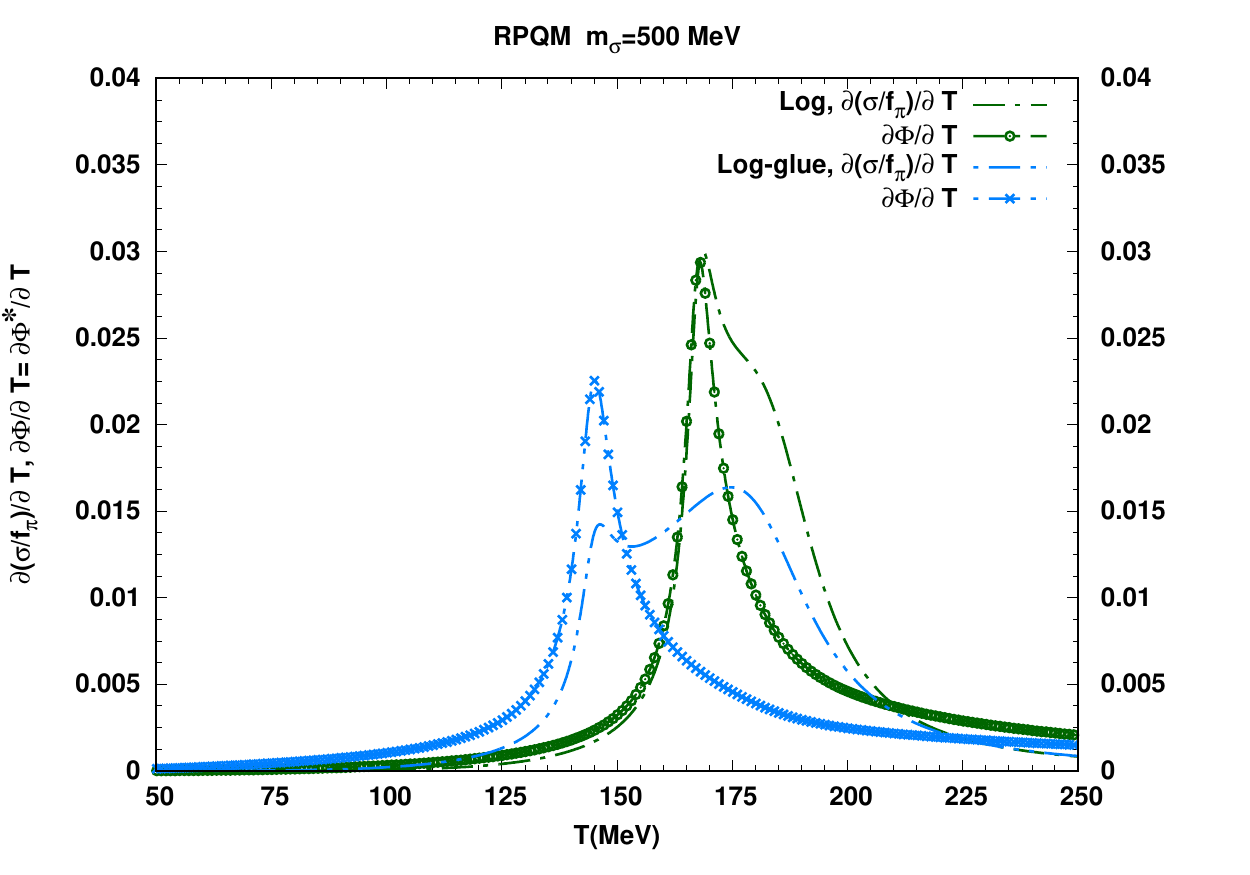}
\end{minipage}}
\hfill
\subfigure[\ ]{
\label{fig4b} 
\begin{minipage}[b]{0.32\textwidth}
\centering \includegraphics[width=\linewidth]{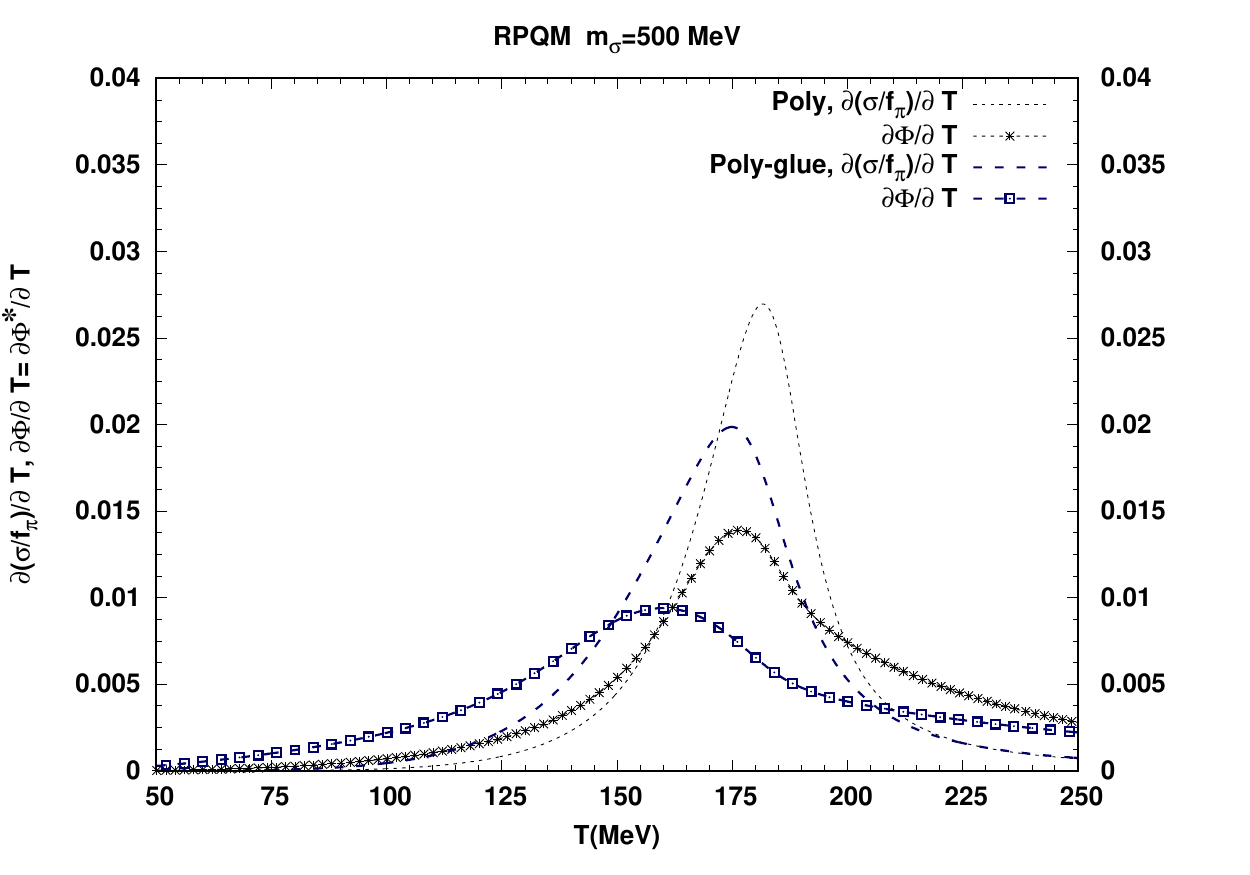}
\end{minipage}}
\hfill
\subfigure[\ ]{
\label{fig4c} 
\begin{minipage}[b]{0.32\textwidth}
\centering \includegraphics[width=\linewidth]{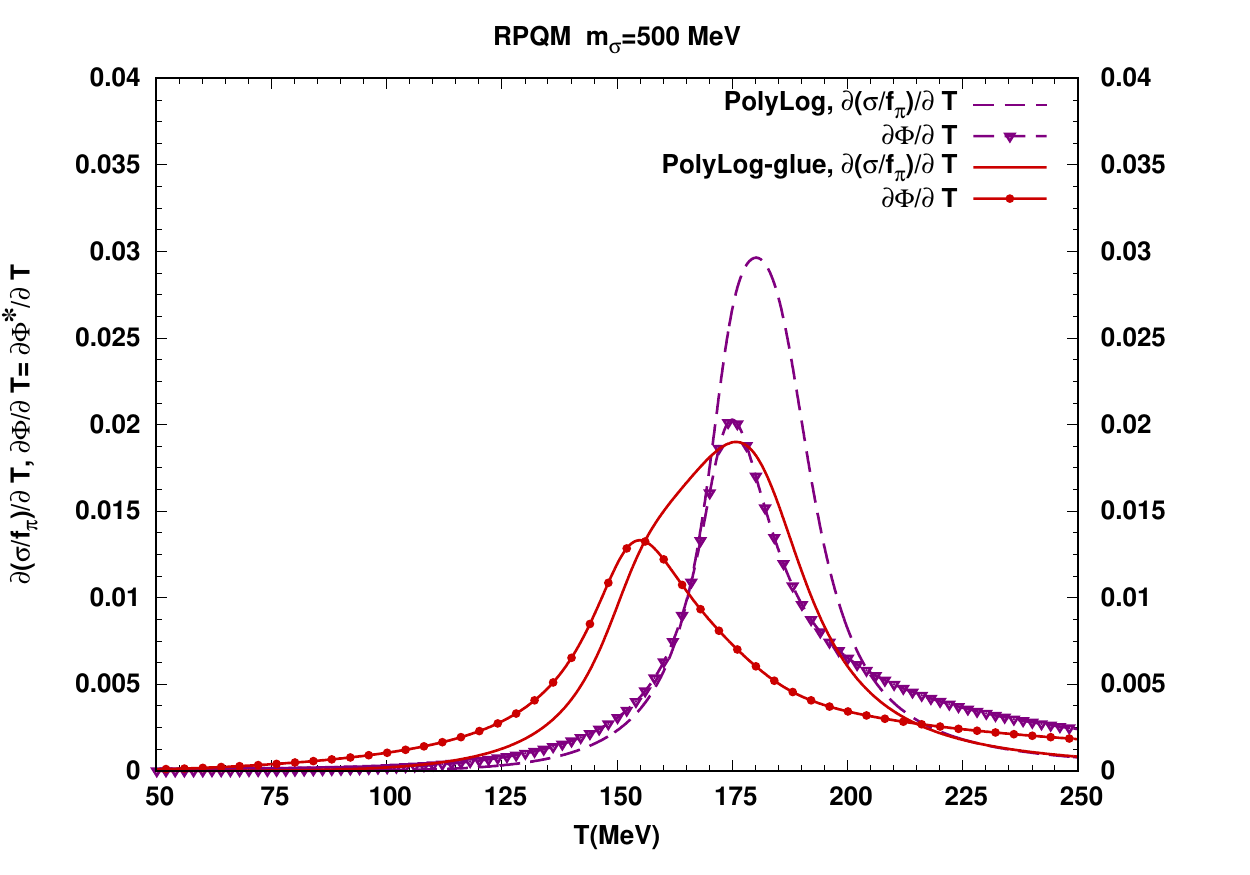}
\end{minipage}}
\caption{The derivatives of the normalized chiral and Polyakov-loop order parameters at $\mu=$0 and $T^{\rm glue}_c = T_0=208 \ \text{MeV} \ $.}
\label{fig:mini:fig4} 
\end{figure*}

The Fig.~\ref{fig3a} and the Fig.~\ref{fig3b} show how the respective temperature variations of the $\sigma/f_\pi$ and $\Phi$, are influenced by the different forms of the Polyakov-loop potential put in combination with the consistent formulation of the chiral sector physics in the RPQM model.~The falling patterns of the $\sigma/f_\pi$ in the Fig.~\ref{fig3a}, shift early on the temperature scale because of the quark back-reaction in the unquenched forms of the Polyakov-loop potentials namely the Log-glue, Poly-glue and PolyLog-glue.~The lattice QCD data for the temperature variation of  the $\Phi$ has also been plotted in the Fig.~\ref{fig3b}.~One can see that the RPQM model Polyakov-loop condensate temperature variations for the Log and the PolyLog potential,~are closer to the \cite{Zantow, Wupertal2010} lattice results when the $T<170$ MeV.~For higher temperatures,~the $\Phi$ variation of the Log, the Log-glue and the PolyLog-glue RPQM model, stand closer to the two flavor LQCD data of the Ref~\cite{Zantow}.

\begin{figure*}[!htbp]
\subfigure[\ ]{
\label{fig5a} 
\begin{minipage}[b]{0.48\textwidth}
\centering \includegraphics[width=\linewidth]{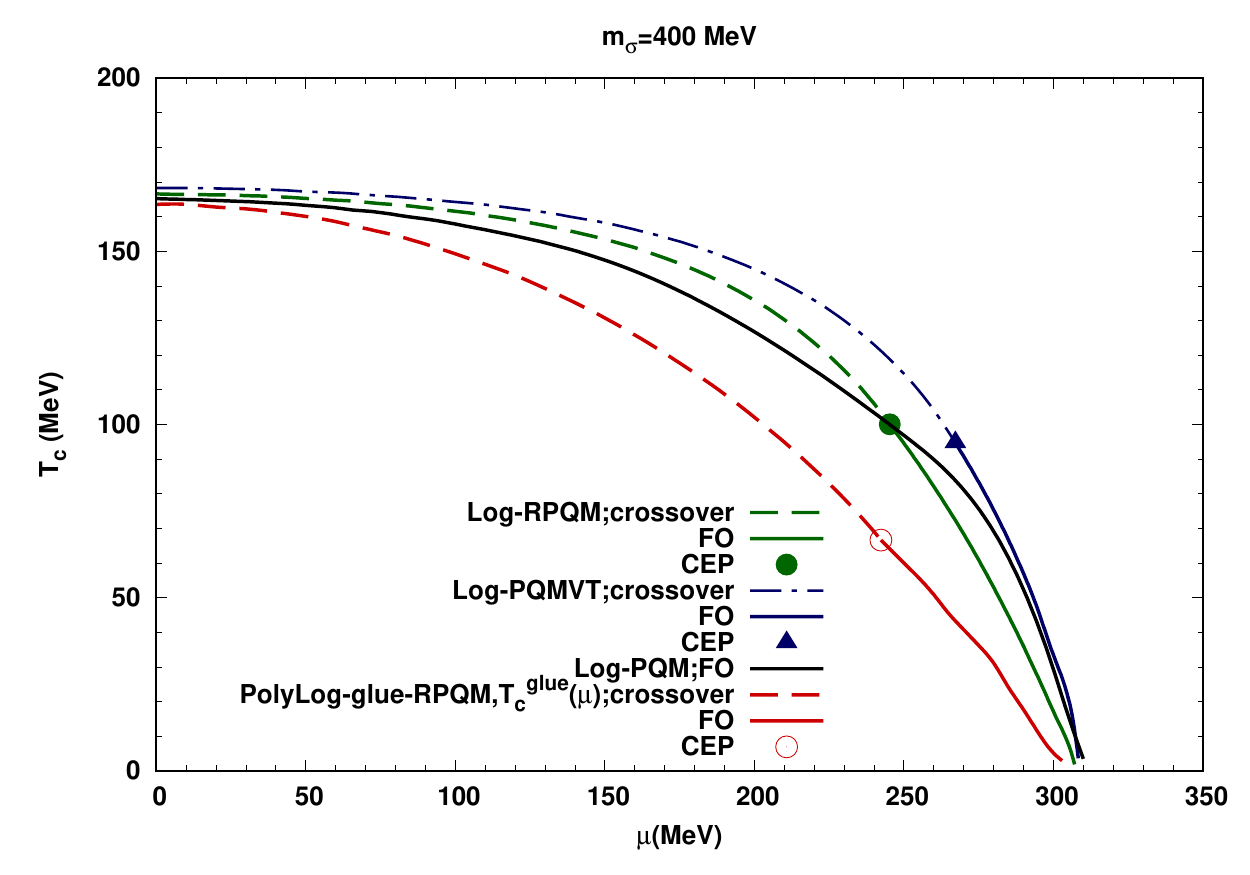}
\end{minipage}}
\hfill
\subfigure[\ ]{
\label{fig5b} 
\begin{minipage}[b]{0.48\textwidth}
\centering \includegraphics[width=\linewidth]{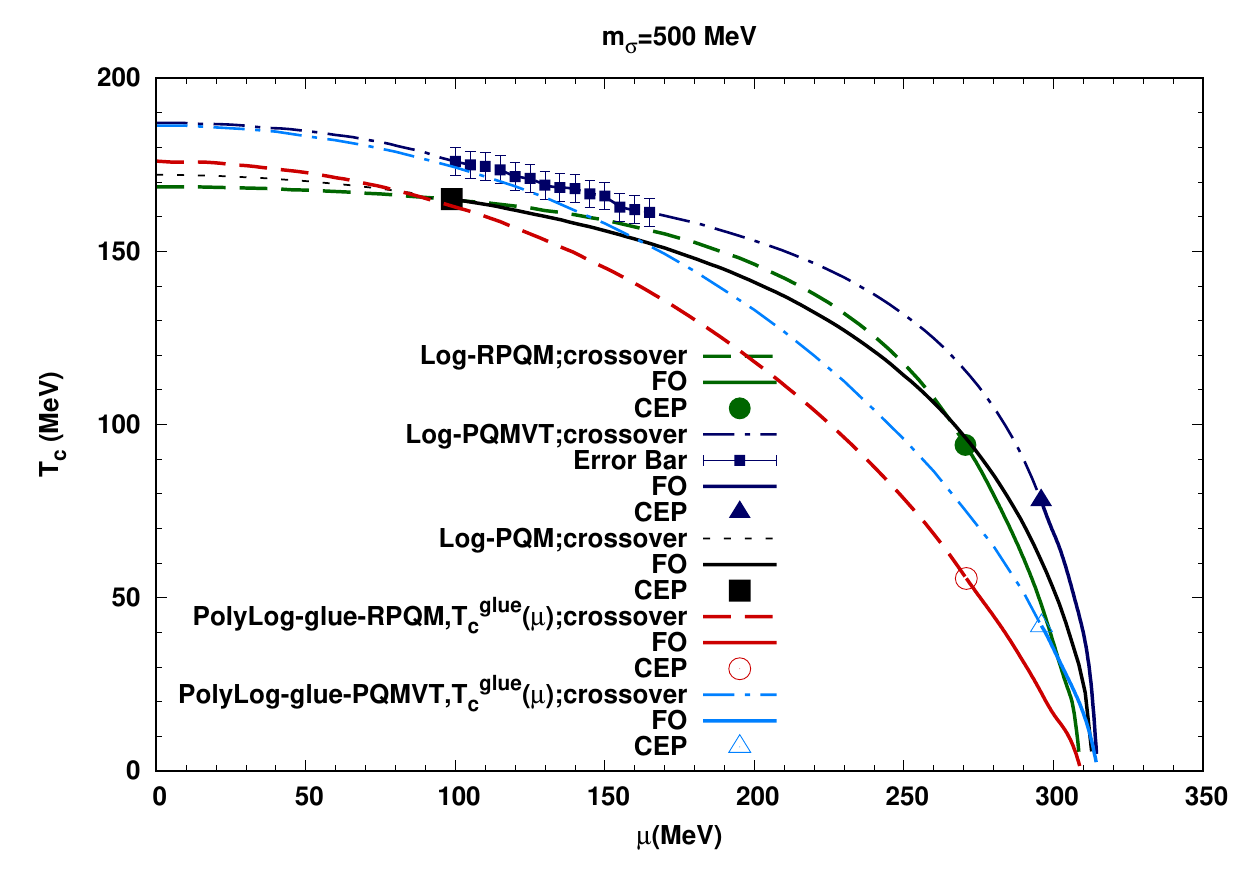}
\end{minipage}}
\caption{The PQM, PQMVT and RPQM model phase diagrams for different forms of the Polyakov-loop potentials. The error bars of the $\pm 4$ MeV on the Log PQMVT model crossover transition line, have been calculated as in the Ref.~\cite{guptiw}.    }
\label{fig:mini:fig5} 
\end{figure*}

The Polyakov-loop potential in the Log form  has stronger influence as one notes that the quite sharp peaks of the temperature variations of the 
$\frac{\partial (\sigma/f_{\pi})}{\partial {T}}$ and the $\frac{\partial \Phi}{\partial {T}}$ are  of almost the same height in the Log-RPQM model in the Fig.~\ref{fig4a} while the corresponding peaks are round and smoother respectively for the polynomial and polynomial combined with the Log form of the Polyakov-loop potential in the Fig.~\ref{fig4b} and Fig.~\ref{fig4c}.~Since the height of the  $\frac{\partial (\sigma/f_{\pi})}{\partial {T}}$ peaks are larger than that of the $\frac{\partial \Phi}{\partial {T}}$ peaks in the Fig.~\ref{fig4b} and Fig.~\ref{fig4c}, the chiral order parameter temperature variation has stronger influence in the Poly-RPQM and PolyLog-RPQM models.~Even though the height of the $\frac{\partial \Phi}{\partial {T}}$ peak is reduced and its sharpness gets moderated due to the quark back-reaction in the Log-glue RPQM model in the Fig.~\ref{fig4a},~the Log contribution makes the influence of the Ployakov loop potential stronger which gives rise to a double peak structure in the temperature variation of the $\frac{\partial (\sigma/f_{\pi})}{\partial {T}}$.~The quark back-reaction has significant smoothing effect on both the order parameters as the $\frac{\partial (\sigma/f_{\pi})}{\partial {T}}$ and the $\frac{\partial \Phi}{\partial {T}}$ temperature variations become more flat and rounded with reduced heights in the 
the Fig.~\ref{fig4b} and Fig.~\ref{fig4c} respectively for the Poly-gule and the PolyLog-glue RPQM model.~It is worth reminding ourselves that the  separation ($T^{\chi}_c-T^{\Phi}_c$) between the pseudo-critical temperatures for the chiral transition and 
the confinement-deconfinement transition is very small of only 0.7 MeV for the Log Polyakov-loop potential and it increases to 5.0 MeV when one has either the Poly or the Poly-Log form of the Ployakov-loop potential. One can see from the Table~\ref{tab:tablefig2} that the quark back-reaction causes the largest separation of 29.4 MeV in the Log-glue form while the separation becomes 15 MeV for the Poly-glue form and 21 MeV for the PolyLog-glue form of the unquenched Polyakov-loop potential in the RPQM model.

\subsection{Sigma mass and model dependence of the phase diagrams and CEPs}

\label{sec:VB}

We have plotted and compared the phase boundaries in the chemical potential and temperature $\mu-T$ plane for the chiral symmetry breaking-restoring phase transitions in the Polyakov-loop enhanced chiral models namely the  PQM, PQMVT and RPQM having different forms of parametrization  for the Polyakov-loop potential. It is well known that the first order transition line shrinks and the crossover line gets extended with the increase of sigma meson mass and therefore the CEP shifts rightwards in the phase diagram.~Below,we are presenting phase diagrams for different sigma meson masses.~Line types in all the phase diagrams are labeled and explained in the Figs. 

\begin{figure*}[!htbp]
\subfigure[\ ]{
\label{fig6a} 
\begin{minipage}[b]{0.48\textwidth}
\centering \includegraphics[width=\linewidth]{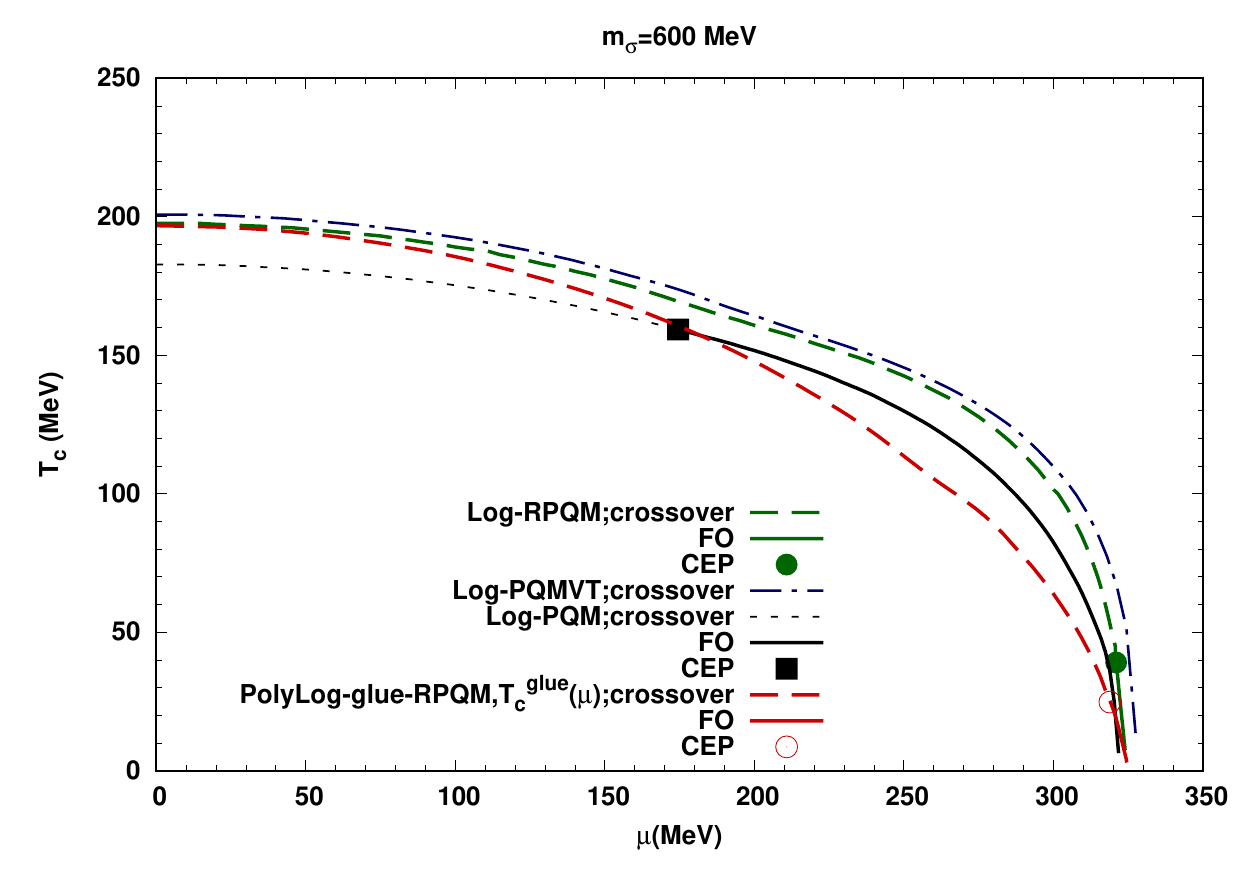}
\end{minipage}}
\hfill
\subfigure[\ ]{
\label{fig6b} 
\begin{minipage}[b]{0.48\textwidth}
\centering \includegraphics[width=\linewidth]{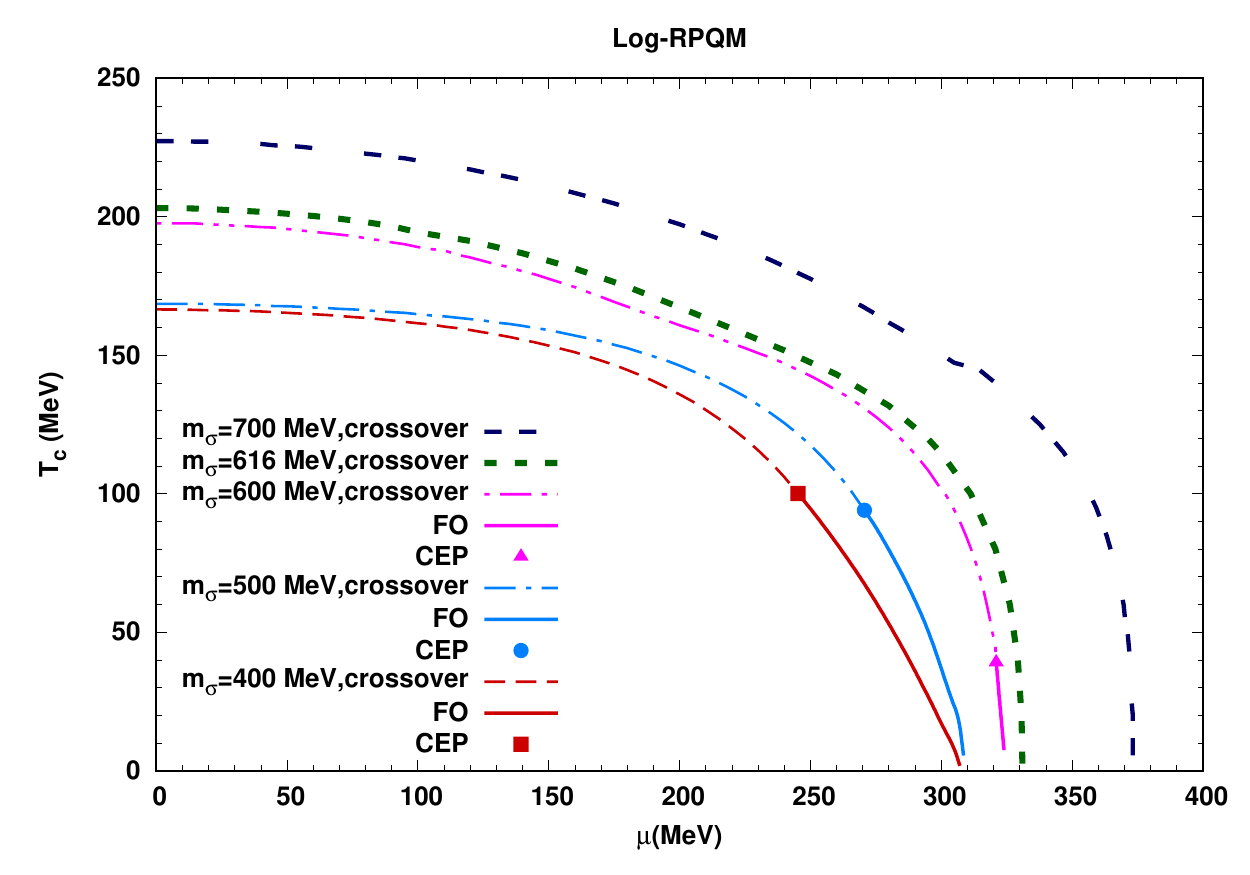}
\end{minipage}}
\caption{Phase diagrams in the (a) are for the PQM, PQMVT and RPQM model and in the (b) are for the different $m_\sigma$ in the Log RPQM model. }
\label{fig:mini:fig6} 
\end{figure*}

The Fig~\ref{fig5a} presents the phase diagram for the $m_\sigma=$ 400 MeV.~The complete phase boundary for the Log PQM model is a first order phase transition line.~Since the quark one-loop vacuum correction with the curvature mass parametrization in the Log PQMVT model generates excessively smooth chiral transition as also reported in earlier works \cite{schafwag12,guptiw,vkkr12}, one gets a longer length of the line depicting the chiral crossover transition which terminates at the critical end point (CEP) at $\mu_{\cep}$ = 267.2 MeV, $T_{\cep}$ = 94.77 MeV and the phase boundary becomes the first order transition line afterwards.~The on-shell parameter fixing for the Log-RPQM model with the consistent renormalization of the quark one-loop vacuum correction, gives rise to a relatively moderate smoothing effect on the chiral transition and the CEP 
gets located higher up in the $\mu-T$ plane at the $T_{\cep}$ = 100.1 MeV and the $\mu_{\cep}$ = 245.3 MeV. When we consider the chemical potential dependence of the parameter  $T_{0} \equiv T_{0}(\mu)=T^{glue}_{c}(\mu)$ together with the unquenched PolyLog-glue form of the Ployakov-loop potential in the RPQM model, the quark back-reaction generates an additional robust smoothing influence on the chiral transition due to which the CEP shifts downwards and gets located at  $\mu_{\cep}$ = 242.3 MeV and $T_{\cep}$ = 66.6 MeV.~Note that the influence of the quark back-reaction is quite strong in the temperature direction as the PolyLog-glue RPQM model $T_{\cep}$  shifts down by 33.5 MeV  
when compared with the $T_{\cep}$ of the Log RPQM model while the corresponding shift in the chemical potential $\mu_{\cep}$ is only 3 MeV.~This effect gives rise to the increased curvature of the phase transition line as first reported and discussed in the Ref. \cite{BielichP}.~We point out that in our recent work \cite{RaiTiw} in the on-shell renormalized quark-meson (RQM) model where the effect of the Polyakov-loop potential is absent, the CEP gets located in the bottom right of the $\mu-T$ plane at $T_{\cep}$ = 38.2 MeV and $\mu_{\cep}$ = 253.5 MeV when the $m_{\sigma}=400$ MeV.~We see that the presence of Polyakov-loop potential either in the Log or in the PolyLog-glue form in the RPQM model leads to significant upward shift of the CEP in the 
$\mu-T$ plane.  

\begin{figure*}[!htb]
\subfigure[\ ]{
\label{fig7a} 
\begin{minipage}[b]{0.48\textwidth}
\centering \includegraphics[width=\linewidth]{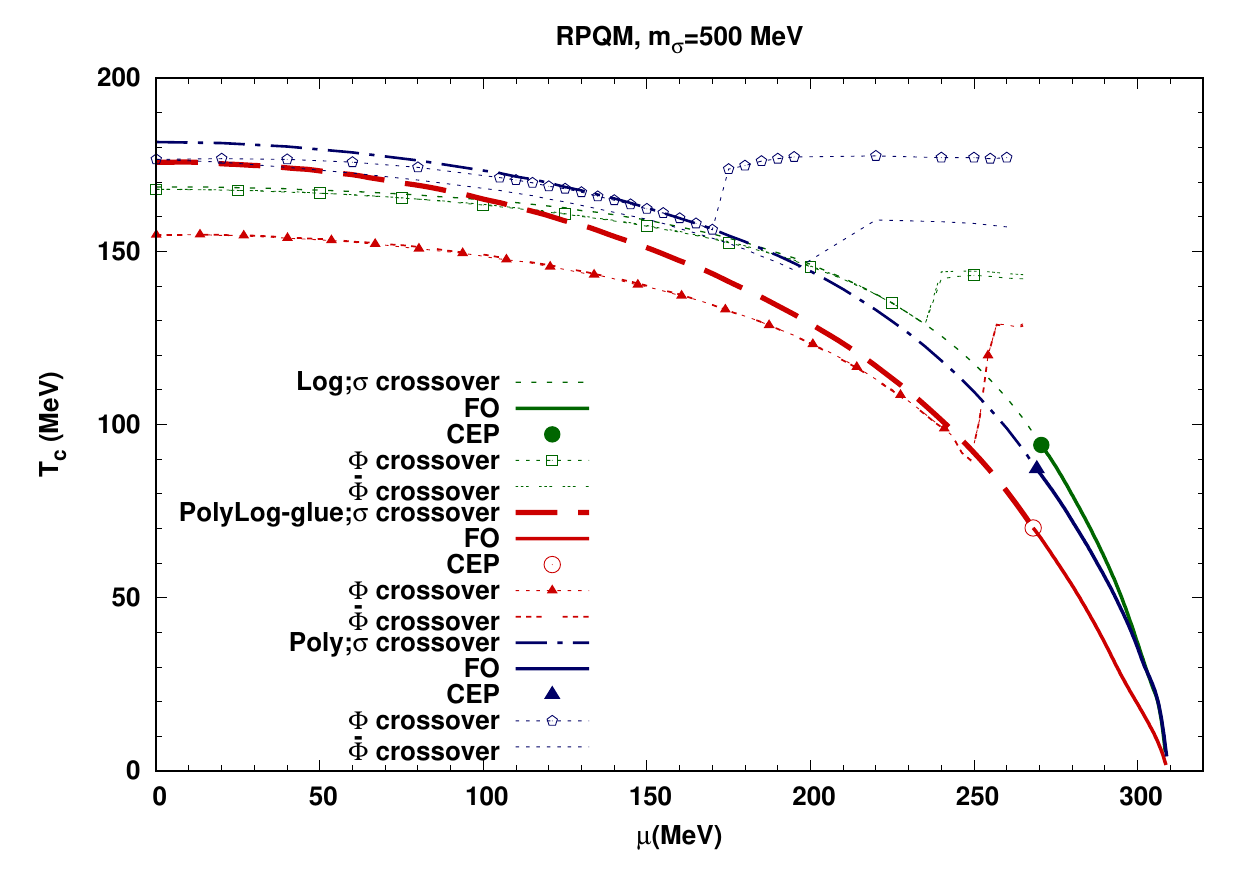}
\end{minipage}}
\hfill
\subfigure[\ ]{
\label{fig7b} 
\begin{minipage}[b]{0.48\textwidth}
\centering \includegraphics[width=\linewidth]{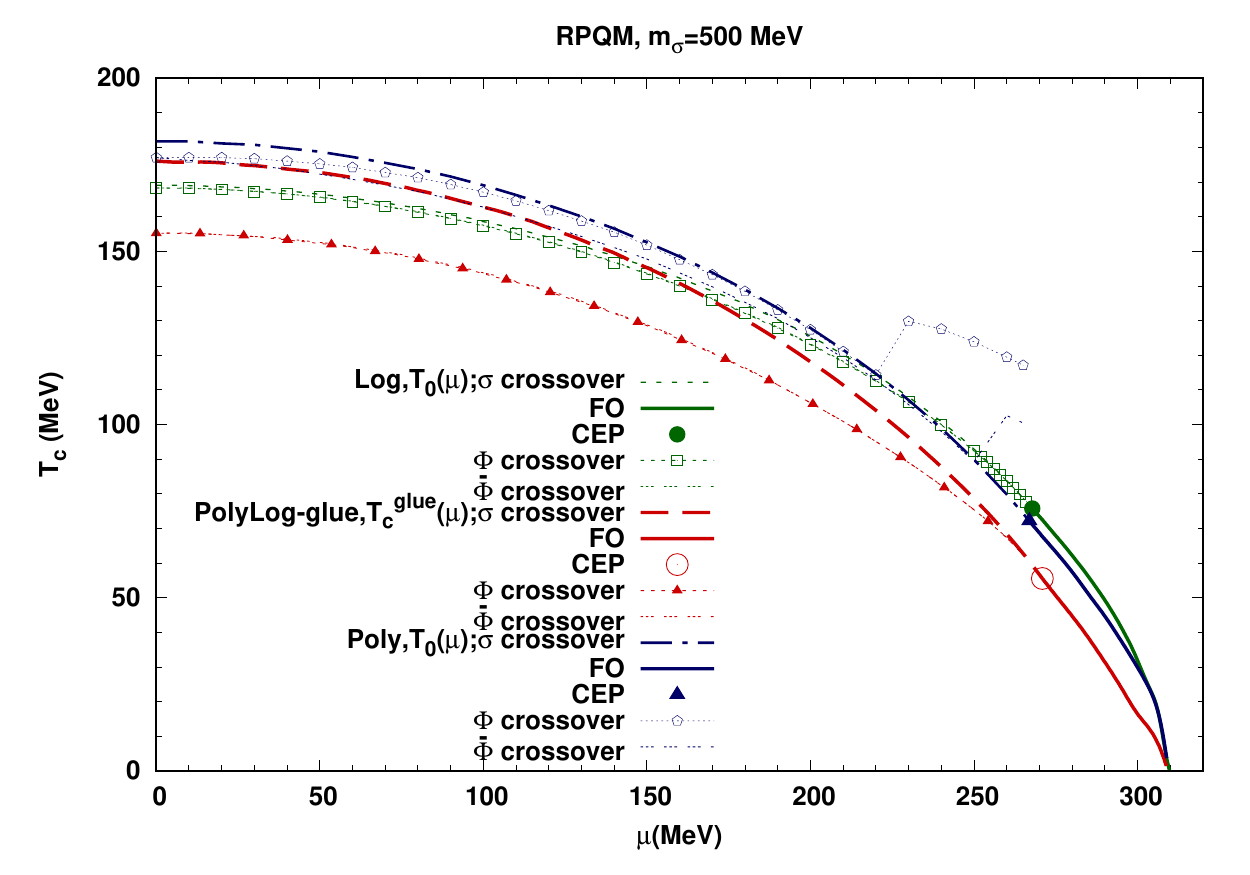}
\end{minipage}}
\caption{Deconfinement and chiral transition phase boundaries for the Log, Poly and PolyLog-glue form of the Polyakov-loop potentials in the RPQM model, (a) for the constant $T_0=208$ MeV and (b) for the $\mu$ dependent $T_0\equiv T_0(\mu)=T^{glue}_{c}(\mu) $.}
\label{fig:mini:fig7} 
\end{figure*}

The phase diagram for the $m_\sigma$ = 500 MeV case has been plotted in the Fig.~\ref{fig5b}.~We get a small crossover line ending in the critical end point at $T_{\cep}$ = 165.1 MeV and $\mu_{\cep}$ = 98.8 MeV and a quite long first order line for the Log-PQM model.~In the Log-PQMVT model, similar to the $m_{\sigma}=400$ MeV case, the crossover line becomes very large at the expense of significantly shrunk first order region and the CEP gets located at $T_{\cep}$ = 78.0 MeV, $\mu_{\cep}$ = 295.9 MeV.~When we consider the unquenched PolyLog-glue form for the Ployakov-loop potential with the chemical potential dependent parameter  $T_{0} \equiv T_{0}(\mu)=T^{glue}_{c}(\mu)$,~the quark back-reaction in the PolyLog-glue PQMVT model causes 36.2 MeV reduction in the temperature axis location of the CEP when compared with the $T_{\cep}$ of the Log-PQMVT model as it gets located at the $T_{\cep}$ = 41.8 MeV while the chemical potential location remains almost the same at the $\mu_{\cep}$ = 296.0 MeV.~Note that the uncertainty bar of the Log PQMVT model for finding  the $T_{c}^{\chi}$ in the chemical potential range 100-165 MeV, also disappears in the PolyLog-glue PQMVT model and we find a well defined crossover transition line with increased curvature.~When compared to the CEP of the Log PQMVT model,~the CEP in the Log RPQM model, gets located higher up on the temperature axis at $T_{\cep}$ = 94.1 MeV with smaller chemical potential at $\mu_{\cep}$ = 270.6 MeV.~The above result is expected because the smoothing influence of the  on-shell renormalized quark one-loop vacuum fluctuation on the chiral transition remains moderate also when the $m_{\sigma}=500$ MeV \cite{RaiTiw}.~The CEP of the Log RPQM model shifts down in temperature by 38.5 MeV due to the effect of quark back-reaction in the unquenched PolyLog-glue RPQM model with $T_{0} \equiv T_{0}(\mu)=T^{glue}_{c}(\mu)$ and gets located at the $T_{\cep}$ = 55.6 MeV while the corresponding chemical potential gets located at the $\mu_{\cep}$ = 270.9 MeV with a negligible shift.~When the $\mu$ dependence of the $T_{0}$ is switched off in the PolyLog-glue RPQM model as shown later in the Fig.~\ref{fig7a},~the CEP gets located at the $T_{\cep}$ = 70.1 MeV and the $\mu_{\cep}$ = 268.1 MeV with a moderate temperature axis shift of 24.0 MeV in reference to the CEP of the Log RPQM model.~The curvature of the phase transition line increases significantly for all the cases where the unquenching of the Polyakov-loop potential has been considered.~In our work we get confirmation of the observation of Ref. \cite{BielichP} that the quark back-reaction due to the unquenching of the Polyakov-loop potential, links the chiral and deconfinement phase transitions also at small temperatures and large chemical potentials.~Here it is also relevant to recall that for the $m_{\sigma}=500$ MeV case,~the CEP for the two flavor RQM model, lies at the $T_{\cep}$ = 36.2 MeV and the $\mu_{\cep}$ = 277.3 MeV  \cite{RaiTiw} in the absence of the Polyakov-loop effect.~Thus whatever be the form of Polyakov-loop potential when  combined with the chiral physics, it generates a noteworthy upward shift of the CEP in the $\mu-T$ plane. 

Fig.~\ref{fig6a} depicts the phase diagram for the $m_{\sigma}=600$ MeV case.~In comparison to the lower $\sigma$ meson masses,~the CEP in the Log PQM model shifts rightwards at $T_{\cep}$ = 159.3 MeV and $\mu_{\cep}$ = 174.5 MeV.~The entire phase boundary for the Log PQMVT model becomes a crossover transition line.~The position of  the CEP for the Log RPQM model is found  at $T_{\cep}$ = 39.1 MeV and $\mu_{\cep}$ = 321.0 MeV.~In the PolyLog-glue RPQM model with the $T_{0} \equiv T_{0}(\mu)=T^{glue}_{c}(\mu)$,~due to the effect of the quark back-reaction,~the CEP gets located at the $T_{\cep}$ = 24.8 MeV and the $\mu_{\cep}$ = 318.9 MeV.~In comparison to the $m_{\sigma}=400 \ \text{and} \ 500$ MeV cases,~here for the  $m_{\sigma}=600$ MeV, we note that the $T_{\cep}$ registers a smaller downwards shift of 14.3 MeV with respect to the Log RPQM model  $T_{\cep}$.


Fig.~\ref{fig6b} presents the comparison of phase diagrams in the Log RPQM model for the different $\sigma$ meson masses of 400, 500, 600, 616, and 700 MeV.~The location of the CEP  
at $T_{\cep}$= 100.1 MeV and $\mu_{\cep}$= 245.3 MeV for the $m_\sigma$ = 400 MeV shifts slightly right to $T_{\cep}$= 94.1 MeV and $\mu_{\cep}$= 270.6 MeV for the $m_\sigma$ = 500 MeV.~The CEP registers a significant rightward shift when the $m_\sigma$ = 600 MeV and gets located at $T_{\cep}$= 39.1 MeV and $\mu_{\cep}$= 321.0 MeV.~For the $m_{\sigma}\ge 616$ MeV, the entire phase boundary depicts a crossover transition.~Two crossover transition lines for the $m_{\sigma}= 616 \ \text{and} \ 700$ MeV have also been shown in the Fig.

\subsection{Quarkyonic phase and its disappearance}
\label{sec:VC}

One can define two pseudo-critical temperatures, the $T_{c}^{\Phi}$ and the $T_{c}^{\bar{\Phi}}$ for the deconfinement crossover transition of the fields $\Phi$ and $\bar{\Phi}$ by identifying the respective peaks in the temperature variations of the $\frac{\partial \Phi}{\partial T}$ and $\frac{\partial \bar{\Phi}}{\partial T}$.~Locus of the different $T_{c}^{\Phi}$ and $T_{c}^{\bar{\Phi}}$ at different chemical potentials gives the respective phase boundaries for the crossover transition for the fields $\Phi$ and $\bar{\Phi}$.~Apart from the phase boundaries for the chiral crossover transition,~the Fig.~\ref{fig7a} presents the plots of the deconfinement crossover transition phase boundaries also for the fields $\Phi$ and $\bar{\Phi}$.~Phase diagrams for the RPQM model with Poly, Log and PolyLog-glue forms of the Polyakov-loop potential with the constant $T_{0}=208$ MeV, have been plotted in the Fig.~\ref{fig7a} while the corresponding plots when the parameter $T_{0}$ has the chemical potential dependence i.e. $T_{0} \equiv T_{0}(\mu)$, have been plotted in the Fig.~\ref{fig7b}.

In the Fig.~\ref{fig7a},~when the $\mu=0$ to 130 MeV in the Poly RPQM model for the constant $T_{0}$,~the
deconfinement crossover transition line for the $\bar{\Phi}$ lies below the corresponding line for the $\Phi$ and both of these lines get placed below the chiral crossover transition line i.e. $T_{c}^{\bar{\Phi}}$ and $T_{c}^{\Phi} \ < T_{c}^{\chi}$.~The $\Phi$ crossover line after remaining almost coincident with the chiral crossover line in the $\mu=130-170$ MeV range, branches out of it and the deconfinement transition temperature $T_{c}^{\Phi}$ jumps to 173.8 MeV while the $T_{c}^{\chi}=$ 154.6 MeV when the $\mu$ becomes 175 MeV.~Thus we are getting a region in the phase diagram where the deconfinement crossover transition line for the field $\Phi$ lies significantly above the chiral crossover line and the $T_{c}^{\Phi} \ > T_{c}^{\chi}$.~The abovementioned portion of the phase diagram, where the chiral symmetry is restored but the quarks are still confined, has been identified as the quarkyonic phase in the literature \cite{kahara,McLerran:2007npa,McLerran:2009npa,Dutra:2013}.~Here in the Poly RPQM model,~a significantly large region of the quarkyonic phase is obtained for the field $\Phi$ as it starts early when the $\mu \ge \ $ 175 MeV.~The $\bar{\Phi}$ crossover line remains below the chiral transition line uptill the $\mu=205$ MeV and crosses it between $\mu=205 -210$ MeV as the $T_{c}^{\bar{\Phi}}$ jumps to 157.6 MeV while the $T_{c}^{\chi}$ stays at 139.1 MeV and the $T_{c}^{\bar{\Phi}} \ > T_{c}^{\chi}$.~Here the extent of the so called quarkyonic phase for the field $\bar{\Phi}$ is moderate as it starts from the $\mu \ge \ $ 210 MeV.~When we consider the $\mu$ dependence of the $T_{0} \equiv T_{0}(\mu)$ for the Poly RPQM model in the Fig.~\ref{fig7b},~the extent of quarkyonic phase for the $\Phi$ and the $\bar{\Phi}$, is considerably reduced.~The deconfinement transition phase boundary for the $\Phi$ branches out of the chiral transition line for the $\mu>$ \ 225 MeV while the corresponding line for the $\bar{\Phi}$ forks out from the chiral crossover line for the $\mu>$ \ 255 MeV.~Thus the quarkyonic phase for the $\Phi$ begins from the $\mu>$ \ 225 MeV while it begins from the $\mu>$ \ 255 MeV for the $\bar{\Phi}$. 

\begin{figure*}[!htbp]
\subfigure[\ ]{
\label{fig8a} 
\begin{minipage}[b]{0.49\textwidth}
\centering \includegraphics[width=\linewidth]{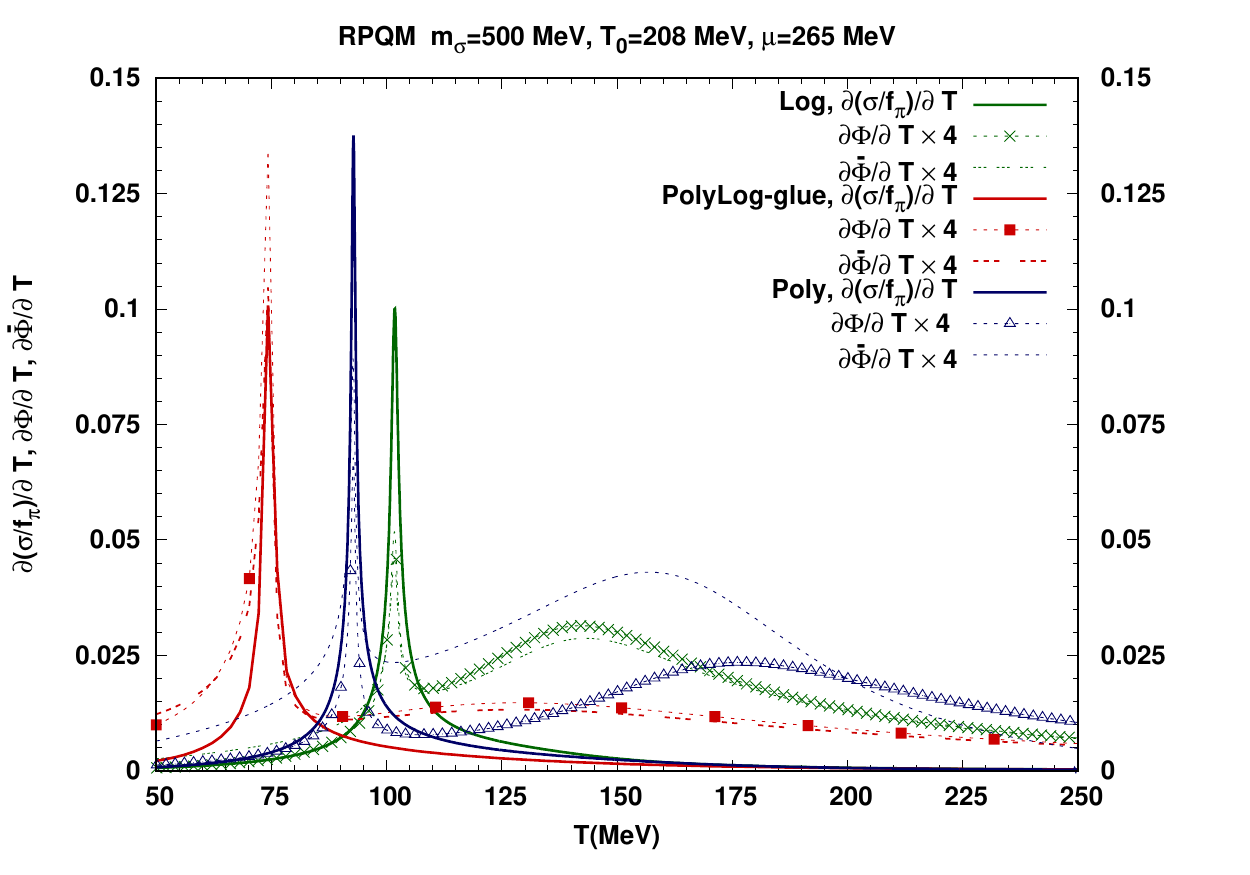}
\end{minipage}}
\hfill
\subfigure[\ ]{
\label{fig8b} 
\begin{minipage}[b]{0.49\textwidth}
\centering \includegraphics[width=\linewidth]{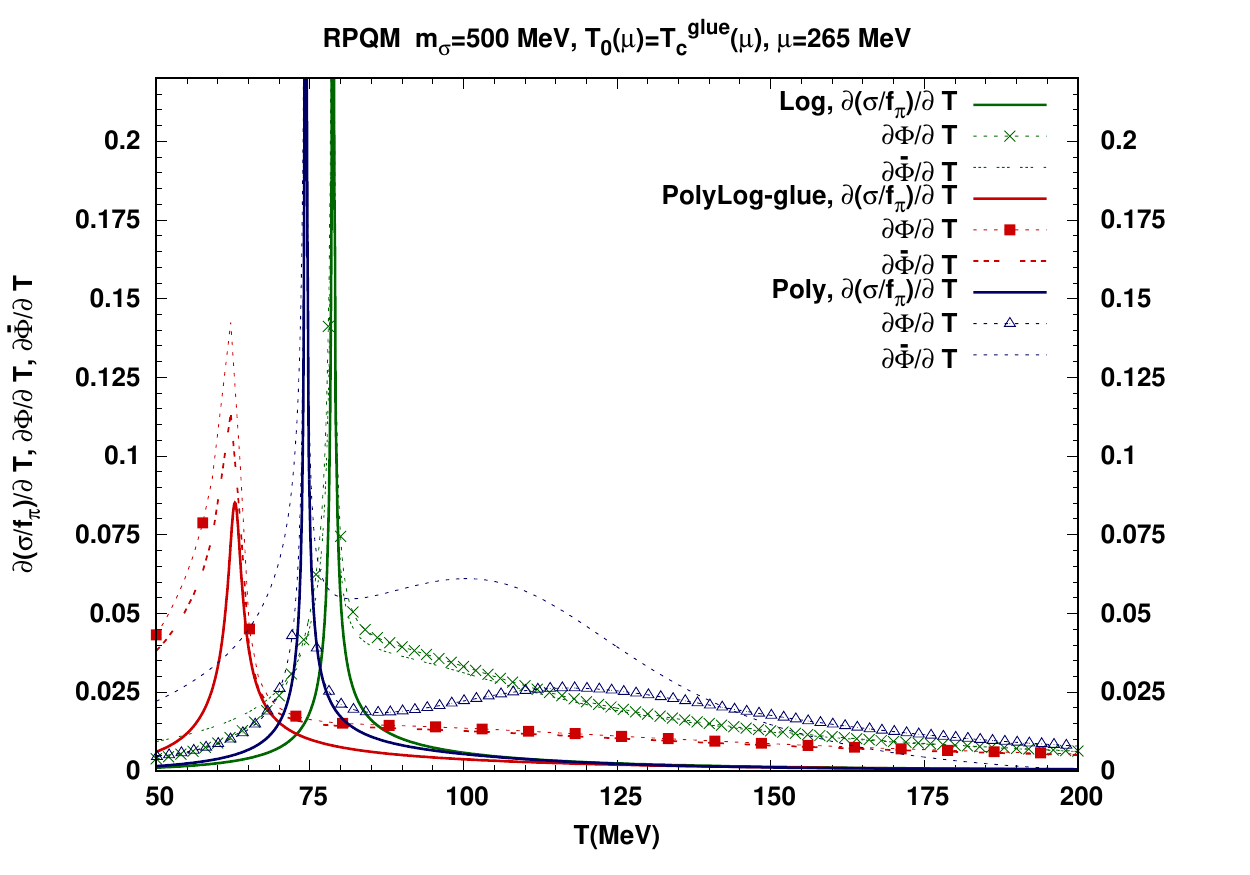}
\end{minipage}}
\caption{The temperature variation of $\frac{\partial (\sigma/f_{\pi})}{\partial T}$, $\frac{\partial \Phi}{\partial T}$ and $\frac{\partial \bar{\Phi}}{\partial T}$ for the Log, Poly and PolyLog-glue form of the Polyakov-loop potentials in the RPQM model at $\mu=$ 265 MeV,  (a) for the constant $T_0=208$ MeV and (b) for the $\mu$ dependent $T_0\equiv T_0(\mu)=T^{glue}_{c}(\mu) $. }
\label{fig:mini:fig8} 
\end{figure*}

~The two deconfinement crossover transition lines for the $\Phi$ and $\bar{\Phi}$ in the Log RPQM model are coincident in the Fig.~\ref{fig7a}.~Lying below the chiral crossover transition line,~these lines keeping a small temperature difference from it in the chemical potenial range 0 -- 220 MeV,~get merged with it at the $\mu=220$ MeV and remain so uptill $\mu=235$ MeV.~Both of the deconfinement crossover lines fork out of the chiral crossover transition line together at $\mu=$235 MeV by registering a significant jump in the deconfinement crossover temperatures for the $\Phi$ and $\bar{\Phi}$ as the $T_{c}^{\Phi}$ becomes 142.1 MeV and the $T_{c}^{\bar{\Phi}}$ becomes 144.1 MeV at $\mu=240.0$ MeV while the chiral crossover temperature remains at the $T_{c}^{\chi}=125.6$ MeV.~Here in the Log RPQM model with the constant $T_{0}$ case,~one gets the simultaneous onset of   the quarkyonic phase for both the fields $\Phi$ and $\bar{\Phi}$ and its region is considerably smaller when compared to the Poly RPQM model as it starts from the chemical potential $\mu > \ 240$ MeV.~When we see the plots for the Log RPQM model with chemical potential dependent $T_{0}(\mu)$ in the Fig.~\ref{fig7b}, we find that the two coincident $\Phi$ and $\bar{\Phi}$ deconfinement crossover transition lines do not fork out of the chiral crossover transition line at higher chemical potentials and remain merged with it.~Thus the quarkyonic phase disappears in the Log RPQM model for the chemical potential dependent $T_{0} \equiv T_{0}(\mu)$.

For the PolyLog-glue RPQM model in the Fig.~\ref{fig7a} at the constant $T_{0}$,~the deconfinement crossover transition phase boundaries for the $\Phi$ and $\bar{\Phi}$,~are completely degenerate and get placed at lower temperatures than the chiral crossover transition line in the  $\mu=$0 to 250 MeV range,~ i.e. $T_{c}^{\Phi}\ =\ T_{c}^{\bar{\Phi}} \ < T_{c}^{\chi}$.~The coincident deconfinement crossover lines are crossing the chiral transition phase boundary near the $\mu=250$ MeV and take a robust jump such that the $T_{c}^{\Phi}=T_{c}^{\bar{\Phi}}=$125.3 MeV at the $\mu=255$ MeV while the $T_{c}^{\chi}$ stays at 86.6 MeV.~Thus one gets the quarkyonic phase and the region of its existence is found to be the smallest in the PolyLog-gule RPQM model as it sets in for the $\mu >255$ MeV.~When we examine the PolyLog-gule RPQM model plot for the chemical potential dependent $T_{0} \equiv T_{0}(\mu)=T^{glue}_{c}(\mu)$ case,~we do not find any region of quarkyonic phase as the degenerate deconfinement crossover lines, completely merge with the chiral crossover transition line for the $\mu \ \ge \ 265$ MeV.~Here we recall that Schaefer et. al.\cite{SchaPQM2F,THerbst2} have argued that the incidence of quarkyonic phase presents an unphysical scenario because the deconfinement temperature should be smaller or equal to the chiral transition temperature.~Similar to our Log RPQM model finding,~they have also reported the coincidence of the chiral and deconfinement transition lines for the entire phase diagram when the parameter $T_{0}$ is taken as the $\mu$ dependent $T_{0}(\mu)$. 

~In order to see how the deconfinement transition temperatures $T_{c}^{\Phi}$ and $T_{c}^{\bar{\Phi}}$ in one model setting, differ from the corresponding chiral transition temperature $T_{c}^{\chi}$ in the region of quarkyonic phase, the temperature variations of the $\frac{\partial (\frac{\sigma}{f_{\pi}})}{\partial T}$, $\frac{\partial \Phi}{\partial T}$ and $\frac{\partial \bar{\Phi}}{\partial T}$, have been plotted for the $\mu=265$ MeV respectively in the Fig.~\ref{fig8a} for the constant $T_{0}$ case and in the Fig.~\ref{fig8b} for the $\mu$ dependent case $T_{0} \equiv T_{0}(\mu)$.~When the double peak structure emerges in the temperature variations of the $\frac{\partial \Phi}{\partial T}$ and the $\frac{\partial \bar{\Phi}}{\partial T}$ at higher chemical potentials, one identifies the quarkyonic phase.~Since the chiral transition dynamics is the driver of the first peak formation at a lower temperature,~its location coincides with the chiral crossover temperature $T_{c}^{\chi}$.~The second peak gets located at a higher temperature respectively for the $\frac{\partial \Phi}{\partial T}$ and the $\frac{\partial \bar{\Phi}}{\partial T}$ temperature variations and one gets  $T_{c}^{\Phi} \ > \ T_{c}^{\chi} $ and $T_{c}^{\bar{\Phi}} \ > \ T_{c}^{\chi} $.~The $\frac{\partial \Phi}{\partial T}$ and $\frac{\partial \bar{\Phi}}{\partial T}$ temperature variations in the Fig.~\ref{fig8a},~show the clear structure of the double peaks in the respective red, green and blue line plots for the PolyLog-glue, Log and Poly RPQM model where the quarkyonic phase exists for the constant $T_{0}$ case.~The first peak of the preceding plots coincide with the peak of the $\frac{\partial (\frac{\sigma}{f_{\pi}})}{\partial T}$ temperature variations for the corresponding model.~The temperature variations of the $\frac{\partial \Phi}{\partial T}$ and $\frac{\partial \bar{\Phi}}{\partial T}$ in the Fig.~\ref{fig8b}, do not show any double peak structure in the respective red and green line plots of the PolyLog-glue and the Log RPQM model because the quarkyonic phase disappears for the chemical potential dependent $T_{0} \equiv T_{0}(\mu)$.~The corresponding blue line plots for the Poly RPQM model, show the distinct double peaks due the presence of the quarkyonic phase.

\begin{figure*}[!htp]
\subfigure[\ ]{
\label{fig9a} 
\begin{minipage}[b]{0.48\textwidth}
\centering \includegraphics[width=\linewidth]{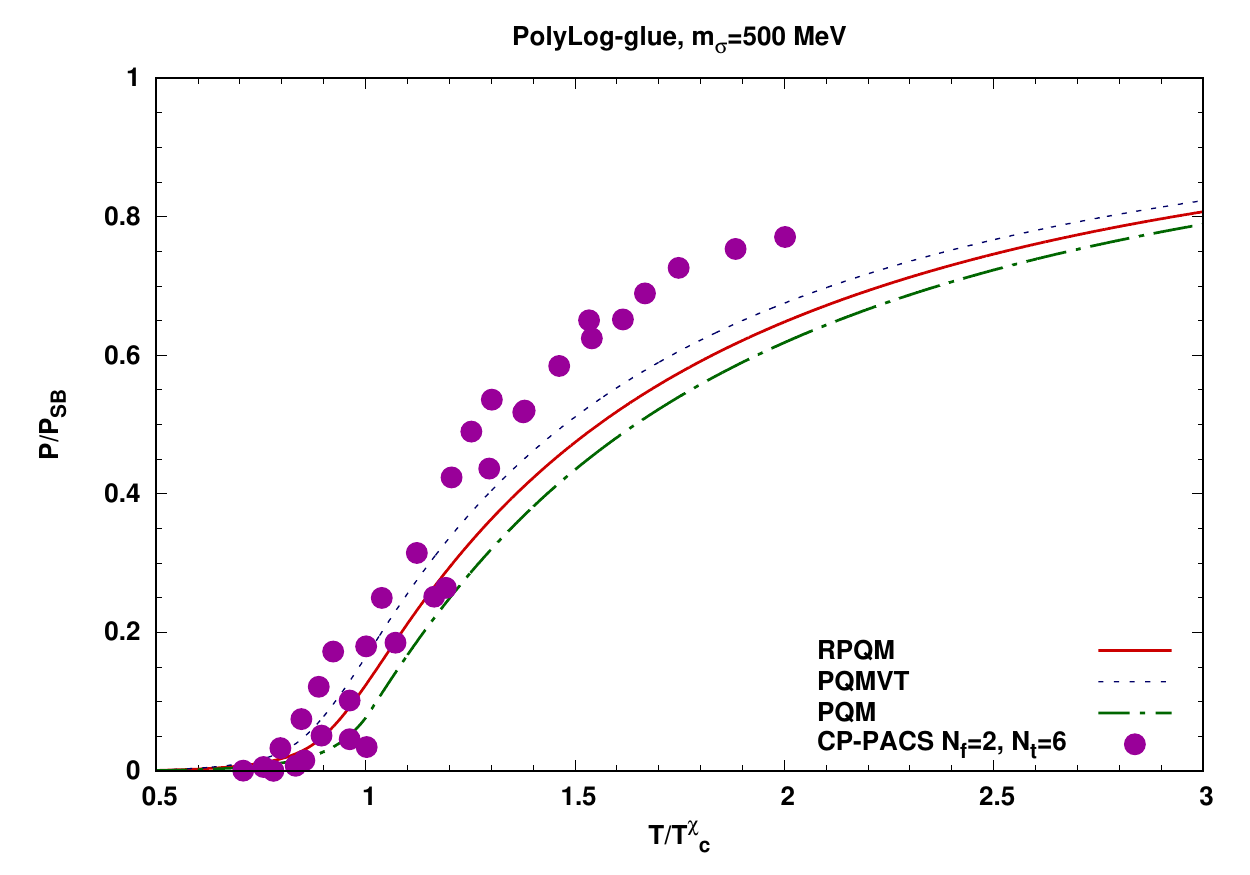}
\end{minipage}}
\hfill
\subfigure[\ ]{
\label{fig9b} 
\begin{minipage}[b]{0.48\textwidth}
\centering \includegraphics[width=\linewidth]{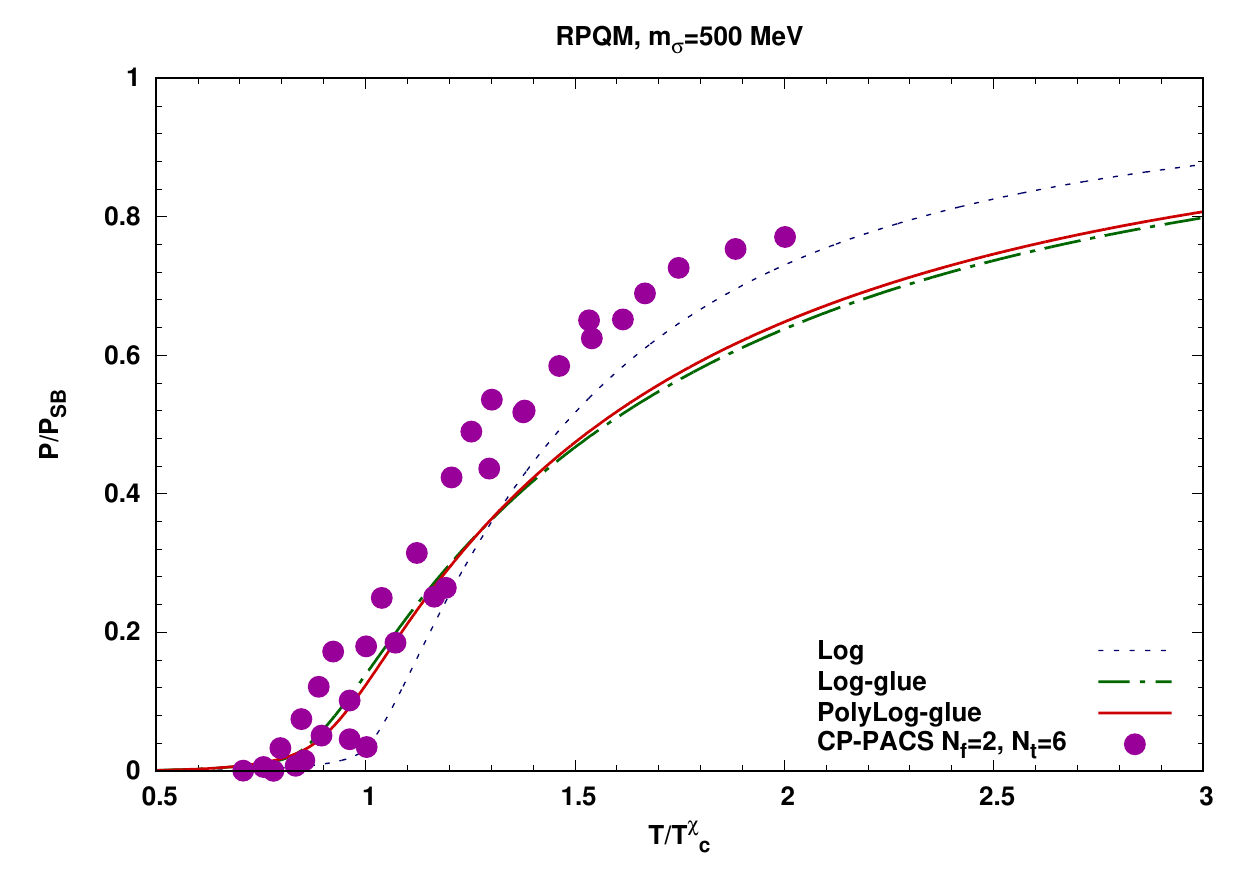}
\end{minipage}}
\caption{Reduced temperature scale variation of the pressure normalized to its Stefan-Boltzmann limit for the (a) RPQM, PQMVT and PQM model with the PolyLog-glue form of the Polyakov-loop potential, (b) RPQM model with the Log, Log-glue and PolyLog-glue form of the Polyakov-loop potential. The two flavor LQCD data of the pressure have been taken from the Ref.~\cite{AliKhan:2001ek}.}
\label{fig:mini:fig9} 
\end{figure*}
\subsection{Pressure, entropy density, energy density and interaction measure}
\label{sec:VD}
The thermodynamic observables pressure, entropy, energy density and interaction measure are sensitive to the QCD phase transition.~Considering the PolyLog-glue ansatz for the Polyakov-loop potential,~the abovementioned observables have been computed and compared to see,~how the PQM model results are changed by the quark one-loop vacuum correction in the on-shell parametrization of the RPQM model versus the curvature mass parametrization of the PQMVT model.~It is also worthwhile to compare,~how the thermodynamic observables are influenced when one takes different form of the Polyakov-loop potential namely the Log, Log-glue or PolyLog-glue for combining it with the consistent chiral physics of the RPQM model.

    The negative grand potential gives us the pressure, 
\bqa
P(T,\mu)=-\Omega_{MF}(T,\mu),\;
\eqa
the pressure in vacuum has been normalized to zero, i.e. $P(0,0)=0$.~When the 
$\mu=0$,~the ideal gas Stefan-Boltzmann limit (SB) pressure is defined as,
\bqa
\frac{P_{SB}}{T^4}=(N^2_c-1)\frac{\pi^2}{45}+N_cN_f\frac{7\pi^2}{180}.\;
\eqa
\begin{figure*}[!htbp]
\subfigure[\ ]{
\label{fig10a} 
\begin{minipage}[b]{0.48\textwidth}
\centering \includegraphics[width=\linewidth]{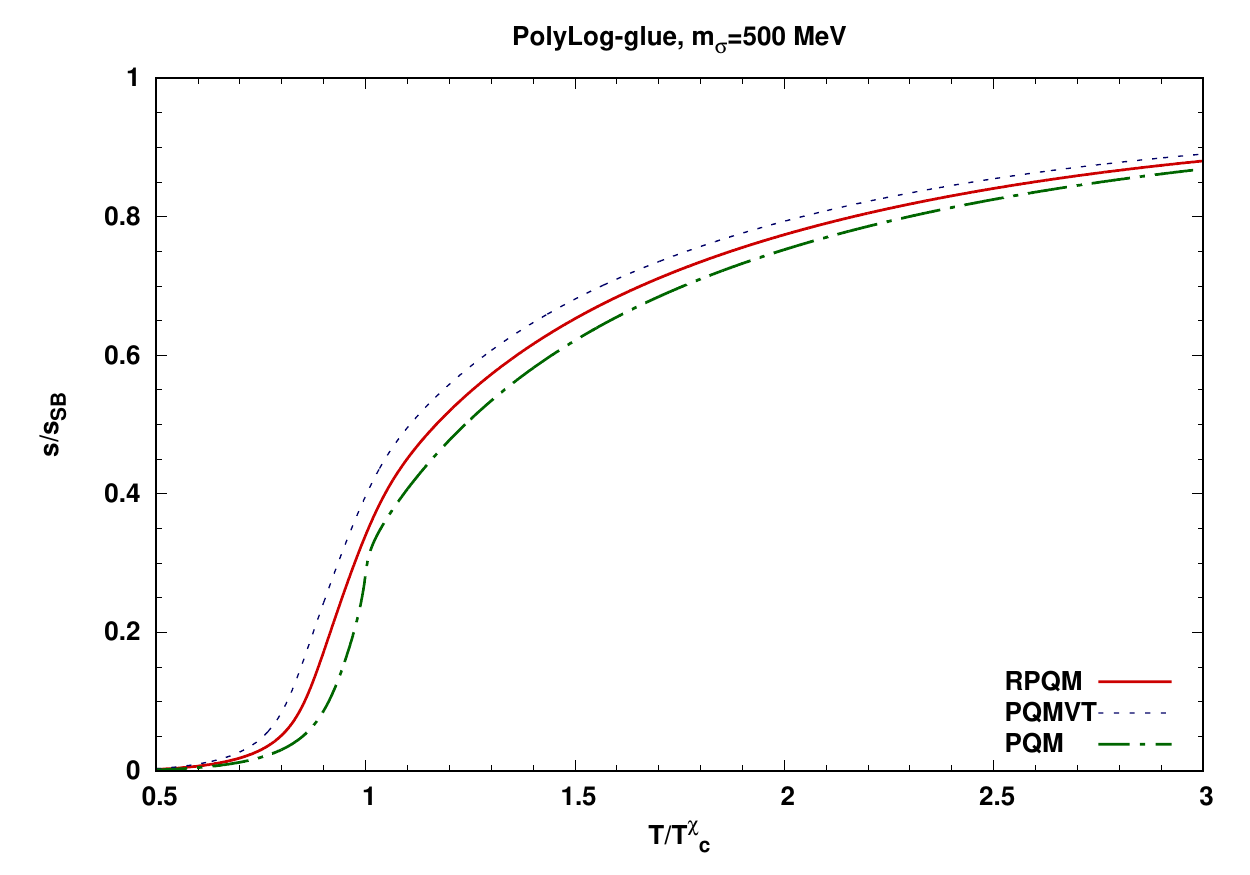}
\end{minipage}}
\hfill
\subfigure[\ ]{
\label{fig10b} 
\begin{minipage}[b]{0.48\textwidth}
\centering \includegraphics[width=\linewidth]{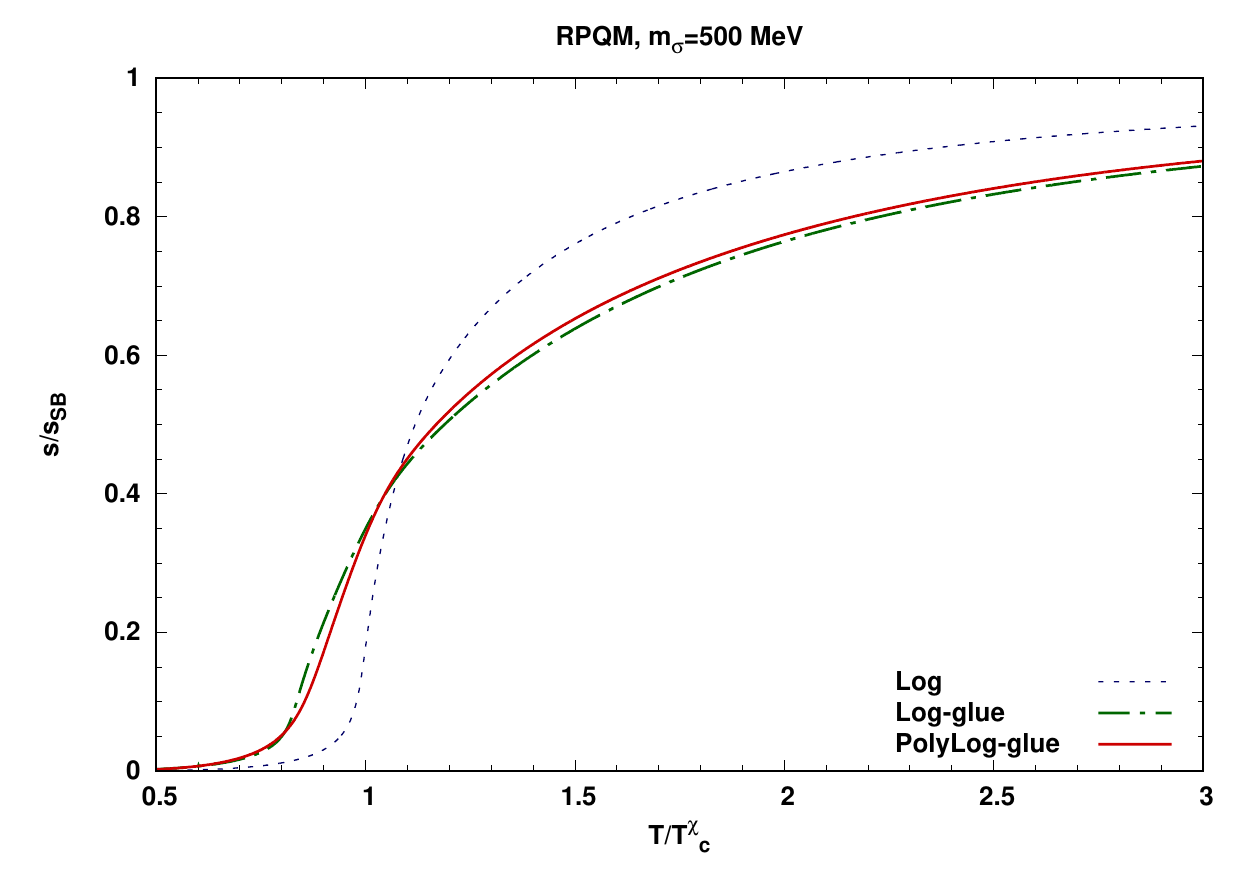}
\end{minipage}}
\caption{Reduced temperature scale variation of the entropy density normalized to its Stefan-Boltzmann limit for the (a) RPQM, PQMVT and PQM model with the PolyLog-glue form of the Polyakov-loop potential, (b) RPQM model with the Log, Log-glue and PolyLog-glue form of the Polyakov-loop potential.}
\label{fig:mini:fig10} 
\end{figure*}
Fig.~\ref{fig9a} presents the comparison of the PQM, RPQM and PQMVT model plots of the normalized pressure ($\frac{p}{p_{SB}}$) on the reduced temperature scale when the $m_{\sigma}=500$ MeV and the PolyLog-glue form of the unquenched Polyakov-loop potential has been taken.~The two flavor lattice QCD data of pressure normalized with the corresponding SB limit (on the discretized space-time) \cite{AliKhan:2001ek} has also been presented for the comparison.~The normalized pressure for the  RPQM model, shows a decent agreement with the lattice data in the range 0.7 to 1.25 of the $T_{c}^{\chi}$.~The quark one-loop vacuum correction leads to the increased pressure and the largest increase of pressure is noticed in the PQMVT model.~In the Fig.~\ref{fig9b},~one can see that the rise of the pressure in the Log RPQM model near the chiral transition temperature,~gets significantly modified in the Log-glue and PolyLog-glue RPQM model due to the quark back-reaction and unquenching of the Polyakov-loop potential.~The rise in the pressure is caused by the melting of the constituent quark masses and it saturates near 80 percent of the SB limit. 

\begin{figure*}[!htbp]
\subfigure[\ ]{
\label{fig11a} 
\begin{minipage}[b]{0.48\textwidth}
\centering \includegraphics[width=\linewidth]{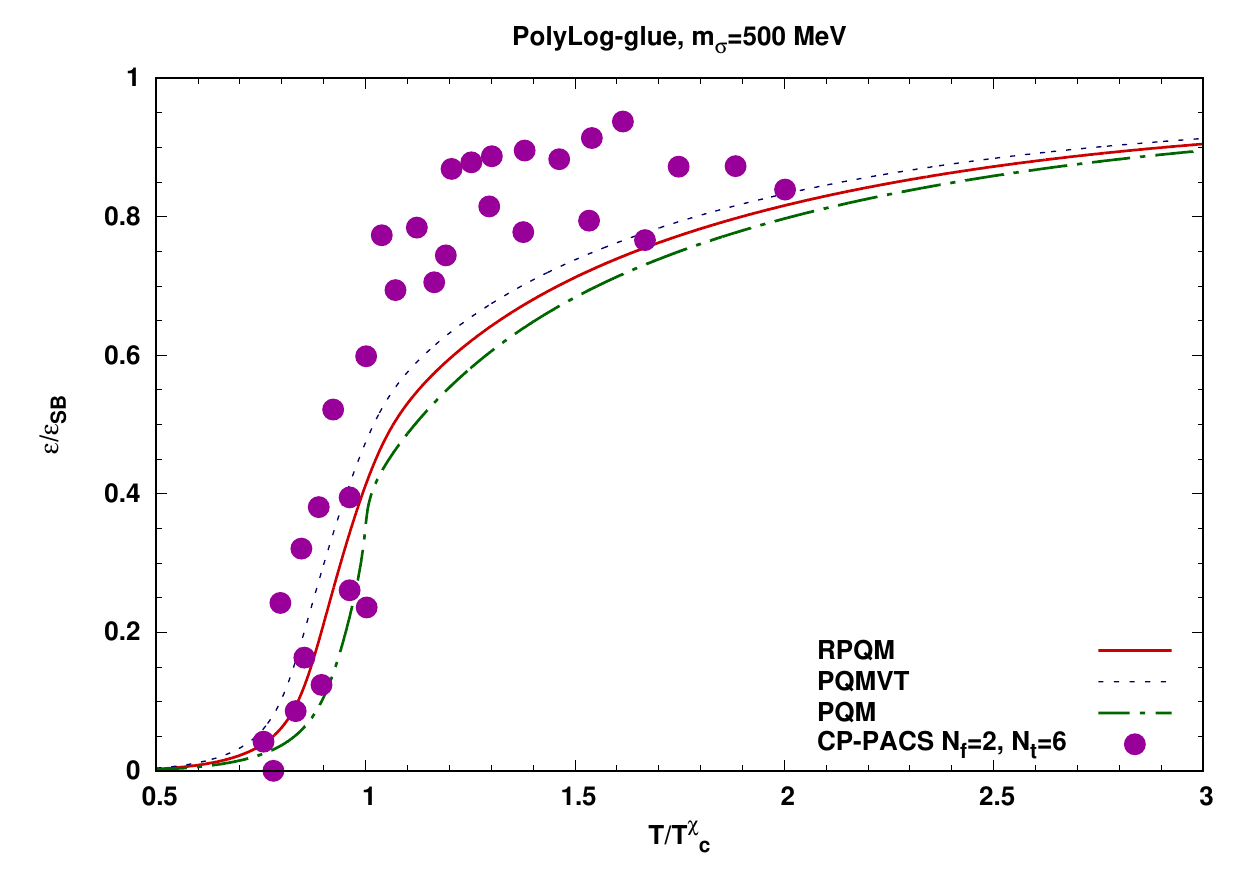}
\end{minipage}}
\hfill
\subfigure[\ ]{
\label{fig11b} 
\begin{minipage}[b]{0.48\textwidth}
\centering \includegraphics[width=\linewidth]{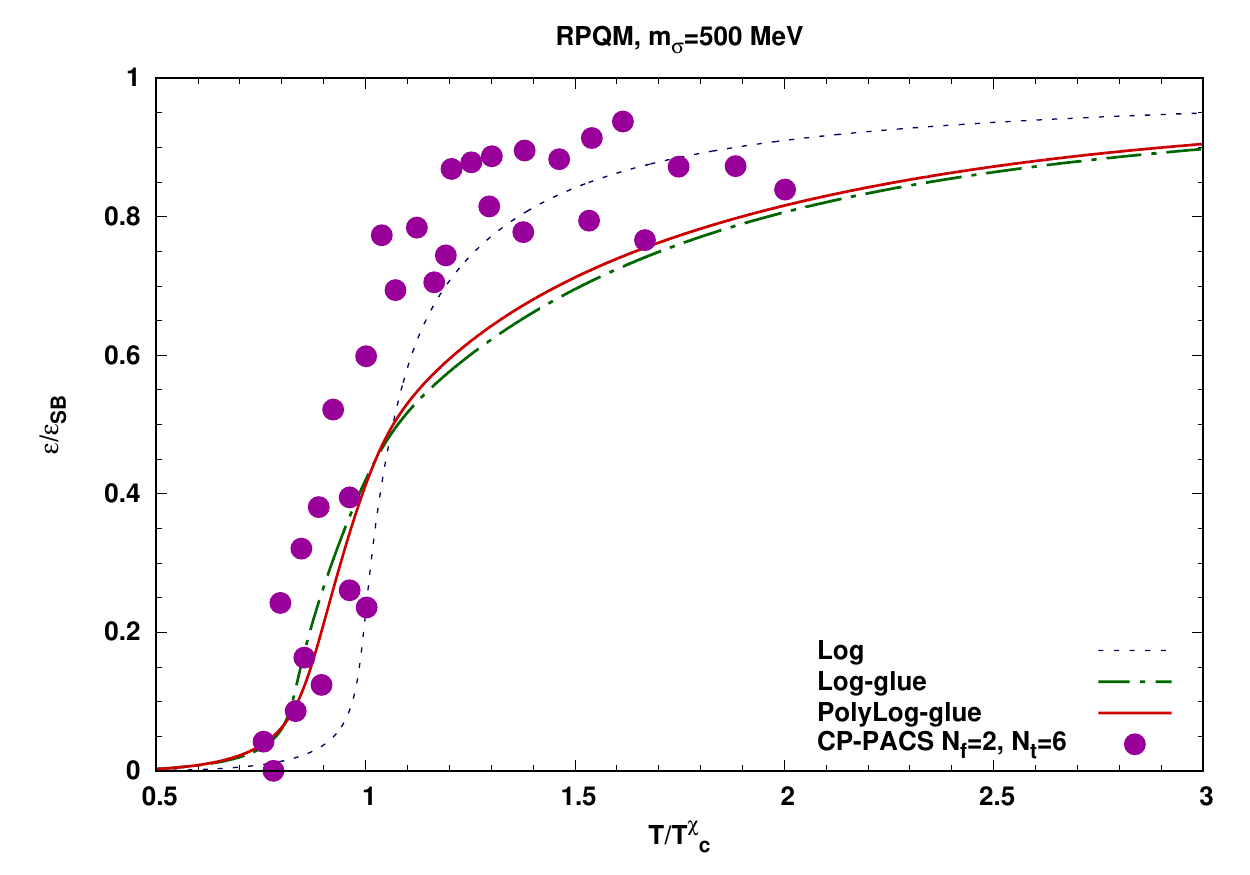}
\end{minipage}}
\caption{Reduced temperature scale variation of the energy density normalized to its Stefan-Boltzmann limit for the (a) RPQM, PQMVT and PQM model with the PolyLog-glue form of the Polyakov-loop potential, (b) RPQM model with the Log, Log-glue and PolyLog-glue form of the Polyakov-loop potential. The two flavor lattice QCD data of the energy density have been taken from the Ref.~\cite{AliKhan:2001ek}.}
\label{fig:mini:fig11} 
\end{figure*}
\begin{figure*}[!htbp]
\subfigure[\ ]{
\label{fig12a} 
\begin{minipage}[b]{0.48\textwidth}
\centering \includegraphics[width=\linewidth]{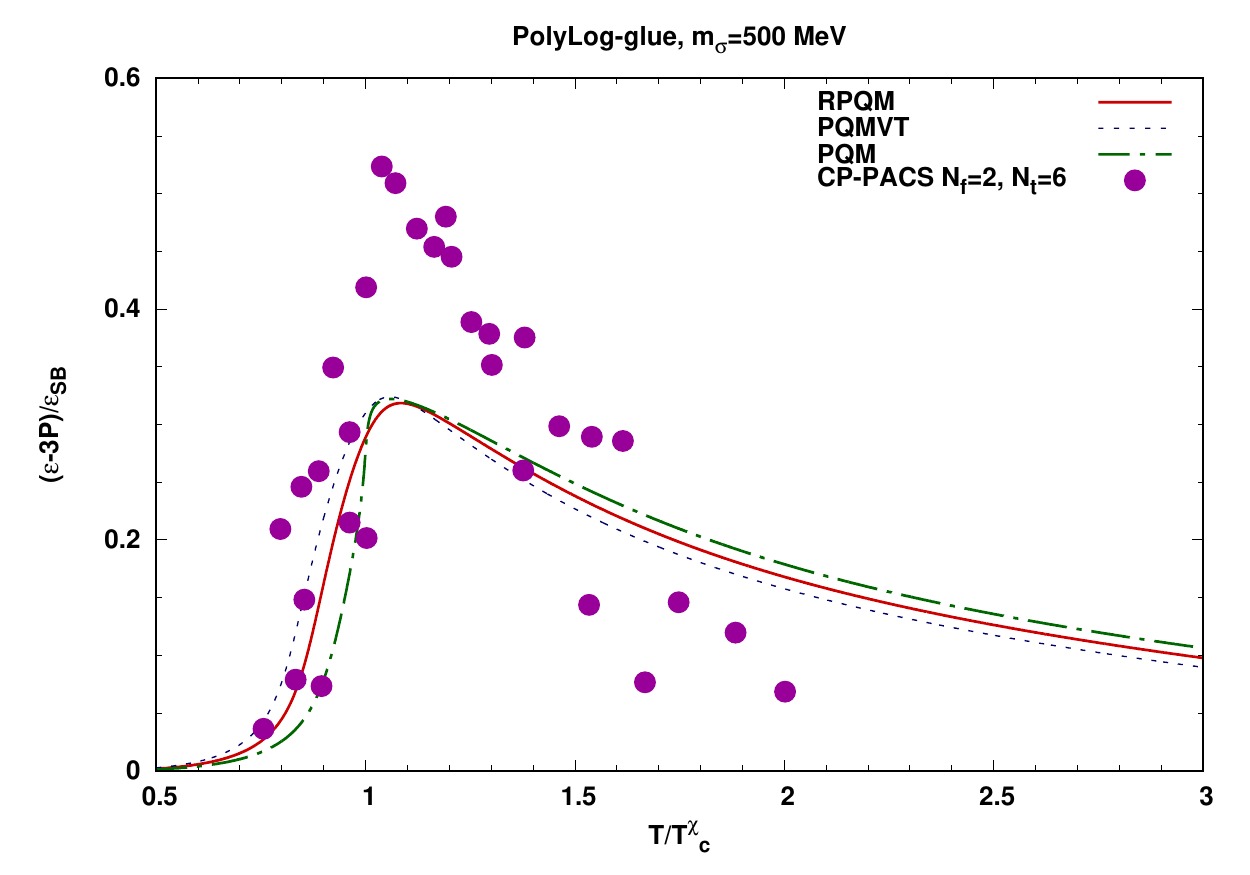}
\end{minipage}}
\hfill
\subfigure[\ ]{
\label{fig12b} 
\begin{minipage}[b]{0.48\textwidth}
\centering \includegraphics[width=\linewidth]{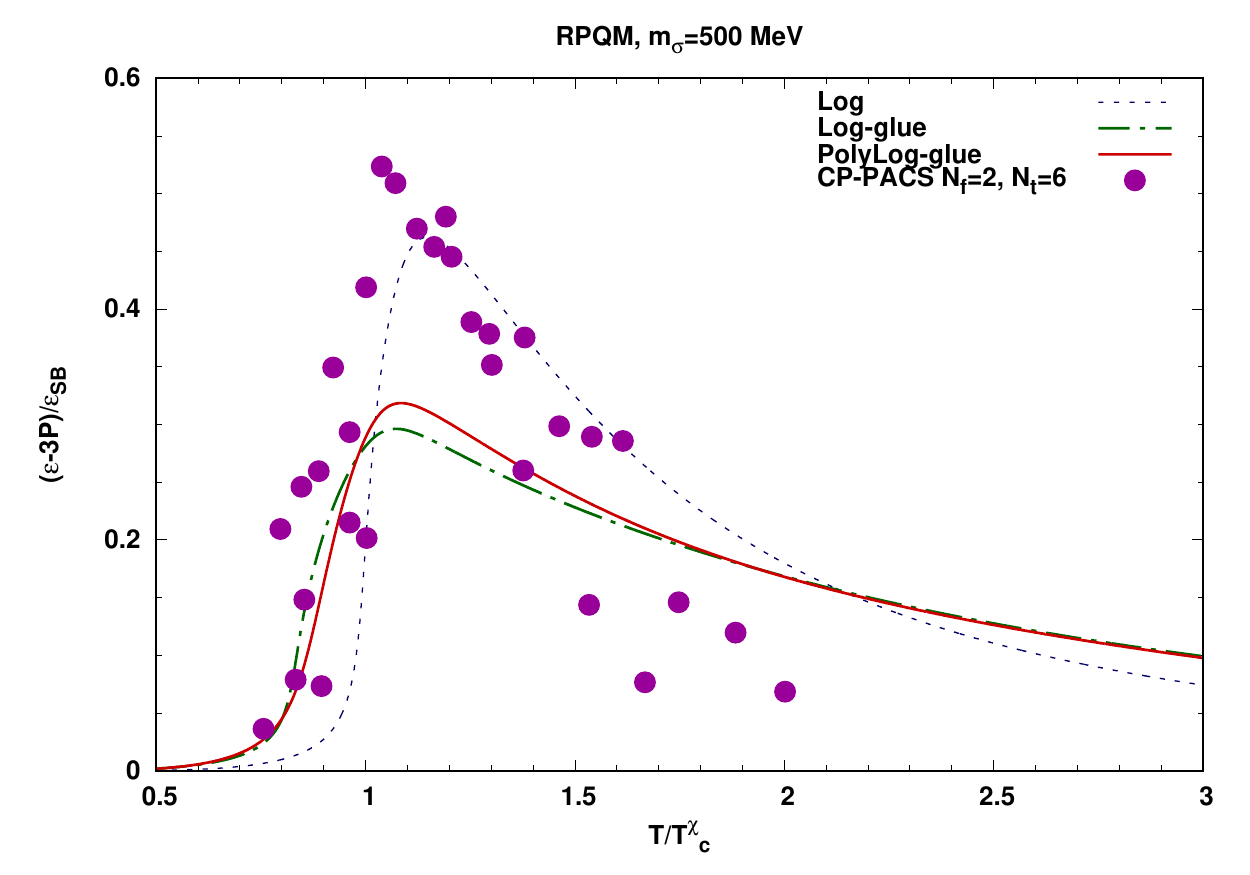}
\end{minipage}}
\caption{Reduced temperature scale variation of the interaction measure normalized to the Stefan-Boltzmann limit of energy density for the (a) RPQM, PQMVT and PQM model with  the PolyLog-glue form of the Polyakov-loop potential, (b) RPQM model with the Log, Log-glue and PolyLog-glue form of the Polyakov-loop potential. The two flavor lattice QCD data of the interaction measure have been taken from the Ref.~\cite{AliKhan:2001ek}. }
\label{fig:mini:fig12} 
\end{figure*}

The entropy density $s$, energy density $\epsilon$ and interaction measure $\Delta$ are defined as

\bqa
s&=&-\frac{\partial\Omega}{\partial T}\;,\\
\epsilon &=& -P + Ts\;,\\
\Delta&=&\epsilon-3P\;.
\eqa
\begin{figure*}[!htbp]
\subfigure[\ ]{
\label{fig13a} 
\begin{minipage}[b]{0.48\textwidth}
\centering \includegraphics[width=\linewidth]{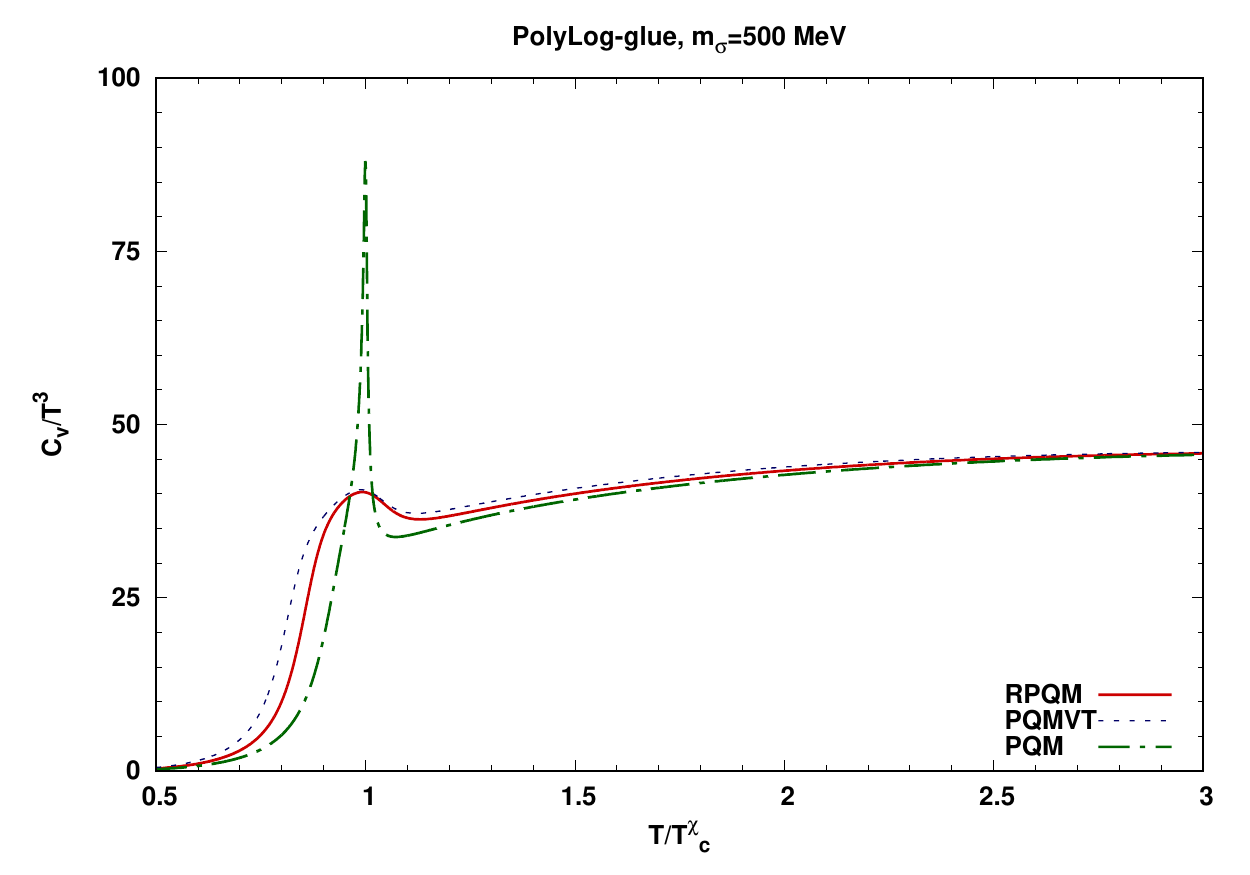}
\end{minipage}}
\hfill
\subfigure[\ ]{
\label{fig13b} 
\begin{minipage}[b]{0.48\textwidth}
\centering \includegraphics[width=\linewidth]{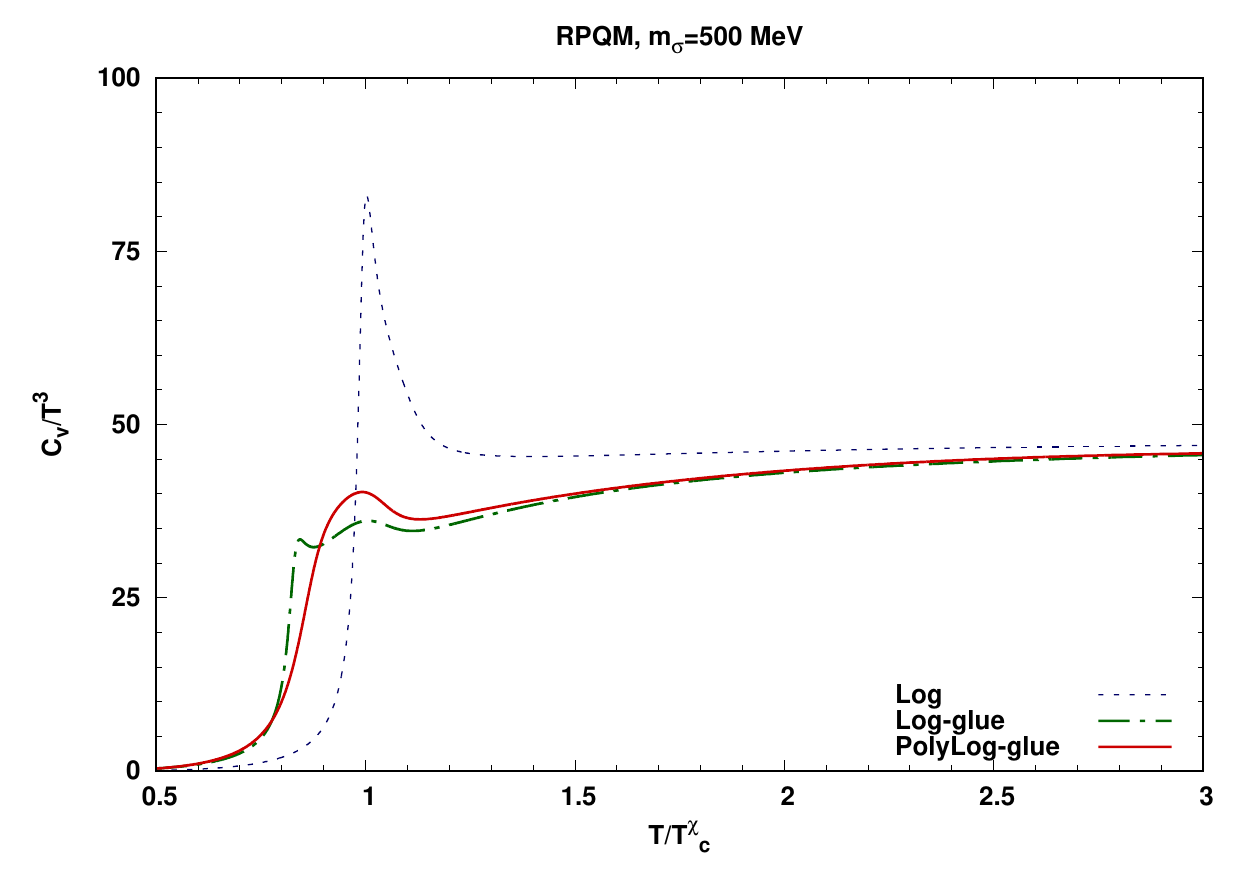}
\end{minipage}}
\caption{Reduced temperature scale variation of the specific heat at constant volume normalized with $T^3$ for the (a) RPQM, PQMVT and PQM model with the PolyLog-glue form of the Polyakov-loop potential, (b) RPQM model with the Log, Log-glue and PolyLog-glue form of the Polyakov-loop potential.}
\label{fig:mini:fig13} 
\end{figure*}

Fig~\ref{fig10a} presents the reduced temperature scale plot of the entropy density normalized to its SB limit.~Similar to the pressure,~the quark one-loop vacuum correction leads to the increased entropy density $s/s_{SB}$.~When the entropy density plot for the  PolyLog-glue RPQM and PQMVT model are compared with that of the PQM Model,~one gets largest entropy density increase in the PQMVT model while the corresponding increase in the RPQM model is moderate.~Even though the smoothing effect of the quark back-reaction is present in the PolyLog-glue PQM model,~the entropy density shows a small kink at the chiral transition temperature ($T/T_{c}^{\chi}=1$).~This kink gets smoothed out in the RPQM and PQMVT plots.~The quark back-reaction in the Log-glue and PolyLog-glue RPQM model, causes a significant smoothing change in the rapidly increasing entropy density plot of the Log RPQM model in the Fig~\ref{fig10b}.~The plots of the normalized energy density on the reduced temperature scale in the Fig~\ref{fig11a} and the Fig~\ref{fig11b} show the same trend that we get for the normalized entropy density respectively in the Fig~\ref{fig10a} and Fig~\ref{fig10b}.~The energy density plot for the PolyLog-glue RPQM model,~shows better agreement with the lattice QCD data \cite{AliKhan:2001ek} in the temperature range $(0.7-1.1) \ T^{\chi}_c$ and   $(1.6-2.1) \ T^{\chi}_c$ .

\begin{figure*}[!htbp]
\subfigure[\ ]{
\label{fig14a} 
\begin{minipage}[b]{0.48\textwidth}
\centering \includegraphics[width=\linewidth]{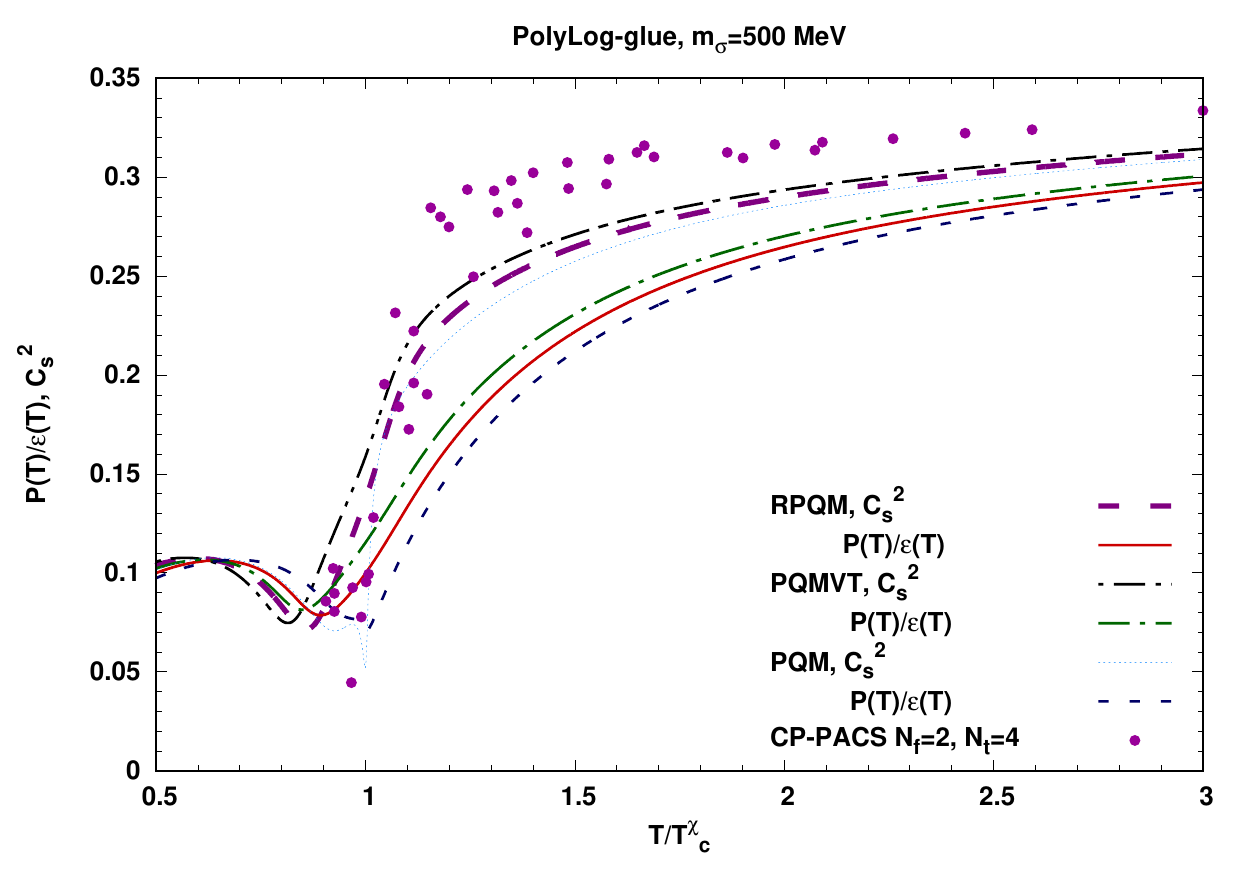}
\end{minipage}}
\hfill
\subfigure[\ ]{
\label{fig14b} 
\begin{minipage}[b]{0.48\textwidth}
\centering \includegraphics[width=\linewidth]{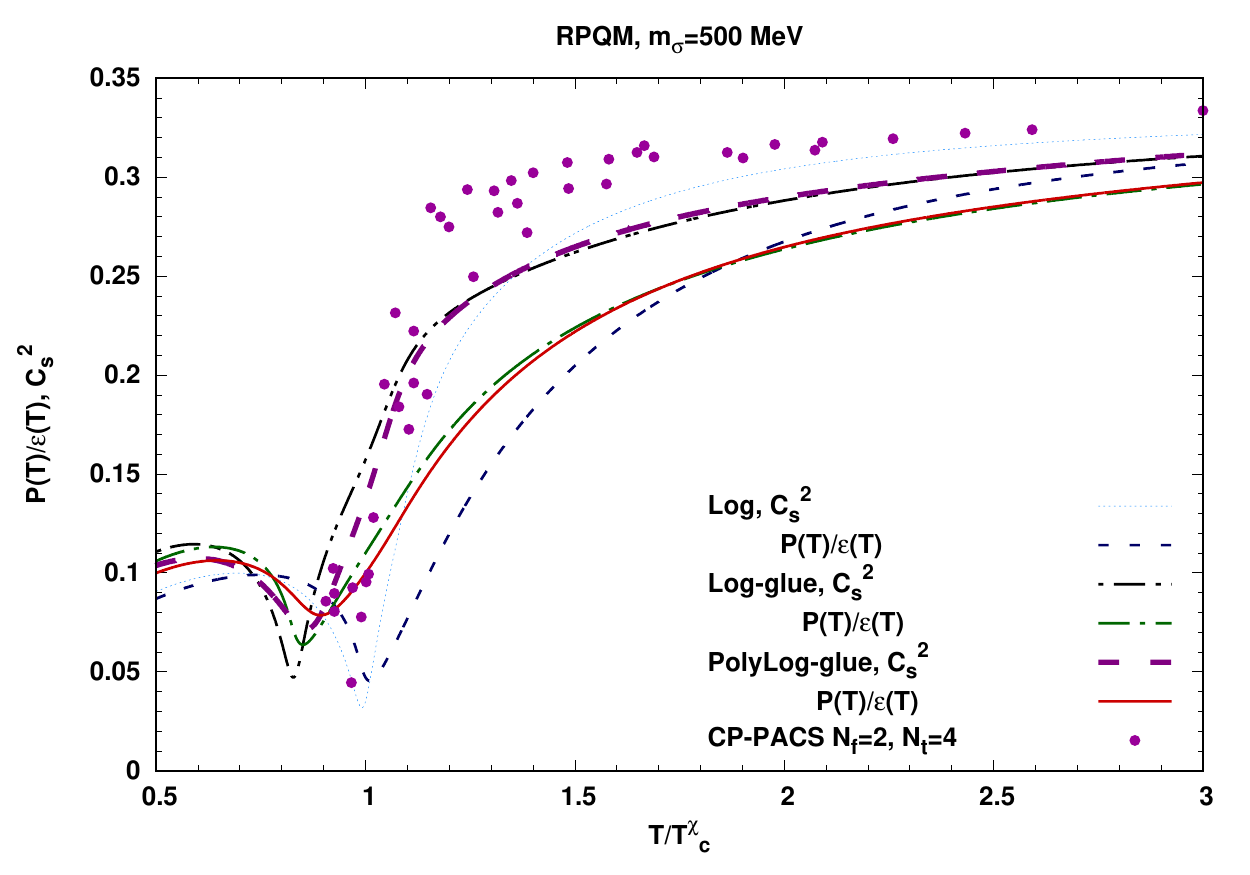}
\end{minipage}}
\caption{Reduced temperature scale variation of the  square of speed of sound and the ratio of pressure with energy density for the (a) RPQM, PQMVT and PQM model with  the PolyLog-glue form of the Polyakov-loop potential, (b) RPQM model with the Log, Log-glue and PolyLog-glue form of the Polyakov-loop potential. The two flavor lattice QCD data of the $C^2_s$ have been taken from the Ref.~\cite{AliKhan:2001ek}.}
\label{fig:mini:fig14} 
\end{figure*}

Fig~\ref{fig12a},~shows the plots of the normalized interaction measure $\Delta/\epsilon_{\stb}$  in the RPQM, PQMVT and PQM model having the PolyLog-glue form for the Polyakov-loop potential.~
The peak of the RPQM model plot,~stands a little right to that of the PQMVT and PQM model.~The  RPQM model plot agrees well with the lattice QCD data points upto $T\sim T^{\chi}_c$ and a point near $1.4 \ T^{\chi}_{c}$ .~Fig~\ref{fig12b} presents  the normalized interaction measure $\Delta/\epsilon_{\stb}$ plot for the RPQM model using Log, Log-glue and PolyLog-glue forms for the Polyakov-loop potential.~Several lattice data points at the peak of the interaction measure lie on the line depicting the Log RPQM model result. 

\subsection{Specific heat and speed of sound }
\label{sec:VE}

The specific heat capacity at constant volume is defined by 
\bqa
C_{V}=\frac{\partial \epsilon}{\partial T}\bigg{\rvert}_V=-T\frac{\partial^2 \Omega}{\partial T^2}\bigg{\rvert}_V
\eqa

Fig~\ref{fig13a} shows the variation of the normalized specific heat capacity at $\mu$ = 0 as a  function of the reduced temperature when the Polylog-glue form has been taken for the Polyakov-loop potential in the RPQM, PQMVT and PQM model.~The highest and sharpest PQM model specific heat peak at the $T_{c}^{\chi}$,~gets significantly reduced and smoothed out in the temperature range $(0.7-1.2) \ T^{\chi}_{c}$ due to the quark one-loop vacuum fluctuations in the PQMVT and RPQM model.~The very sharp and high specific heat peak at the $T_{c}^{\chi}$ for the Log RPQM model in the Fig~\ref{fig13b},~becomes quite smooth, reduced and round due to the robust effect of the quark back-reaction when the Log-glue or the PolyLog-glue form has been taken for the unquenching of the Polyakov-loop potential in the RPQM model.~We point out that the two peak structure in the Log-glue RPQM model plot for the specific heat is the consequence of the fact that the temperature variation of the  chiral order parameter has two peaks as one can see in the Fig~\ref{fig4a}.~The specific heat plots for all the cases approach the corresponding SB-limit at high temperatures.~Since the gluon degrees of freedom contribute differently \cite{SchaPQM3F}, 
the specific heat plot for the Log RPQM model in the Fig~\ref{fig13b}
does not merge completely with the Log-glue and PolyLog-glue RPQM model plots even when  $T=3 \ T_{c}^{\chi}$.  
 
The speed of sound is fundamental property of the strongly interacting medium.~The square of the speed of sound at constant entropy density is defined as

\bqa
C^2_s=\frac{\partial P}{\partial \epsilon}\bigg{\rvert}_s=\frac{\partial P}{\partial T}\bigg{\rvert}_V\bigg{/}\frac{\partial \epsilon}{\partial T}\bigg{\rvert}_V=\frac{s}{C_V}
\eqa

The speed of sound and the ratio $P/\epsilon$, have been plotted in the Fig.~\ref{fig:mini:fig14} on the reduced temperature scale.~The $C_s^2$ is very close to the $P/\epsilon$ for the $T < 0.7 \ T_{c}^{\chi}$ .~While approaching its ideal gas limit of $1/3$ for the $T > 2.5 \ T_{c}^{\chi}$,~the $C_s^2$ comes close to the $P/\epsilon$ again.~Similar to the results of Ref. \cite{schafwag12,guptiw},~the $C^2_{s}$ plot lies sufficiently above the $P/\epsilon$ plot for the temperature interval $ 0.7 \ T_{c}^{\chi} < T < 2.5 \ T_{c}^{\chi}$ except near the $T_{c}^{\chi}$.~The very sharp and a relatively less sharp dip,~noticed at the $T_{c}^{\chi}$ respectively in the $C_s^2$ and  the $P/\epsilon$ plot for the PolyLog-glue PQM model in the Fig.~\ref{fig14a},~becomes very smooth and moderately smooth for the respective cases of  the PQMVT and RPQM model.~In the PolyLog-glue RPQM model,~the speed of sound $C_s^2$ plot,~shows a better agreement with the lattice QCD data around the $T_{c}^{\chi}$ and the minimum value 0.078 for the ratio $P/\epsilon$ is closest to the corresponding minimum value of 0.075 obtained from the lattice QCD in Ref. \cite{Cheng:08}.~The Fig.~\ref{fig14b} shows that the sharp $C_s^2$ and  $P/\epsilon$ temperature variations for the Log RPQM model,~becomes quite smooth due to the quark back-reaction in the Log-glue and the PolyLog-glue RPQM model where the Ployakov-loop potential is unquenched.~The Log RPQM model plot for the speed of sound $C^2_{s}$, lies  closest to the lattice QCD data points at higher temperatures ($T > 1.4 \ T_{c}^{\chi}$).~Similar to the  findings of the Ref.~\cite{schafwag12,guptiw},~the $C_s^2$ value is found to be less than 0.1 below the $0.5 \ T^{\chi}_c$ (not shown in the figure) in the all of our model results.~This stands in contrast to the results of the Ref.~\cite{bedanga} where a confinement model has been applied and the value of
 $C^2_s \sim$ 0.2 in the vicinity of the $0.5 \ T^{\chi}_c$ and $C^2_s \sim$ 0.15 near the
 $T_{c}^{\chi}$.
 
\section{Summary and Conclusion}
\label{sec:VI}

Computing the phase diagrams and the thermodynamic quantities for the PQM, PQMVT and RPQM model,~we have compared how,~the quark one-loop vacuum correction in combination of the Polyakov-loop potential,~influences the dynamics of both the chiral as well as the confinement-deconfinement transition.~Taking different parametric form for the Polyakov-loop potential with/without the quark back-reaction and the $\mu$ dependence of the parameter $T_{0}$,~we have also made a detailed comparison of how the chiral and Polyakov-loop condensates,~their derivatives,~phase diagrams and the thermodynamic quantities, get affected in the RPQM model which has the exact chiral effective potential.~The results obtained for the different thermodynamic quantities have also been compared with the available lattice QCD data.

The sharp temperature variation of the chiral condensate and its temperature derivative in the PQM model for the $\mu=$0 and $m_{\sigma}=500$ MeV,~become  excessively smooth in the PQMVT model while the corresponding smoothness turns out to be moderate when the effect of the quark one-loop vacuum correction is computed using the  exact on-shell renormalized RPQM model.~The deconfinement corssover transition occurs earlier than the chiral transition, i.e. $T^{\Phi}_c<T^{\chi}_c$, except for the cases of the PolyLog-glue PQM and the Log RPQM model where the $T^{\Phi}_c \sim T^{\chi}_c$.~The presence of the quark back-reaction (represented by the glue form of the Ployakov-loop potential), in all the three chiral models,~generates a significant smoothing effect on the temperature variations of the chiral condensate, the Polyakov-loop condensate and their derivatives.~The combined effect of the quark back-reaction and the one-loop vacuum correction for the PolyLog-glue PQMVT model,~generates the highest separation of the $(T^{\chi}_c-T^{\Phi}_c)=$ 30.2 MeV  between the deconfinement and the chiral crossover transition while the corresponding separation generated for the PolyLog-glue RPQM model is $(T^{\chi}_c-T^{\Phi}_c)=$ 21.0 MeV.

~The phase boundary for the RPQM model stands closer to the PQM model phase boundary when compared to that of the PQMVT model.~Since the Log PQMVT model generates excessively smooth chiral transition similar to the earlier findings \cite{guptiw,schafwag12,chatmoh1,vkkr12,chatmoh2,RaiTiw},~the critical end point (CEP) respectively for the $m_{\sigma}=400$ and 500 MeV cases, gets located at the ($T_{\cep}$ = 94.77 MeV, \ $\mu_{\cep}$ = 267.2 MeV) and the ( $T_{\cep}$ = 78.0 MeV, \ $\mu_{\cep}$ = 295.9 MeV) positions in the right lower corner of the phase diagram.~The on-shell renormalized Log-RPQM model,~gives rise to a relatively moderate smoothing effect on the chiral transition, hence the CEP gets located higher up in the phase diagram respectively at the ($T_{\cep}$ = 100.1 MeV, \ $\mu_{\cep}$ = 245.3 MeV) and the ($T_{\cep}$ = 94.1 MeV, \ $\mu_{\cep}$ = 270.6 MeV) for the $m_{\sigma}=400$ and 500 MeV.

When compared to the CEP in the Log RPQM model for the $m_{\sigma}=500$ MeV,~the CEP for the PolyLog-glue RPQM model with the $T_{0} \equiv T_{0}(\mu)$, shifts down in temperature by 38.5 MeV due to the significant smoothing effect of the quark back-reaction in the temperature direction  and gets located at the $T_{\cep}$ = 55.6 MeV while the chemical potential registers a negligible shift as the $\mu_{\cep}$ = 270.9 MeV.~Due to the above effect, the curvature of the phase transition line increases significantly for all the cases where the unquenching of the Polyakov-loop potential has been considered.~In our work,~the observation of the Ref. \cite{BielichP} gets confirmed that the unquenching of the Polyakov-loop potential,~links the chiral and deconfinement phase transitions at all temperatures and chemical potentials.~Comparing the results of the Polyakov-loop enhanced calculations with the 
two flavor RQM model results \cite{RaiTiw} where the effect of Polyakov-loop is absent,~we note that the presence of Polyakov-loop potential either in the Log or in the PolyLog-glue form in the RPQM model,~leads to significant upward shift of the CEP in the $\mu-T$ plane.  

The occurrence of the so called quarkyonic phase,~where the chiral symmetry is restored but the quarks and anti-quarks are confined,~depends on the form taken for the Polyakov loop potential.~It occurs in a large region of the phase diagram for the Poly RPQM model while its region gets reduced for the Log RPQM model.~Its region gets significantly reduced by the quark back-reaction in the unquenched Polyakov-loop potential.~It altogether disappears when the parameter $T_{0}$ becomes chemical potential dependent i.e. $T_{0} \equiv T_{0} (\mu)$ in the Log or the PolyLog-glue form of the Polyakov-loop potential in the RPQM model while it occurs with a significantly reduced extent for the corresponding case of the Polynomial form of the Polyakov-loop potential in the RPQM model.

The thermodynamic quantities namely the pressure,~entropy density,~energy density,~interaction measure,~specific heat,~the speed of sound and the $P/\epsilon$ ratio have been computed for the $\mu=$0 and the $m_{\sigma}=$500 MeV.~As expected their variations in the PQMVT model, are smoother than that of the RPQM Model.~The sharp reduced temperature scale variation in the Log RPQM model,~becomes quite smooth due to the quark back-reaction in the Log-glue and the PolyLog-glue RPQM model.~Comparing the results with the available lattice QCD data,~we find that near the chiral transition $(0.7-1.2)\ T^{\chi}_c$,~the Polylog-glue RPQM model results for all the thermodynamic quantities, are in best agreement with the lattice QCD results.~The Log RPQM model results are close to the lattice QCD results at higher temperatures.~In the PolyLog-glue RPQM model,~the minimum value 0.078 for the ratio $P/\epsilon$ is closest to the corresponding minimum value of 0.075 obtained from the lattice QCD in Ref. \cite{Cheng:08}.


\section*{Acknowledgments}
The authors would like to thank Shinji Ejiri for providing the lattice data of CP-PACS collaboration \cite{AliKhan:2001ek}.
\appendix
\section{THE QMVT PARAMETER FIXING} 
\label{appenA}
The tree level expression of the curvature masses of mesons for the QM model are given by the mass matrix evaluated in Ref. \cite{Roder}.~In this work,~the above mass matrix is renamed as $(m^m_{\alpha,ab})^2$ where superscript $m$ stands for the contribution of the pure mesonic potential.In the QMVT model,the meson curvature masses get modified by the quark one-loop vacuum contribution.The total expression of the meson curvature masses in the QMVT model is written as  
\bqa
\label{m2meson}
m^2_{\alpha,ab}&=&(m^m_{\alpha,ab})^2+(\delta m^v_{\alpha,ab})^2
\eqa
where $\alpha=s,p$; ``$s$'' stands for the scalar and ``$p$'' stands for the pseudoscalar mesons and $a,b=0,1,2,3$.
$m^2_{s,00}\equiv m^2_\sigma$; $m^2_{s,11}=m^2_{s,22}=m^2_{s,33}\equiv m^2_{a_0}$ and $m^2_{p,00}\equiv m^2_\eta$; $m^2_{p,11}=m^2_{p,22}=m^2_{p,33}\equiv m^2_{\pi}$.The $(m^m_{\alpha,ab})^2$ and $(\delta m^v_{\alpha,ab})^2$  are defined in the similar fashion.Superscript ``$v$'' stands for the quark/antiquark vacuum contribution to the curvature masses. It is written as : 

\bqa
\label{mvac}
\nonumber
(\delta m^v_{\alpha,ab})^2&=&\left.\frac{\partial^2 \Omega^{vac}_{q\bar q}}{\partial \xi_{\alpha,a} \partial \xi_{\alpha,b}}\right|_{min}\;,\\ 
\nonumber
&=&\sum_{q=u,d} \frac{2N_c}{(4\pi)^2}\left[\left\lbrace m^2_{q,\alpha a}m^2_{q,\alpha b}+m^2_qm^2_{q,\alpha ab}\right\rbrace\right. \\
&&\left.\left\lbrace 1+\ln\left(\frac{\Lambda^2}{m^2_q}\right)\right\rbrace-m^2_{q,\alpha a}m^2_{q,\alpha b}\right] \;.
\eqa
where $m^2_{q,\alpha a}=\frac{\partial m^2_q}{\partial \xi_{\alpha,a}}$ and $m^2_{q,\alpha ab}=\frac{\partial m^2_{q,\alpha a}}{\partial \xi_{\alpha,b}}$.
When one computes the second derivative of the Eq.(\ref{vacdiv}) 
for the quark contribution, the full dependence of all the scalar
and pseudo-scalar meson fields, cf. Eq. (\ref{mesmat}), in the quark
masses has to be considered. The resulting quark mass 
matrix is diagonalized similar to the three flavor case given in Ref. \cite{Schaefer:09}. In all the quark mass derivatives with respect to
the meson fields, the meson fields are replaced by the non vanishing vacuum
expectation value $\overline{\sigma}$ and the final values are collected in the Table \ref{tab:table4}.

\begin{table}[!htbp]
  \begin{center}
    \caption{Expressions of the curvature masses $(m^m_{\alpha,ab})^2$ are calculated from the second derivative of the pure mesonic potential as has been evaluated in Ref.\cite{Roder}.}
    \label{tab:table3}
    \begin{tabular}{p{2cm} p{2cm} p{3.7cm}}
      \toprule 
      $(m^m_{\alpha,ab})^2$&& Meson mass found from the pure mesonic potential  \\
      \hline 
      \hline
      $(m^m_{s,00})^2$&$(m^m_\sigma)^2$& $m^2-c+3\left(\lambda_1+\frac{\lambda_2}{2}\right)\overline{\sigma}^2$\;,\\
      $(m^m_{s,11})^2$&$(m^m_{a_0})^2$&$m^2+c+\left(\lambda_1+\frac{3\lambda_2}{2}\right)\overline{\sigma}^2$\;,\\
      $(m^m_{p,00})^2$&$(m^m_\eta)^2$&$m^2+c+\left(\lambda_1+\frac{\lambda_2}{2}\right)\overline{\sigma}^2$\;,\\
      $(m^m_{p,11})^2$&$(m^m_\pi)^2$& $m^2-c+\left(\lambda_1+\frac{\lambda_2}{2}\right)\overline{\sigma}^2$\;,\\
      \hline 
    \end{tabular}
  \end{center}
\end{table}

\begin{table}[!htbp]
  \begin{center}
    \caption{Squared quark mass derivatives with respect to the meson fields evaluated at the minimum.The last two columns present the first and second derivative of the squared quark mass summed over two quark flavor.}
    \label{tab:table4}
    \begin{tabular}{p{1cm} p{1cm} p{1cm} p{2.5cm} p{2.5cm}}
      \toprule 
      $s/p$&$a$&$b$  & $m^2_{q,\alpha a}m^2_{q,\alpha b}/g^4$ & $m^2_{q,\alpha ab}/g^2$\\
      \hline 
      \hline
      $s$& $0$ &$0$ & $\frac{1}{2}{\overline{\sigma}}^2$ & 1\\
      $s$& $1$ &$1$ & $\frac{1}{2}{\overline{\sigma}}^2$ & 1\\
      $p$& $0$ &$0$ & 0 & 1\\
      $p$& $1$ &$1$ & 0 & 1\\
      \hline 
    \end{tabular}
  \end{center}
\end{table}
Using the Table-\ref{tab:table4} in the Eq.(\ref{mvac}) we get vacuum contributions of meson masses as,
\bqa
\label{vacmass1}
(\delta m^v_\sigma)^2&\equiv&(\delta m^v_{s,00})^2=\dfrac{N_cg^4\overline{\sigma}^2}{2(4\pi)^2}\left[1+3\ln\left(\dfrac{\Lambda^2}{m^2_q}\right)\right]\;, \
\eqa
\bqa
(\delta m^v_{a_0})^2&\equiv&(\delta m^v_{s,11})^2=\dfrac{N_cg^4\overline{\sigma}^2}{2(4\pi)^2}\left[1+3\ln\left(\dfrac{\Lambda^2}{m^2_q}\right)\right]\;, \
\eqa
\bqa
(\delta m^v_\eta)^2&\equiv&(\delta m^v_{p,00})^2=\dfrac{N_cg^4\overline{\sigma}^2}{2(4\pi)^2}\left[1+\ln\left(\dfrac{\Lambda^2}{m^2_q}\right)\right]\;, \
\eqa
\bqa
(\delta m^v_\pi)^2&\equiv&(\delta m^v_{p,11})^2=\dfrac{N_cg^4\overline{\sigma}^2}{2(4\pi)^2}\left[1+\ln\left(\dfrac{\Lambda^2}{m^2_q}\right)\right]\;. \
\label{vacmass4}
\eqa
We get $(m^m_\sigma)^2$,$(m^m_\eta)^2$,$(m^m_{a_0})^2$ and $(m^m_\eta)^2$ after substitution of the Eqs.(\ref{vacmass1})--(\ref{vacmass4}) into the Eq.(\ref{m2meson}) as,
\bqa
\label{mesonm1}
(m^m_\sigma)^2&=&m^2_\sigma-\dfrac{N_cg^4\overline{\sigma}^2}{2(4\pi)^2}\left[1+3\ln\left(\dfrac{\Lambda^2}{m^2_q}\right)\right]\;,
\eqa
\bqa
(m^m_{a_0})^2&=&m^2_{a_0}-\dfrac{N_cg^4\overline{\sigma}^2}{2(4\pi)^2}\left[1+3\ln\left(\dfrac{\Lambda^2}{m^2_q}\right)\right]\;,
\eqa
\bqa
(m^m_\eta)^2&=&m^2_\eta-\dfrac{N_cg^4\overline{\sigma}^2}{2(4\pi)^2}\left[1+\ln\left(\dfrac{\Lambda^2}{m^2_q}\right)\right]\;,
\eqa
\bqa
(m^m_\pi)^2&=&m^2_\pi-\dfrac{N_cg^4\overline{\sigma}^2}{2(4\pi)^2}\left[1+\ln\left(\dfrac{\Lambda^2}{m^2_q}\right)\right]\;.
\label{mesonm4}
\eqa

The parameters in vacuum are obtained as,

\bqa
\label{paraqmvt1}
\lambda_1&=&(m^m_\sigma)^2+(m^m_\eta)^2-(m^m_{a_0})^2-(m^m_\pi)^2\over 2 f^2_\pi\;,
\eqa
\bqa
\lambda_2&=&(m^m_{a_0})^2-(m^m_\eta)^2\over f^2_\pi\;,
\eqa
\bqa
m^2&=&(m^m_\pi)^2+\frac{(m^m_\eta)^2-(m^m_\sigma)^2}{2}\;,
\eqa
\bqa
c&=&\frac{(m^m_\eta)^2-(m^m_\pi)^2}{2}\;.
\label{paraqmvt4}
\eqa
we get the parameters of the QMVT on substitution of the Eqs.(\ref{mesonm1})--(\ref{mesonm4}) into the Eqs.(\ref{paraqmvt1})--(\ref{paraqmvt4}) and found that $\lambda_1$,$c$ of the QMVT are same with respect  to $\lambda_1$,$c$ of the QM. We observe change in $\lambda_2$ and $m^2$ as,
\bqa
\lambda_2&=&\lambda_{2s}-\dfrac{N_cg^4}{(4\pi)^2}\ln\left(\dfrac{4\Lambda^2}{g^2f^2_\pi}\right)\;,
\eqa
\bqa
m^2&=&m^2_{s}-\dfrac{N_cg^4f^2_\pi}{2(4\pi)^2}\;.
\eqa
where $\lambda_{2s}$ and $m^2_s$ are same old $\lambda_{2}$ and $m^2$ parameters of the QM model.Putting the value of the new parameters $\lambda_2$ and $m^2$ in Eq. (\ref{mesonm4})
one can write the expression of pion mass independent of renormalization scale as,
\bqa
\nonumber
m^2_\pi&=&m^2_s-\frac{N_cg^4}{2(4\pi)^2}(f^2_\pi-\overline{\sigma}^2)-c+\left(\lambda_1+\frac{\lambda_{2s}}{2}\right)\overline{\sigma}^2\; \\
&&+\frac{N_cg^4}{2(4\pi)^2}\log\left(\frac{f^2_\pi}{\overline{\sigma}^2}\right)\overline{\sigma}^2
\eqa

\section{Fixing of the parameters in the RQM model}
\label{appenB}
In the RQM model, the divergence of the first term of the Eq.~(\ref{vac1}) (as rewritten in the Eq.~(\ref{vacdiv})) , is removed by the renormalization of its parameters. This Appendix presents the relation between the physical quantities and the parameters of the Lagrangian~(\ref{lag}) using the $\overline{\text{MS}}$ and on-shell renormalization schemes \cite{RaiTiw}. One introduces the counterterms $\delta m^{2} $, $\delta g^{2} $, $\delta \lambda_{1} $,  $\delta \lambda_{2} $, $\delta c$ and $\delta h$ for the parameters and the wave function/field counterterms $\delta Z_\sigma $,$\delta Z_{a_0} $,$\delta Z_\eta $,$\delta Z_\pi $ and $\delta Z_\psi $ in the Lagrangian (\ref{lag}) to define the renormalized fields and couplings. 

\bqa
\label{ctrm1}
\sigma_b=\sqrt{Z_\sigma}\sigma, \ \eta_b=\sqrt{Z_\eta}\eta, \ a^i_{0b}=\sqrt{Z_{a_{0}}}a_{0}
\\ \pi^i_b=\sqrt{Z_\pi}\pi, \ \psi_b=\sqrt{Z_\psi}\psi, \ m^2_b=Z_m m^2 \\
\lambda_{1b}=Z_{\lambda_1} \lambda_1, \ \lambda_{2b}=Z_{\lambda_2} \lambda_2, \ g_b=\sqrt{Z_g}g
\\ h_b=Z_h h,\ \ \ c_b=Z_c c
\label{ctrm5}
\eqa
Where $Z_ {(\sigma,a_0,\eta,\pi,\psi )}=1+\delta Z_{(\sigma,a_0,\eta,\pi,\psi )} $, denote  the field strength renormalization constant while $Z_ {(m,\lambda_1,\lambda_2,g,h,c )}=1+\delta Z_{(m,\lambda_1,\lambda_2,g,h,c )} $ denote the mass and coupling renormalization constant.

The scalar ($\sigma,a_0 $) and pseudo-scalar
($\pi,\eta $) meson inverse propagators with self-energy correction are written as
\bqa
p^2-m_{\sigma,a_0,\pi,\eta}^2-i\Sigma_{\sigma,a_0,\pi,\eta}(p^2)
{\rm +counterterms}
\;.
\label{definv}
\eqa
Implementing the on-shell scheme, one puts the physical mass equal to the renormalized mass in the Lagrangian i.e.\ $m=m_{\rm pole}$\footnote{The imaginary parts of the self-energies are ignored for defining the mass.} and writes
\bqa
\Sigma(p^2=m_{\sigma,a_0,\pi,\eta}^2)
{\rm +counterterms}
&=&0
\label{pole}
\;.
\eqa
The on-shell scheme demands that the propagator residue becomes unity and one gets 
\bqa
\label{res}
{\partial\over\partial p^2}\Sigma_{\sigma,a_0,\pi,\eta}(p^2)\Big|_{p^2=m_{\sigma,a_0,\pi,\eta}^2}
{\rm +counterterms}
&=&0\;.
\eqa

The quark one-loop correction to the one-point function and the tadpole counterterm can be written as 
\bqa
\delta\Gamma^{(1)}
&=&-
4 N_c g m_q \mathcal{A}(m_q^2)+i\delta t
\;,
\eqa
when the one-point function $\Gamma^{(1)}=it=i(h-m_{\pi}^2 \  \overline \sigma)$ vanishes, one gets the equation of motion $t=0$ at  tree level. This should hold also at one-loop level, which results into the renormalization condition 
$\delta\Gamma^{(1)}=0$. 

The counterterms of the two-point functions are
\bqa
\label{count1}
\Sigma_{\sigma}^{\rm ct1}(p^2)&=&i\left[\delta Z_{\sigma}(p^2-m_{\sigma}^2)-\delta m_{\sigma}^2\right]\;,
\eqa
\bqa
\label{count2}
\Sigma_{a_0}^{\rm ct1}(p^2)&=&i\left[\delta Z_{a_0}(p^2-m_{a_0}^2)-\delta m_{a_0}^2\right]\;,
\eqa
\bqa
\label{count3}
\Sigma_{\pi}^{\rm ct1}(p^2)&=&i\left[\delta Z_{\pi}(p^2-m_{\pi}^2)-\delta m_{\pi}^2\right]\;,
\eqa
\bqa
\label{count4}
\Sigma_{\eta}^{\rm ct1}(p^2)&=&i\left[\delta Z_{\eta}(p^2-m_{\eta}^2)-\delta m_{\eta}^2\right]\;,
\eqa
\bqa
\label{count5}
\nonumber
\Sigma^{\rm ct2}_{\sigma}&=&3\Sigma^{\rm ct2}_{a_0}=3\Sigma^{\rm ct2}_{\pi}=3\Sigma^{\rm ct2}_{\eta}\;,
\\ \nonumber
&=&-{24(\lambda_1+\mbox{$\lambda_2\over 2$}) g \overline{\sigma}N_cm_q\over m_{\sigma}^2}\mathcal{A}(m_q^2)\;,
\\
&=&-6i(\lambda_1+\mbox{$\lambda_2\over 2$})\overline{\sigma} \delta t\over m_{\sigma}^2\;,\\
\delta t
&=&-4i N_c g m_q \mathcal{A}(m_q^2)  \;.
\eqa

The respective tadpole contributions to the $\sigma, a_0$ and $\pi,\eta$ self-energies are cancelled by the counterterms in Eq.~(\ref{count5}). The on-shell evaluation of the self-energies and their derivatives give all the  renormalization constants. Combining the Eqs.~(\ref{pole}), (\ref{res}) and (\ref{count1})--(\ref{count4}), one obtains the following set of equations.

\bqa
\delta m_{\sigma}^2&=&-i\Sigma_{\sigma}(m_{\sigma}^2)\;;\delta Z_\sigma =
i{\partial\over\partial p^2}\Sigma_\sigma(p^2)\Big|_{p^2=m_\sigma^2}\;,
\eqa
\bqa
\delta m_{a_0}^2&=&-i\Sigma_{a_0}(m_{a_0}^2)\;;\delta Z_{a_0} =
i{\partial\over\partial p^2}\Sigma_{a_0}(p^2)\Big|_{p^2=m_{a_0}^2}\;,
\eqa
\bqa
\delta m_{\pi}^2&=&-i\Sigma_{\pi}(m_{\pi}^2)\;;\delta Z_\pi= i{\partial\over\partial p^2}\Sigma_\pi(p^2)\Big|_{p^2=m_\pi^2}\;,
\eqa
\bqa
\delta m_{\eta}^2&=&-i\Sigma_{\eta}(m_{\eta}^2)\;;\delta Z_\eta =i{\partial\over\partial p^2}\Sigma_\eta(p^2)\Big|_{p^2=m_\eta^2}\;.
\eqa

Using the above equations and the expressions of meson self-energies given in Ref.~\cite{RaiTiw}, we can write 
\bqa
\label{countf1}
\delta m_{\sigma}^2
&=&
2ig^2N_c\left[\mathcal{A}(m_q^2)-\mbox{$1\over2$}(m_{\sigma}^2-4m_q^2)\mathcal{B}(m_{\sigma}^2)
\right]\;,
\eqa
\bqa
\delta m_{a_0}^2
&=&
2ig^2N_c\left[\mathcal{A}(m_q^2)-\mbox{$1\over2$}(m_{a_0}^2-4m_q^2)\mathcal{B}(m_{a_0}^2)
\right]\;,
\eqa
\bqa
\delta m_{\pi}^2
&=&
2ig^2N_c\left[\mathcal{A}(m_q^2)-\mbox{$1\over2$}m_{\pi}^2\mathcal{B}(m_{\pi}^2)
\right]\;,
\eqa
\bqa
\delta m_{\eta}^2
&=&
2ig^2N_c\left[\mathcal{A}(m_q^2)-\mbox{$1\over2$}m_{\eta}^2\mathcal{B}(m_{\eta}^2)
\right]\;,
\eqa
\bqa
\delta Z_{\sigma}&=&
ig^2N_c\left[\mathcal{B}(m_{\sigma}^2)+(m_{\sigma}^2-4m_q^2)\mathcal{B}^{\prime}(m_{\sigma}^2)
\right]\;,
\eqa
\bqa
\delta Z_{a_0}&=&
ig^2N_c\left[\mathcal{B}(m_{a_0}^2)+(m_{a_0}^2-4m_q^2)\mathcal{B}^{\prime}(m_{a_0}^2)
\right]\;,
\eqa
\bqa
\delta Z_{\pi}&=&
ig^2N_c\left[\mathcal{B}(m_{\pi}^2)+m_{\pi}^2\mathcal{B}^{\prime}(m_{\pi}^2)
\right]\;,
\eqa
\bqa
\label{countf8}
\delta Z_{\eta}&=&
ig^2N_c\left[\mathcal{B}(m_{\eta}^2)+m_{\eta}^2\mathcal{B}^{\prime}(m_{\eta}^2)
\right]\;,
\eqa
where $\mathcal{A}(m^2_q)$ and $\mathcal{B}(p^2)$ are defined in the Appendix~\ref{appenC}.

The counterterms $\delta m^{2} $,$\delta \lambda_{1} $,  $\delta \lambda_{2} $, $\delta c$ and $\delta g^{2} $, $\delta h$ can be expressed in terms of the counterterms $\delta m^{2}_{\sigma} $, $\delta m^{2}_{a_0} $, $\delta m^{2}_{\eta} $, $\delta m^{2}_{\pi} $ and  $\delta m_{q} $, $\delta \overline \sigma^2 $. Using Eqs.~(\ref{ch5_m1})--(\ref{ch5_m4}) together with Eqs.~(\ref{ctrm1})--(\ref{ctrm5}), we can write
\bqa
\label{delta:lambda_1}
\delta \lambda_{1}&=&\frac{\delta m^2_\sigma+\delta m^2_\eta-\delta m^2_{a_0}-\delta m^2_\pi}{2 \ \overline \sigma^2 }-\lambda_1 \frac{\delta  \overline \sigma^2}{ \overline \sigma^2}\;,
\\
\label{delta:lambda_2}
\delta \lambda_{2}&=&\frac{\delta m^2_{a_0}-\delta m^2_\eta}{\overline \sigma^2}-\lambda_2 \frac{\delta  \overline \sigma^2}{ \overline \sigma^2}\;,
\\
\delta c&=&\frac{\delta m^2_\eta-\delta m^2_\pi}{2}\;,
\\
\delta m^2&=&\delta m^2_\pi+\frac{\delta m^2_\eta-\delta m^2_\sigma}{2}\;,
\\
\frac{\delta g^2}{4}&=&\frac{\delta m^2_q}{\overline \sigma^2}-g^2\frac{\delta  \overline \sigma^2}{4 \ \overline \sigma^2}
\label{delta:g}
\eqa

The one loop correction at the pion-quark vertex is of order $N_{c}^0$. Hence $Z_\psi=1$ and the quark self energy correction $\delta m_q=0$ at this order. In consequence, we get $Z_\psi \sqrt{Z_{g^2} \ g^2}\sqrt{Z_{\pi}}\approx g(1+\frac{1}{2}\frac{\delta g^2}{g^2} \ + \frac{1}{2}\delta Z_\pi)=g$, thus $\frac{\delta g^2}{g^2} \ + \delta Z_\pi=0$. Furthermore the $\delta m_q=0$ implies that $\delta g \ \overline \sigma/2 + g \ \delta \overline \sigma/2 =0 $. Eq.~(\ref{delta:g}) gives 
\bqa
\label{Zpi}
\frac{\delta  \overline \sigma^2}{ \overline \sigma^2}&=&-\frac{\delta g^2}{g^2}=\delta Z_\pi,
\eqa
Now we can rewrite Eq.~(\ref{delta:lambda_1}), (\ref{delta:lambda_2}) as
\bqa
\label{delta:lambda_1n}
\delta \lambda_{1}&=&\frac{\delta m^2_\sigma+\delta m^2_\eta-\delta m^2_{a_0}-\delta m^2_\pi}{2 \ \overline \sigma^2}-\lambda_1 \delta Z_\pi,
\\
\label{delta:lambda_2n}
\delta \lambda_{2}&=&\frac{\delta m^2_{a_0}-\delta m^2_\eta}{\overline \sigma^2}-\lambda_2 \delta Z_\pi.
\eqa
The equation $h=t+m_{\pi}^2 \  \overline \sigma $ enables the writing of the
counterterm $\delta h$ in terms of the tadpole counterterm  $\delta t$
\bqa
\label{delta:h}
\delta h&=&m^2_\pi \ \delta \overline \sigma +\delta m^2_\pi \ \overline \sigma +\delta t.
\eqa
Using Eq.~(\ref{Zpi}) we can write
\bqa
\label{delta:hn}
\delta t&=&-\frac{1}{2}m^2_\pi \ \overline \sigma  \delta Z_\pi-\delta m^2_\pi \ \overline \sigma +\delta h.
.
\eqa
Exploiting the Eqs. (\ref{countf1})--(\ref{countf8}) together with the Eqs. (\ref{delta:lambda_1}) --(\ref{Zpi}) and (\ref{delta:hn}), we find the following expessions for the counterterms in the on-shell scheme.
\begin{widetext}
 \bqa \nonumber
 \label{os1}
\delta\lambda_{1\os}&=&\frac{iN_cg^2}{\overline{\sigma}}\left[-\frac{1}{2}(m^2_\sigma-4m^2_q)\mathcal{B}(m^2_\sigma)-\frac{1}{2}m^2_\eta \mathcal{B}(m^2_\eta)+\frac{1}{2}(m^2_{a_0}-4m^2_q)\mathcal{B}(m^2_{a_0})+\frac{1}{2}m^2_\pi \mathcal{B}(m^2_\pi)\right]-\lambda_1ig^2N_c\left[\mathcal{B}(m_{\pi}^2)\right. \\ \nonumber
&&\left.+m_{\pi}^2\mathcal{B}^{\prime}(m_{\pi}^2)\right]=\delta \lambda_{1\text{div}}+\dfrac{N_cg^2}{(4\pi)^2}\left[2\lambda_1\log\left(\frac{\Lambda^2}{m_q^2}\right)+\lambda_1(\mathcal{C}(m^2_\pi)+m^2_\pi\mathcal{C}^\prime(m^2_\pi))\right. \\
&&\left.+\frac{(m^2_\sigma-4m^2_q)\mathcal{C}(m^2_\sigma)+m^2_\eta \mathcal{C}(m^2_\eta)-(m^2_{a_0}-4m^2_q)\mathcal{C}(m^2_{a_0})-m^2_\pi \mathcal{C}(m^2_\pi)}{2\overline{\sigma}^2}\right]\;,
\\
\nonumber
\delta\lambda_{2\os} 
&=&\frac{iN_cg^2}{\overline{\sigma}^2}\left[-(m^2_{a_0}-4m^2_q)\mathcal{B}(m^2_{a_0})+m^2_\eta \mathcal{B}(m^2_\eta)\right]-\lambda_2ig^2N_c\left[\mathcal{B}(m_{\pi}^2)+m_{\pi}^2\mathcal{B}^{\prime}(m_{\pi}^2)\right]\;,
\\
&=&\delta \lambda_{2\text{div}}+\dfrac{N_cg^2}{(4\pi)^2}\left[\left(2\lambda_2-g^2\right)\log\left(\frac{\Lambda^2}{m_q^2}\right)+\frac{(m^2_{a_0}-4m^2_q)\mathcal{C}(m^2_{a_0})-m^2_\eta \mathcal{C}(m^2_\eta)}{{\overline{\sigma}^2}}+\lambda_2(m^2_\pi \mathcal{C}^{\prime}(m^2_\pi)+\mathcal{C}(m^2_\pi))\right]\;,
\\
\nonumber
\delta m^2_{\os}&=&2iN_cg^2\left[\mathcal{A}(m^2_q)-\frac{1}{2}m^2_\pi \mathcal{B}(m^2_\pi)\right]+iN_cg^2\left[-\frac{1}{2}m^2_\eta \mathcal{B}(m^2_\eta)+\frac{1}{2}(m^2_\sigma-4m^2_q)\mathcal{B}(m^2_\sigma)\right]\;
\\
&&=\delta m^2_{\text{div}}+\dfrac{N_cg^2}{(4\pi)^2}\left[m^2\log\left(\frac{\Lambda^2}{m_q^2}\right)
+m^2_\pi \mathcal{C}(m^2_\pi)+\dfrac{m^2_{\eta}\mathcal{C}(m^2_{\eta})-(m^2_\sigma-4m^2_q) \mathcal{C}(m^2_\sigma)}{2}-2m^2_q\right]\;,
\\
\delta c_{\os}&=&\frac{iN_cg^2}{2}\left[-m^2_\eta \mathcal{B}(m^2_\eta)+m^2_\pi \mathcal{B}(m^2_\pi)\right]=\delta c_{\text{div}}+\dfrac{N_cg^2}{(4\pi)^2}\left[c\log\left(\frac{\Lambda^2}{m_q^2}\right)+m^2_{\eta}\mathcal{C}(m^2_{\eta})-m^2_\pi \mathcal{C}(m^2_\pi)\right]\;,
\\
\delta g^2_{\os}&=&-iN_cg^4\left[m^2_\pi \mathcal{B}^\prime(m^2_\pi)+\mathcal{B}(m^2_\pi)\right]=\delta g^2_{\text{div}}+\dfrac{N_cg^4}{(4\pi)^2}\left[\log\left(\frac{\Lambda^2}{m_q^2}\right)+\mathcal{C}(m^2_\pi)+m^2_\pi \mathcal{C}^{\prime}(m^2_\pi)\right]\;,
\\
\delta \overline{\sigma}^2_{\os}&=&iN_cg^2\overline{\sigma}^2\left[m^2_\pi \mathcal{B}^\prime(m^2_\pi)+\mathcal{B}(m^2_\pi)\right]=\delta \overline{\sigma}^2_{\text{div}}-\dfrac{N_cg^4}{(4\pi)^2}\left[\log\left(\frac{\Lambda^2}{m_q^2}\right)+\mathcal{C}(m^2_\pi)+m^2_\pi \mathcal{C}^{\prime}(m^2_\pi)\right]\;
\\
\label{os7}
\delta h_{\os}
&=&\dfrac{iN_cg^2}{2(4\pi)^2}h\left[m^2_\pi \mathcal{B}^\prime(m^2_\pi)-\mathcal{B}(m^2_\pi)\right]=\delta h_{\text{div}}+\dfrac{N_cg^2}{2(4\pi)^2}h\left[\log\left(\frac{\Lambda^2}{m_q^2}\right)+\mathcal{C}(m^2_\pi)-m^2_\pi \mathcal{C}^{\prime}(m^2_\pi)\right]\;\\
\delta Z^{\os}_{\pi}&=&\delta Z_{\pi,\rm div}-
\frac{N_cg^2}{(4\pi)^2}\left[\log\left(\frac{\Lambda^2}{m_q^2}\right)+\mathcal{C}(m_{\pi}^2)+m_{\pi}^2\mathcal{C}^{\prime}(m_{\pi}^2)
\right]\;
\eqa
The $\mathcal{B}(m^2),\mathcal{B}^{\prime}(m^2)$ and $\mathcal{C}(m^2),\mathcal{C}^{\prime}(m^2)$ are defined in the appendix.The divergent part of the counterterms are $\delta \lambda_{1\text{div}}=\frac{N_cg^22\lambda_1}{(4\pi)^2\epsilon}\;$, \ $ \delta \lambda_{2\text{div}}=\frac{N_cg^2}{(4\pi)^2\epsilon}(2\lambda_2-g^2)\;$, \ $\delta m^2_{\text{div}}=\frac{N_cg^2 m^2}{(4\pi)^2\epsilon}\;$, \ $\delta c_{\text{div}}=\frac{N_cg^2 c}{(4\pi)^2\epsilon}\;$, \ $\delta g^2_{\text{div}}=\frac{N_cg^4}{(4\pi)^2\epsilon}\; $, \ $\delta \overline{\sigma}^2_{\text{div}}=-\frac{N_cg^2\overline{\sigma}^2}{(4\pi)^2\epsilon}\; $, \ $ \delta h_{\text{div}}=\frac{N_cg^2 h}{2(4\pi)^2\epsilon}\;$, $\delta Z_{\pi,\rm div}=-\frac{N_cg^2}{(4\pi)^2\epsilon}$.

For both,the on-shell and the $\overline{\text{MS}}$ schemes, the divergent part of the counterterms are the same, i.e. $\delta \lambda_{1\text{div}}=\delta \lambda_{1\ms}$, \ $\delta \lambda_{2\text{div}}=\delta \lambda_{2\ms}$ etc.
\end{widetext}
Since the bare parameters are independent of the renormalization scheme, we can immediately write down the relations between the renormalized parameters in the on-shell and $\text{MS}$ schemes. We find
\bqa
\lambda_{1\ms}&=&\frac{Z^{\os}_{\lambda_1}}{Z^{\ms}_{\lambda_1}} \lambda_1\approx\lambda_1+\delta \lambda_{1\os}-\delta \lambda_{1\ms}\;,
\\
\lambda_{2\ms}&=&\frac{Z^{\os}_{\lambda_2}}{Z^{\ms}_{\lambda_2}} \lambda_2\approx\lambda_2+\delta \lambda_{2\os}-\delta \lambda_{2\ms}\;,
\\
m^2_{\ms}&=&\frac{Z^{\os}_{m^2}}{Z^{\ms}_{m^2}} m^2\approx m^2+\delta m^2_{\os}-\delta m^2_{\ms}\;,
\eqa
\bqa
c_{\ms}&=&\frac{Z^{\os}_{c}}{Z^{\ms}_{c}} c\approx c+\delta c_{\os}-\delta c_{\ms}\;,
\\
h_{\ms}&=&\frac{Z^{\os}_{h}}{Z^{\ms}_{h}} h\approx h+\delta h_{\os}-\delta h_{\ms}\;,
\\
g^2_{\ms}&=&\frac{Z^{\os}_{g^2}}{Z^{\ms}_{g^2}} g^2\approx g^2+\delta g^2_{\os}-\delta g^2_{\ms}\;,\\
{\overline{\sigma}}^2_{\ms}&=&\frac{Z^{\os}_{{\overline{\sigma}}^2}}{Z^{\ms}_{{\overline{\sigma}}^2}} {\overline{\sigma}}^2\approx {\overline{\sigma}}^2+\delta {\overline{\sigma}}^2_{\os}-\delta {\overline{\sigma}}^2_{\ms}\;.
\eqa
The minimum of the effective potential is at $\overline{\sigma}=f_\pi$ and masses have been measured in vacuum. Applying in Eqs.~(\ref{os1})--(\ref{os7}), we find the $\Lambda$ scale dependent parameters in $\overline{\text{MS}}$ scheme 
\begin{widetext}
\bqa
\label{params1}
\nonumber
\lambda_{1\ms}(\Lambda)&=&\lambda_1+\dfrac{N_cg^2}{(4\pi)^2}\left[\lambda_1\log\left(\frac{\Lambda^2}{m_q^2}\right)+\frac{(m^2_\sigma-4m^2_q)C(m^2_\sigma)+m^2_\eta C(m^2_\eta)-(m^2_{a_0}-4m^2_q)C(m^2_{a_0})-m^2_\pi C(m^2_\pi)}{2f_\pi^2}\right. \\
&&\left.+\lambda_1(\mathcal{C}(m^2_\pi)+m^2_\pi\mathcal{C}^\prime(m^2_\pi))\right],\qquad \quad\;\\
\lambda_{2\ms}(\Lambda)
&=&\lambda_2+\dfrac{N_cg^2}{(4\pi)^2}\left[\left(2\lambda_2-g^2\right)\log\left(\frac{\Lambda^2}{m_q^2}\right)+\frac{(m^2_{a_0}-4m^2_q)\mathcal{C}(m^2_{a_0})-m^2_\eta \mathcal{C}(m^2_\eta)}{f_\pi^2}+\lambda_2(\mathcal{C}(m^2_\pi)+m^2_\pi\mathcal{C}^\prime(m^2_\pi))\right]\;,
\\
m^2_{\ms}(\Lambda)
&=&m^2+\dfrac{N_cg^2}{(4\pi)^2}\left[m^2\log\left(\frac{\Lambda^2}{m_q^2}\right)
+m^2_\pi \mathcal{C}(m^2_\pi)+\dfrac{m^2_\eta C(m^2_\eta)-(m^2_\sigma-4m^2_q)C(m^2_\sigma)}{2}-2m_q^2\right]\;,
\\
c_{\ms}(\Lambda)
&=&c+\dfrac{N_cg^2}{(4\pi)^2}\left[c\log\left(\frac{\Lambda^2}{m_q^2}\right)+\frac{m^2_{\eta}\mathcal{C}(m^2_{\eta})-m^2_\pi \mathcal{C}(m^2_\pi)}{2}\right]\;,
\\
h_{\ms}(\Lambda)
&=&h+\dfrac{N_cg^2}{2(4\pi)^2}h\left[\log\left(\frac{\Lambda^2}{m_q^2}\right)+\mathcal{C}(m^2_\pi)-m^2_\pi \mathcal{C}^{\prime}(m^2_\pi)\right]\;,
\\
\label{params6}
g^2_{\ms}(\Lambda)
&=&g^2+\dfrac{N_cg^4}{(4\pi)^2}\left[\log\left(\frac{\Lambda^2}{m_q^2}\right)+\mathcal{C}(m^2_\pi)+m^2_\pi \mathcal{C}^{\prime}(m^2_\pi)\right]\;,
\\
\label{params7}
\overline{\sigma}^2_{\ms}(\Lambda)
&=&\overline{\sigma}^2-\dfrac{4N_cm^2_q}{(4\pi)^2}\left[\log\left(\frac{\Lambda^2}{m_q^2}\right)+\mathcal{C}(m^2_\pi)+m^2_\pi \mathcal{C}^{\prime}(m^2_\pi)\right]\;.
\eqa
\end{widetext}
Where the physical on-shell parameters are related with meson and quark mass at vacuum given by Eqs.(\ref{para1})--(\ref{para5}).

In the large-$N_c$ limit the parameters $\lambda_{1\ms}$,$\lambda_{2\ms}$,$m^2_{\ms}$,$c_{\ms}$,$h_{\ms}$ and $g^2_{\ms}$ are running with the scale $\Lambda$ and a set of simultaneous renormalization group equations are satisfied , which are
\bqa
\label{diffpara1}
\dfrac{d\lambda_{1\ms}(\Lambda)}{d\log(\Lambda)}&=&\dfrac{2N_c}{(4\pi)^2}g^2_{\ms}\lambda_{1\ms}\;,
\\
\dfrac{d\lambda_{2\ms}(\Lambda)}{d\log(\Lambda)}&=&\dfrac{2N_c}{(4\pi)^2}\left[2\lambda_{2\ms}g^2_{\ms}-g^4_{\ms}\right]\;,
\\
\dfrac{dm^2_{\ms}(\Lambda)}{d\log(\Lambda)}&=&\dfrac{2N_c}{(4\pi)^2}g^2_{\ms}m^2_{\ms}\;,
\\
\dfrac{d c_{\ms}(\Lambda)}{d\log(\Lambda)}&=&\dfrac{2N_c}{(4\pi)^2}g^2_{\ms}c_{\ms}\;,
\eqa
\bqa
\dfrac{d h_{\ms}(\Lambda)}{d\log(\Lambda)}&=&\dfrac{N_c}{(4\pi)^2}g^2_{\ms}h_{\ms}\;,
\\
\dfrac{d g^2_{\ms}}{d\log(\Lambda)}&=&\dfrac{2N_c}{(4\pi)^2}g^4_{\ms}\;,
\\
\label{diffpara7}
\dfrac{d \overline{\sigma}^2_{\ms}}{d\log(\Lambda)}&=&-\dfrac{2N_c}{(4\pi)^2}g^2_{\ms}\overline{\sigma}^2_{\ms}
\eqa
solutions of Eq.~(\ref{diffpara1})--(\ref{diffpara7}) are
\bqa
\label{para01}
\lambda_{1\ms}(\Lambda)&=&\frac{\lambda_{10}}{1-\frac{N_c g^2_0}{(4\pi)^2}\log\left(\dfrac{\Lambda^2}{\Lambda_0^2}\right)}\;,
\\
g^2_{\ms}(\Lambda)&=&\frac{g^2_0}{1-\dfrac{N_c g^2_0}{(4\pi)^2}\log\left(\dfrac{\Lambda^2}{\Lambda^2_0}\right)}\;,
\eqa
\bqa
\lambda_{2\ms}(\Lambda)&=&\frac{\lambda_{20}-\dfrac{N_c g^4_0}{(4\pi)^2}\log\left(\dfrac{\Lambda^2}{\Lambda^2_0}\right)}{\left(1-\dfrac{N_c g^2_0}{(4\pi)^2}\log\left(\dfrac{\Lambda^2}{\Lambda^2_0}\right)\right)^2}\;,  
\\
m^2_{\ms}(\Lambda)&=&\frac{m^2_0}{1-\dfrac{N_c g^2_0}{(4\pi)^2}\log\left(\dfrac{\Lambda^2}{\Lambda^2_0}\right)}\;,
\\
c_{\ms}(\Lambda)&=&\frac{c_0}{1-\dfrac{N_c g^2_0}{(4\pi)^2}\log\left(\dfrac{\Lambda^2}{\Lambda^2_0}\right)}\;,
\\
\label{para06}
h_{\ms}(\Lambda)&=&\frac{h_0}{\sqrt{1-\dfrac{N_c g^2_0}{(4\pi)^2}\log\left(\dfrac{\Lambda^2}{\Lambda^2_0}\right)}}\;,
\\
\label{para07}
\overline{\sigma}^2&=&f^2_\pi\left[1-\frac{N_cg^2_0}{(4\pi)^2}\log\left(\frac{\Lambda^2}{\Lambda^2_0}\right)\right]\;.
\eqa
Where the parameters $\lambda_{10}$,\ $\lambda_{20}$,\ $g^2_0$,\ $m^2_0$,\ $c_0$ and $h_0$, are the $\Lambda$ dependent parameters at the value $\Lambda_0$, we consider $\Lambda_0$ to satisfy the relation
\bqa
\log\left(\frac{\Lambda^2_0}{m_q^2}\right)+\mathcal{C}(m^2_\pi)+m^2_\pi \mathcal{C}^{\prime}(m^2_\pi)&=&0\;.
\eqa
\subsection{Effective Potential}
The vacuum effective potential in the $\overline{\text{MS}}$ scheme is given by
\bqa
\label{omegarqm}
\Omega_{vac}&=&U(\overline{\sigma}_{\ms})+\Omega^{q,vac}_{\ms}+\delta U(\overline{\sigma}_{\ms})\;,
\eqa
\begin{widetext}
where 
\bqa
\label{omegams1}
\nonumber
U(\overline{\sigma}_{\ms})&=&\frac{m_{\ms}^2(\Lambda)}{2}\overline{\sigma}^2_{\ms}-\frac{c_{\ms}(\Lambda)}{2}\overline{\sigma}^2_{\ms}+\frac{1}{4}\left(\lambda_{1\ms}(\Lambda)+\frac{\lambda_{2\ms}(\Lambda)}{2}\right)\overline{\sigma}^4_{\ms}-h_{\ms}(\Lambda)\overline{\sigma}_{\ms}\;,
\eqa $\delta U(\overline{\sigma}_{\ms})=-\frac{N_cg^4_{\ms}\overline{\sigma}^4_{\ms}}{8(4\pi)^2}\frac{1}{\epsilon}$ and $\Omega^{q,vac}_{\ms}=\frac{N_cg^4_{\ms}\overline{\sigma}^4_{\ms}}{8(4\pi)^2}\left[\frac{1}{\epsilon}+\frac{3}{2}+\ln\left(\frac{4\Lambda^2}{g^2_{\ms}\overline{\sigma}^2_{\ms}}\right)\right]\;$ as in Ref.~\cite{RaiTiw}. One can define the scale $\Lambda$ independent parameter $\Delta=\frac{g_{\ms}\overline{\sigma}_{\ms}}{2}$ using the Eq.~(\ref{params6}) and Eq.~(\ref{params7}) and rewrite the vacuum effective potential in terms of it as

\bqa
\label{rqmefsc}
\Omega_{vac}(\Delta)&=&2\left(\frac{m^2_0}{g^2_0}-\frac{c_0}{g^2_0}\right)\Delta^2+4\left(\frac{\lambda_{10}}{g^4_0}+\frac{\lambda_{20}}{2g^4_0}\right)\Delta^4-2\frac{h_0}{g_0} \Delta+\frac{2N_c\Delta^4}{(4\pi)^2}\left[\frac{3}{2}+\ln\left(\frac{\Lambda^2}{\Delta^2}\right)\right]\;.
\eqa
Expressing the couplings and mass parameter in terms of the Yukawa coupling, pion decay constant and physical meson masses, one can write 
\bqa
\label{rqmeff}
\nonumber
\Omega_{vac}(\Delta)
&=&\frac{(3m^2_\pi-m^2_\sigma)f^2_{\pi}}{4}\left\lbrace 1-\frac{N_cg^2}{(4\pi)^2}\left(\mathcal{C}(m^2_\pi)+m^2_\pi\mathcal{C}^\prime(m^2_\pi)\right)\right\rbrace\frac{\Delta^2}{m^2_q}+\frac{N_cg^2f^2_\pi}{2(4\pi)^2}\left\lbrace\frac{3m^2_\pi\mathcal{C}(m^2_\pi)-(m^2_\sigma-4m^2_q)\mathcal{C}(m^2_\sigma)}{2}-2m^2_q\right\rbrace\frac{\Delta^2}{m^2_q}\;\\ \nonumber
&&+\frac{(m^2_\sigma-m^2_\pi)f^2_\pi}{8}\left\lbrace1-\frac{N_cg^2}{(4\pi)^2}\left(\mathcal{C}(m^2_\pi)+m^2_\pi\mathcal{C}^\prime(m^2_\pi)\right)\right\rbrace\frac{\Delta^4}{m^4_q}+\frac{N_cg^2f^2_\pi}{(4\pi)^2}\left[(m^2_\sigma-4m^2_q)\mathcal{C}(m^2_\sigma)-m^2_\pi\mathcal{C}(m^2_\pi)\over 8\right]\frac{\Delta^4}{m^4_q}\;\\
&&+\frac{2N_c\Delta^4}{(4\pi)^2}\left\lbrace\frac{3}{2}-\ln\left(\frac{\Delta^2}{m^2_q}\right)\right\rbrace-m^2_\pi f^2_\pi\left\lbrace 1-\frac{N_cg^2}{(4\pi)^2}m^2_\pi\mathcal{C}^\prime(m^2_\pi)\right\rbrace\frac{\Delta}{m_q}\;
\eqa
\end{widetext}
Here it is relevant to remind ourselves that the Yukawa coupling and the pion decay constant  both get renormalized in the vacuum  due to the dressing of the meson propagator in the RQM model.~However, the Eq. (\ref{params6}) gives $g_{\ms}=g_{ren}=g$  and the Eq. (\ref{params7}) gives $\overline{\sigma}_{\ms}= f_{\pi, ren}=f_{\pi}$ at the scale $\Lambda_0$.~Applying the stationarity condition  $\frac{\partial \Omega_{vac}(\Delta)}{\partial \Delta}$ to the Eq.(\ref{rqmefsc}),~one gets  $h_{0}=m_{\pi,c}^2 \ \overline{\sigma}_{\ms} =m^2_\pi \left\lbrace 1-\frac{N_cg^2}{(4\pi)^2}m^2_\pi\mathcal{C}^\prime(m^2_\pi)\right\rbrace f_\pi$.~Note that the pion curvature mass $ m_{\pi,c}$ differs from its pole mass $m_{\pi} $ on account of the consistent parameter fixing and we have $ m_{\pi,c}^2 =m^2_\pi \left\lbrace 1-\frac{N_cg^2}{(4\pi)^2}m^2_\pi\mathcal{C}^\prime(m^2_\pi)\right\rbrace$ as in Ref. \cite{fix1}.~The  minimum of the vacuum effective potential lies at $\overline{\sigma}_{\ms}=f_{\pi}$.
\section{INTEGRALS}
\label{appenC}
The divergent loop integrals are regularized by encorporating dimensional regularization.
\bqa
\int_p=\left(\frac{e^{\gamma_E}\Lambda^2}{4\pi}\right)^\epsilon\int \frac{d^dp}{(2\pi)^d}\;,
\eqa
where $d=4-2\epsilon$ , $\gamma_E$ is the Euler-Mascheroni constant, and $\Lambda$ is renormalization scale associated with the $\overline{\text{MS}}$.
\bqa
\nonumber
\mathcal{A}(m^2_q)&=&\int_p \frac{1}{p^2-m^2_q}=\frac{i m^2_q}{(4\pi)^2}\left[\frac{1}{\epsilon}+1\right. \\
\nonumber
&&\left.+\log(4\pi e^{-\gamma_E})+\log\left(\frac{\Lambda^2}{m^2_q}\right)\right]\;
\eqa
we rewrite this after redefining $\Lambda^2\longrightarrow \Lambda^2\frac{e^{\gamma_E}}{4\pi}$.
\bqa
\label{aint1}
\mathcal{A}(m^2_q)&=&\frac{i m^2_q}{(4\pi)^2}\left[\frac{1}{\epsilon}+1+\log\left(\frac{\Lambda^2}{m^2_q}\right)\right]
\\
\label{bint1}
\nonumber
\mathcal{B}(p^2)&=&\int_k \frac{1}{(k^2-m^2_q)[(k+p)^2-m^2_q)]} \\
&=&\frac{i}{(4\pi)^2}\left[\frac{1}{\epsilon}+\log\left(\frac{\Lambda^2}{m^2_q}\right)+C(p^2)\right]
\\
\label{bprimeint1}
\mathcal{B}^\prime(p^2)&=&\frac{i}{(4\pi)^2}C^\prime(p^2)
\eqa
\begin{widetext}
\begin{equation}
\label{eq:cp1}
\mathcal{C}(p^2)=2-2\sqrt{\dfrac{4 m^2_f}{p^2}-1}\arctan\left(\dfrac{1}{\sqrt{\dfrac{4 m^2_f}{p^2}-1}}\right); \quad  \mathcal{C}^{\prime}(p^2)=\frac{4 m^2_f}{p^4\sqrt{\dfrac{4 m^2_f}{p^2}-1}}\arctan\left(\dfrac{1}{\sqrt{\dfrac{4 m^2_f}{p^2}-1}}\right)-\frac{1}{p^2}\;,
\end{equation}
\begin{equation}
\label{eq:cp2}
\mathcal{C}(p^2)=2+\sqrt{1-\dfrac{4 m^2_f}{p^2}}\ln\left(\dfrac{1-\sqrt{1-\dfrac{4 m^2_f}{p^2}}}{1+\sqrt{1-\dfrac{4 m^2_f}{p^2}}}\right); \quad \mathcal{C}^{\prime}(p^2)=\frac{2 m^2_f}{p^4\sqrt{\dfrac{4 m^2_f}{p^2}-1}}\ln\left(\dfrac{1-\sqrt{1-\dfrac{4 m^2_f}{p^2}}}{1+\sqrt{1-\dfrac{4 m^2_f}{p^2}}}\right)-\frac{1}{p^2}\;,
\end{equation}
The Eqs.~(\ref{eq:cp1}) and (\ref{eq:cp2}) are valid with the constraints  ($p^2<4m^2_f$) and ($p^2>4m^2_f$) respectively. 
\end{widetext}


\bibliographystyle{apsrmp4-1}

\end{document}